\documentclass{tlp}

\usepackage{amsmath}
\usepackage{stmaryrd}
\usepackage[latin1]{inputenc}
\usepackage{graphicx}
\usepackage{listings}
\usepackage{latexsym}
\usepackage{epic}
\usepackage{eepic}
\usepackage{environ}

\usepackage[T1]{fontenc}
\usepackage{ae, aecompl} 
\usepackage[scaled=0.80]{beramono} 

\usepackage{tikz}
\usetikzlibrary{arrows,shapes,shadows,calc}

\newtheorem{example}{Example}[section]

\newcommand{\intelpict}[3]{
  \begin{figure}
    \centering
    \includegraphics[width=0.8\linewidth]{clop_eps__plot-default-#1}
    \caption{#2 --- Intel.}
    \label{#3}
  \end{figure}
}

\newcommand{\ppcpict}[3]{
  \begin{figure}[p]
    \centering
    \begin{minipage}[t]{0.49\linewidth}
    \includegraphics[width=\linewidth]{ppc_eps__plot-default-#1}
    \caption{#2 --- PowerPC.}
    \label{#3}
    \end{minipage}
  \end{figure}
}

\newcommand{\ppcpicttwo}[6]{
  \begin{figure}[p]
    \centering
    \begin{minipage}[t]{0.49\linewidth}
    \includegraphics[width=\linewidth]{ppc_eps__plot-default-#1}
    \caption{#2 --- PowerPC.}
    \label{#3}
    \end{minipage}
    \begin{minipage}[t]{0.49\linewidth}
    \includegraphics[width=\linewidth]{ppc_eps__plot-default-#4}
    \caption{#5 --- PowerPC.}
    \label{#6}
    \end{minipage}
  \end{figure}
}

\NewEnviron{ruleset}[2]{%
  \begin{figure}[bt]
    \fbox{
      \scalebox{0.85}{
        \begin{minipage}[t]{1.10\textwidth}
          \BODY
        \end{minipage}
      }}
    \caption{#1}
    \label{#2}
  \end{figure}
}

\newcommand{\rulebegin}[1]{(\textsc{#1})$~$}
\newcommand{\rulename}[1]{(\textsc{#1})}

\newenvironment{quotedcode}{

\hspace{2em}\begin{minipage}[t]{0.8\textwidth}
}{
\end{minipage}

}
\newenvironment{quotedterm}{

\hspace{2em}\begin{minipage}[t]{0.8\textwidth}
}{
\end{minipage}

}

\newcommand{\ip}{\textsf{{imProlog}}}

\newcommand{\lang}[1]{\ensuremath{{\cal L}_{#1}}}
\newcommand{\langrefl}[1]{\ensuremath{{\cal L}_{#1}^{r}}}
\newcommand{\la}{\lang{a}}
\newcommand{\lb}{\lang{b}}
\newcommand{\lc}{\lang{c}}
\newcommand{\lsource}{\lang{s}}
\newcommand{\lsourcerefl}{\langrefl{s}}
\newcommand{\lint}{\lang{i}}
\newcommand{\lintrefl}{\langrefl{i}}
\newcommand{\machall}{\ensuremath{\mathcal{M}}}
\newcommand{\mach}[1]{\ensuremath{\machall_{\it #1}}}

\newcommand{\quoted}[1]{`#1'}

\newcommand{\emudef}{\ensuremath{\mathcal{E}}}

\newcommand{\mgen}{\textit{mgen}}
\newcommand{\emucomp}{\textit{emucomp}}

\newcommand{\papertitle}{Description and Optimization of Abstract
  Machines in a Dialect of Prolog}

\title[Theory and Practice of Logic Programming]
{\papertitle\footnote{This is a significantly extended and revised
    version of the 
    paper ``Towards Description and Optimization of Abstract Machines
    in an Extension of Prolog'' published in the proceedings of the
    \emph{2006 International Symposium on Logic-based Program
      Synthesis and Transformation}
    (LOPSTR'06)~\cite{morales06:improlog-lopstr}.}}

\author[J. F. Morales, M. Carro, and M. Hermenegildo]{
     JOS\'{E} F. MORALES$^1$
  ~~ MANUEL CARRO$^{1,2}$
  ~~ MANUEL HERMENEGILDO$^{1,2}$ \\
  $^1$ IMDEA Software Institute \\
  \email{\{josef.morales,manuel.carro,manuel.hermenegildo\}@imdea.org} \\
  $^2$ School of Computer Science, Technical University of Madrid (UPM)\\
  \email{\{mcarro,herme\}@fi.upm.es}}

\date{}

\submitted{26 January 2007}
\revised{8 July 2009, 20 April 2014}
\accepted{26 June 2014}

\newcommand{\loexpr}[1]{\textsf{#1}}

\begin{document}

\label{firstpage}

\maketitle

\textbf{Note:} To appear in Theory and Practice of Logic Programming (TPLP)\\
\hrule
\begin{abstract}
  In order to achieve competitive performance, abstract machines for
  Prolog and related languages end up being large and intricate, and
  incorporate sophisticated optimizations, both at the design and at
  the implementation levels.  At the same time, efficiency
  considerations make it necessary to use low-level languages in their
  implementation. 
  This makes them laborious to code, optimize, and, especially,
  maintain and extend.  Writing the abstract machine (and ancillary
  code) in a higher-level language can help tame this inherent
  complexity.
  We show how the semantics of most basic components of an
  efficient virtual machine for Prolog can be described using (a
  variant of) Prolog. 
  These descriptions are then compiled to C and assembled to build a
  complete bytecode emulator.  Thanks to the high level of the
  language used and its closeness to Prolog, the abstract machine
  description can be manipulated using standard Prolog compilation
  and optimization techniques with relative ease.  We also show how,
  by applying program transformations selectively, we obtain abstract
  machine implementations whose performance can match and even exceed
  that of state-of-the-art, highly-tuned, hand-crafted emulators.
\end{abstract}


\section{Introduction}
\label{sec:intro}

Abstract machines have proved themselves very useful when
defining theoretical models  
and implementations of software and hardware systems. In particular,
they have been widely used to define execution models and as
implementation vehicles for many languages, most frequently in
functional and logic programming, and more recently also in
object-oriented programming, with the Java abstract
machine~\cite{gosling_jls_3} being a very popular recent example.
There are also early examples of the use of abstract machines in
traditional procedural languages (e.g., the P-Code used as a target in
early Pascal implementations~\cite{nori81:pascal_P}) and in other
realms such as, for example, operating systems (e.g., Dis, the virtual
machine for the Inferno operating system~\cite{dorward97:inferno}).

The practical applications of the indirect execution mechanism that an
abstract machine represents are countless: portability, generally
small executables, simpler security control through sandboxing,
increased compilation speed, etc.
However, unless the abstract machine implementation 
is highly optimized, these benefits can come at the cost of poor 
performance.  Designing and implementing fast, resource-friendly,
competitive abstract machines is a complex task.  This is especially so
in the case of programming languages where there is a wide gap
between many of their basic constructs and features and 
what is available in the underlying off-the-shelf hardware:
term representation vs.\ memory words, unification vs.\ assignment,
automatic vs.\ manual memory management, destructive vs.\
non-destructive updates, backtracking and tabling vs.\ Von
Neumann-style control flow, etc. 
In addition, the extensive code optimizations required to achieve good
performance make development and, especially, maintenance and further
modifications non-trivial.  Implementing or testing new optimizations
is often involved, as decisions previously taken need to be revisited,
and low-level, tedious recoding is often necessary to test a new idea.

Improved performance has been achieved
by post-processing the input program (often called \emph{bytecode}) of
the abstract machine (\emph{emulator}) and generating efficient native
code --- sometimes achieving performance that is very close to that 
of an implementation of the source program directly written in C. 
This is technically challenging and the overall picture is even more
complex when bytecode and native code are mixed, usually by dynamic
recompilation. This in principle
combines the best of both worlds
by deciding
when and how native code (which may be large and/or costly
to obtain) is generated based on runtime analysis of the program
execution, while leaving the rest as bytecode.  Some examples are the
Java HotSpot VM~\cite{Paleczny:hotspot}, the Psyco~\cite{rigo:psyco}
extension for Python, or for logic programming, all-argument predicate
indexing in recent versions of Yap Prolog~\cite{vitor07:indexing}, the
dynamic recompilation of asserted predicates in
BinProlog~\cite{binprolog2006}, etc.  Note, however, that the initial
complexity of the virtual machine and all of its associated
maintenance issues have not disappeared, and emulator
designers and maintainers still have to struggle with thousands of 
lines of low-level code.

In this paper we explore the possibility of 
rewriting most of the runtime and virtual machine in the high-level source
language (or a close dialect of it), and then 
use all the available compilation machinery to
obtain native code from it.  Ideally, this native code should provide
comparable performance to that of hand-crafted code,
while keeping the size of low-level coded parts in the 
system to a minimum.
This is the approach taken herein, where we explore
writing Prolog emulators in a Prolog dialect. As we will 
see later, the approach is interesting not only for simplifying the
task of developers  but also for widening 
the application domain of the language to other kinds of problems which
extend beyond just emulators, such as 
reusing analysis and transformation passes, and making it easier
to automate tedious optimization techniques for emulators, such as
specializing emulator instructions.
The advantages of using a higher-level language are rooted on one
hand in the capability of hiding implementation details that a
higher-level language provides, and on the other hand in its
amenability to transformation and manipulation. 
These, as we will see, are key for our goals, as they reduce
error-prone, tedious programming work, while making it possible to
describe at the same time complex abstract machines in a concise and
correct way.

A similar objective has been pursued elsewhere.  For example, the
JavaInJava~\cite{taivalsaari98:JavaInJava} and
PyPy~\cite{rigo06:pypy_approach} projects have similar
goals.  The initial performance figures reported for these implementations 
highlight how challenging it is to make them 
competitive with existing hand-tuned abstract machines:
JavaInJava started with an initial slowdown of
approximately 700 times w.r.t.\ then-current implementations, and PyPy
started at the 2000$\times$ slowdown level.  Competitive execution times
were only possible after changes in the source language and
compilation tool chain,  by restricting it or adding
annotations. For
example, the slowdown of PyPy was reduced to 3.5$\times$ -- 11.0$\times$
w.r.t.\ CPython~\cite{rigo06:pypy_approach}. 
These results can be partially traced back to the attempt to reimplement
the whole machinery at once, which has the disadvantage of making such
a slowdown almost inevitable.  This makes it difficult to use the
generated virtual machines as ``production'' software (which would
therefore be routinely tested) and, especially, it makes it difficult
to study how a certain, non-local optimization will carry over to a
complete, optimized abstract machine.

Therefore, 
we chose an alternative approach: gradually porting
selected key components (such as, e.g., instruction definitions) 
and combining this port with other emulator generation
techniques~\cite{morales05:generic_eff_AM_implem_iclp}.  At every step
we made sure experimentally\footnote{And also with stronger means: for
some core components we checked that the binary code produced from the
high-level definition of the abstract machine and that coming from the
hand-written one were identical.} that the performance of the original
emulator was maintained throughout the process.  The final result is
an emulator that is 
completely written in a high-level language and which, when
compiled to native code, does not lose any performance w.r.t.\ a
manually written emulator. Also, and as a very relevant byproduct, we 
develop a language (and a compiler for it) which is high-level enough 
to make several 
program transformation and analysis techniques applicable, while
offering the possibility of being compiled into efficient native
code.  While this language is general-purpose and can be used
to implement  arbitrary programs, throughout this paper we will focus
on its use in writing abstract machines and to easily generate
variations of such machines.

The rest of the paper proceeds as follows:
Section~\ref{sec:overv-extend-arch} gives an overview of the different
parts of our compilation architecture and information flow and
compares it with a more traditional setup.  Section~\ref{sec:improlog}
presents the source language with which our abstract
machine is written, and justifies the design decisions (e.g., typing,
destructive updates, etc.) based on the needs of applications which
demand high performance.  Section~\ref{sec:cgen} summarizes how
compilation to efficient native code is done through C.
Section~\ref{sec:generating-emul-main} describes language extensions
which were devised specifically to write abstract machines (in
particular, the WAM) and Section~\ref{sec:automatic-spec-merg}
explores how the added flexibility of the high-level language approach
can be taken advantage of in order to easily generate variants of
abstract machines with different core characteristics. This section
also studies experimentally the performance that can be attained with
these variants. Finally, Section~\ref{sec:conclusions} presents our
conclusions.

\section{Overview of our Compilation Architecture}
\label{sec:overv-extend-arch}

\def\progtext#1#2#3{%
\begin{minipage}{0.20\textwidth}%
\textbf{#2}\\%
#1\\%
#3%
\end{minipage}%
}

\def\progtextq#1#2{%
\begin{minipage}{0.20\textwidth}%
#1\\%
#2%
\end{minipage}%
}

\def\proglang#1#2{%
#1:#2%
}

The compilation architecture we present here uses several
languages, language translators, and program representations which
must be defined and placed in the ``big picture.''
For generality, and since those elements are common to most
bytecode-based systems, we will refer to them by more abstract names
when possible or convenient, although in our initial discussion we
will equate them with the actual languages in our production environment.

The starting point in this work is the 
Ciao system~\cite{hermenegildo11:ciao-design-tplp}, which includes an
efficient, WAM-based~\cite{Warren83,hassan-wamtutorial}, abstract
machine coded in C (an independent evolution which forked from the 
SICStus 0.5/0.7 virtual machine), a
compiler to bytecode with an experimental extension 
to emit optimized C code~\cite{morales04:p-to-c-padl}, and the CiaoPP
preprocessor, a program analysis, specialization and transformation
framework~\cite{aadebug97-informal,prog-glob-an,assrt-theoret-framework-lopstr99,ciaopp-sas03-journal-scp}.

\begin{figure}[t]
\pgfdeclarelayer{background}
\pgfdeclarelayer{foreground}
\pgfsetlayers{background,main,foreground}
\scalebox{0.85}{
\begin{tikzpicture}[
  font=\rm\small, node distance=1.5cm, auto, >=latex', thick
  ]
    \tikzstyle{program} = [draw, thin, text centered, fill=white, drop shadow]
    \tikzstyle{translator} = [draw, fill=black!20, text width=5em,
      text badly centered, rounded corners]
    \tikzstyle{translatorp} = [draw, fill=black!20, line width=1.5pt, text width=5em,
      text badly centered, rounded corners]
    \tikzstyle{conclusion} = [text centered]
    \tikzstyle{line} = [->]
    \node (traditional) {\textsc{Traditional approach}};
    \node[program, below of=traditional, xshift=3cm,yshift=-0.25cm]
    (einterhc) {\progtext{\proglang{E}{\lc}}{\lb\ emulator}{low-level
        code\\(hand written)}};
    \draw[->, line width=2pt] (traditional) -- (einterhc);
    \node[program, below of=traditional, xshift=-3cm,yshift=-0.25cm] (psource)
    {\progtext{\proglang{P}{\lsource}}{user program}{source code\\[1ex] \ }};
    \draw[->, line width=2pt] (traditional) -- (psource);
    \node[translator, below of=psource] (tcomp1)  {Front-end compiler};
    \draw[->] (psource) edge node {} (tcomp1);
    \node[program, below of=tcomp1] (pinter) {\progtextq{\proglang{P}{\lint}}{intermediate code}};
    \draw[->] (tcomp1) edge node {} (pinter);
    \node[translator, below of=pinter, xshift=-2cm]  (tcomp2)  {Bytecode back-end};
    \path[line] (pinter) edge [out=-90, in=90] node {} (tcomp2);
    \node[program, below of=tcomp2] (psym) {\progtextq{\proglang{P}{\la}}{symbolic\\bytecode}};
    \path[line] (tcomp2) edge node {} (psym);
    \node[translator, below of=psym] (tcomp3) {Bytecode assembler};
    \path[line] (psym) edge node {} (tcomp3);
    \node[program, below of=tcomp3] (pbyte) {\progtextq{\proglang{P}{\lb}}{bytecode}};
    \path[line] (tcomp3) edge node {} (pbyte);
    \node[translator, xshift=2.0cm, right of=tcomp2] (tcompnat) {Low-level back-end};
    \path[line] (pinter) edge [out=-90, in=90] node {} (tcompnat);
    \node[program, xshift=2.0cm, right of=psym] (pnat) {\progtextq{\proglang{P}{\lc}}{low-level code}};
    \path[line] (tcompnat) edge node {} (pnat);
    %
    \node[conclusion, right of=pbyte, xshift=4cm] (conclusion)
    {\begin{minipage}{0.5\textwidth}
        Emulation and native compilation must exhibit the same behaviour: \\
        (\proglang{E}{\lc} $\circ$ \proglang{P}{\lb}) $\equiv$ (\proglang{P}{\lc})
    \end{minipage}};
\end{tikzpicture}}
\caption{Traditional compilation architecture.}
\label{fig:lang-levels}
\end{figure}
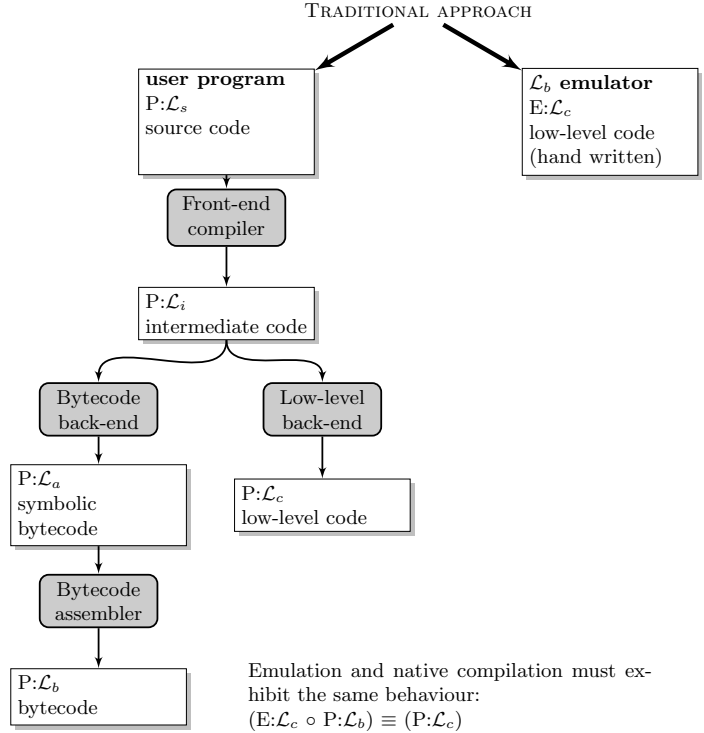

We will denote the source Prolog language
as \lsource, the symbolic WAM code as \la, the byte-encoded WAM code as \lb,
and the C language as \lc.  The different languages and translation processes
described in this section are typically found in most Prolog systems, and
this scheme could be easily applied to other situations.
We will use \proglang{N}{L} to denote the program $N$ written
in language $L$. Thus, we can distinguish in the \emph{traditional
  approach} (Figure~\ref{fig:lang-levels}):

\begin{description}
\item[Front-end compiler:] \proglang{P}{\lsource} is compiled to
  \proglang{P}{\lint}, where \lsource\ is the source language and
  \lint\ is an intermediate representation. For simplicity, we assume 
  that this phase includes any analysis, transformation, and optimization.
\item[Bytecode back-end:] \proglang{P}{\lint} is translated to
  \proglang{P}{\la}, where \la\ is the symbolic representation of the
  bytecode language.
\item[Bytecode assembler:] \proglang{P}{\la} is encoded into
  \proglang{P}{\lb}, where \lb\ defines encoded bytecode streams (a
  sequence of bytes whose design is focused on interpretation speed).
\end{description}

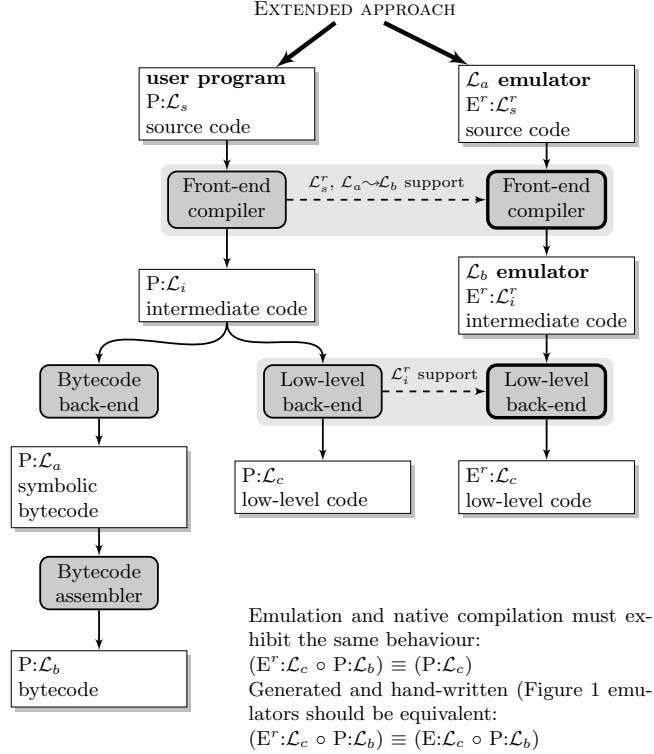
\begin{figure}[t]
\pgfdeclarelayer{background}
\pgfdeclarelayer{foreground}
\pgfsetlayers{background,main,foreground}
\scalebox{0.85}{
\begin{tikzpicture}[
  font=\rm\small, node distance=1.5cm, auto, >=latex', thick
  ]
    \tikzstyle{program} = [draw, thin, text centered, fill=white, drop shadow]
    \tikzstyle{translator} = [draw, fill=black!20, text width=5em,
      text badly centered, rounded corners]
    \tikzstyle{translatorp} = [draw, fill=black!20, line width=1.5pt, text width=5em,
      text badly centered, rounded corners]
    \tikzstyle{conclusion} = [text centered]
    \tikzstyle{line} = [->]
    \node (extended) {\textsc{Extended approach}};
    \node[program, below of=extended, xshift=-2cm] (psource) {\progtext{\proglang{P}{\lsource}}{user program}{source code}};
    \draw[->, line width=2pt] (extended) -- (psource);
    \node[translator, below of=psource] (tcomp1)  {Front-end compiler};
    \draw[->] (psource) edge node {} (tcomp1);
    \node[program, below of=tcomp1] (pinter) {\progtextq{\proglang{P}{\lint}}{intermediate code}};
    \draw[->] (tcomp1) edge node {} (pinter);
    \node[translator, below of=pinter, xshift=-2cm]  (tcomp2)  {Bytecode back-end};
    \path[line] (pinter) edge [out=-90, in=90] node {} (tcomp2);
    \node[program, below of=tcomp2] (psym) {\progtextq{\proglang{P}{\la}}{symbolic\\bytecode}};
    \path[line] (tcomp2) edge node {} (psym);
    \node[translator, below of=psym] (tcomp3) {Bytecode assembler};
    \path[line] (psym) edge node {} (tcomp3);
    \node[program, below of=tcomp3] (pbyte) {\progtextq{\proglang{P}{\lb}}{bytecode}};
    \path[line] (tcomp3) edge node {} (pbyte);
    \node[translator, xshift=2.0cm, right of=tcomp2] (tcompnat) {Low-level back-end};
    \path[line] (pinter) edge [out=-90, in=90] node {} (tcompnat);
    \node[program, xshift=2.0cm, right of=psym] (pnat) {\progtextq{\proglang{P}{\lc}}{low-level code}};
    \path[line] (tcompnat) edge node {} (pnat);
    \node[program, xshift=3.5cm, right of=psource] (esource) {\progtext{\proglang{E$^r$}{\lsourcerefl}}{\la\ emulator}{source code}};
    \draw[->, line width=2pt] (extended) -- (esource);
    \node[program, xshift=3.5cm, right of=pinter] (einter) {\progtext{\proglang{E$^r$}{\lintrefl}}{\lb\ emulator}{intermediate code}};
    \node[translatorp, below of=esource] (tcomp1p) {Front-end compiler};
    \draw[->] (tcomp1) edge [dashed] node {\scriptsize \lsourcerefl, \la$\leadsto\!$\lb\ support} (tcomp1p);
    \path[line] (esource) edge node {} (tcomp1p);
    \path[line] (tcomp1p) edge node {} (einter);
    \node[program, xshift=5.5cm, right of=psym] (elow) {\progtextq{\proglang{E$^r$}{\lc}}{low-level code}};
    \node[translatorp, below of=einter] (tcompnatp) {Low-level back-end};
    \draw[->] (tcompnat) edge [dashed] node {\scriptsize \lintrefl\ support} (tcompnatp);
    \path[line] (einter) edge node {} (tcompnatp);
    \path[line] (tcompnatp) edge node {} (elow);
    \node[conclusion, right of=pbyte, xshift=4cm] (conclusion)
    {\begin{minipage}{0.5\textwidth}
        Emulation and native compilation must exhibit the same behaviour: \\
        (\proglang{E$^r$}{\lc} $\circ$ \proglang{P}{\lb}) $\equiv$ (\proglang{P}{\lc}) \\
        Generated and hand-written (Figure~\ref{fig:lang-levels}
        emulators should be equivalent: \\ 
        (\proglang{E$^r$}{\lc} $\circ$ \proglang{P}{\lb}) $\equiv$ (\proglang{E}{\lc} $\circ$ \proglang{P}{\lb})
    \end{minipage}};
    \begin{pgfonlayer}{background}
      \path (tcomp1.north west)+(-0.1, 0.1) node (a) {};
      \path (tcomp1p.south east)+(0.1, -0.1) node (b) {};
      \path[rounded corners, fill=black!10] (a) rectangle (b);
      \path (tcompnat.north west)+(-0.1, 0.1) node (c) {};
      \path (tcompnatp.south east)+(0.1, -0.1) node (d) {};
      \path[rounded corners, fill=black!10] (c) rectangle (d);
    \end{pgfonlayer}
\end{tikzpicture}}
\caption{Extended compilation architecture.}
\label{fig:lang-levels-new}
\end{figure}

To execute \lb\ programs, a hand-written emulator
\proglang{E}{\lc} (where \lc\ is a lower-level language) is
required, in addition to some runtime code (written in \lc).
Alternatively, it is possible to translate intermediate code by means
of a \textbf{low level back end}, which compiles a program
\proglang{P}{\lint} directly
into its,\proglang{P}{\lc} equivalent.\footnote{Note that in JIT systems
  the low-level code is generated from the encoded bytecode
  representation. For simplicity we keep the figure for static code
  generation, which takes elements of \lint\ as the input.} 
This avoids the need for any emulator if all programs are compiled
to native code, but additional runtime support is usually necessary.

The initial \emph{classical architecture} needs manual coding of a
large and complex piece of code, the emulator, using a low-level
language, often missing the opportunity to reuse compilation
techniques that may have been developed for high-level languages, not
only to improve efficiency, but also program readability, correctness,
etc.

In the \emph{extended compilation scheme} we assume that
we can design a dialect from \lsource\ and write the emulator for
symbolic \la\ code, instead of byte-encoded \lb, in that language.  We
will call \lsourcerefl\ the extended language and
\proglang{E$^r$}{\lsourcerefl} the emulator.  Note that the mechanical
generation of an interpreter for \lb\  from an interpreter for \la\  was
previously described and successfully tested in an \emph{emulator
  generator}~\cite{morales05:generic_eff_AM_implem_iclp}.
Adopting it for this work was perceived as a correct decision, since it
moves low-level implementation (and bytecode representation) issues to
a translation phase, thus reducing the requirements on the language to
preserve emulator efficiency, and in practice making the code
easier to manage.
The new translation pipeline, depicted in the \emph{extended approach}
path (Figure~\ref{fig:lang-levels-new}) shows the following
processes (the dashed lines represent the modifications that some of
the elements undergo with respect to the original ones): 

\begin{description}
\item[Extended front-end compiler:] it compiles \lsourcerefl\ programs
  into \lintrefl\ programs (where \lintrefl\ is the intermediate
  language with extensions required to produce efficient native code).
  This compiler includes the 
  \emph{emulator generator}. That framework makes it possible to write
  instruction definitions for an \la\ emulator, plus separate bytecode
  encoding rules, and process them together to obtain an \lb\
  emulator.  I.e., it translates \proglang{E$^r$}{\lsourcerefl} to an
  \proglang{E$^r$}{\lintrefl} emulator for \lb.
\item[Extended low-level back-end:] it compiles \lintrefl\ programs into
  \lc\ programs.  The resulting \proglang{E$^r$}{\lintrefl} is finally
  translated to \proglang{E$^r$}{\lc}, which should be equivalent (in the
  sense of producing the same output) to \proglang{E}{\lc}.
\end{description}
The advantage of this scheme lies in its flexibility for sharing
optimizations, analyses, and transformations at different compilation
stages (e.g., the same machinery for partial evaluation,
constant propagation, common subexpression elimination, etc. can be
reused), which are normally reimplemented for high- and low-level
languages.

\section{The \ip\ Language}
\label{sec:bottom-up}
\label{sec:improlog}

We describe in this section our \lsourcerefl\ language, \ip, and the
analysis and code generation techniques used to compile it into highly
efficient code.\footnote{The name \ip\ stands for \emph{im}perative
  \emph{Prolog}, because its purpose is to make typically imperative
  algorithms easier to express in Prolog, but minimizing and
  controlling the scope of impurities.}
This Prolog variant is motivated by the following problems:
\begin{itemize}
\item It is hard to guarantee that certain overheads in Prolog that
  are directly related with the language expressiveness (e.g., boxed
  data for dynamic typing, trailing for non-determinism, uninstantiated
  free variables, multiple-precision arithmetic, etc.) will always be
  removed by compile-time techniques. 
\item Even if that overhead could be eliminated, there is also the
  cost of some basic operations, such as modifying a single attribute
  in a custom data structure, which is not constant in the declarative
  subset of Prolog.  For example, the cost of replacing the value of
  an argument in a Prolog structure is, in most straightforward
  implementations, linear w.r.t.\ the number of arguments of the
  structure, since a certain amount of copying of the structure spine
  is typically involved. In contrast, replacing an element in a C
  structure is a constant-time operation. Again, while this overhead
  can be optimized away in some cases during compilation, the
  objective is to define a language layer in which constant-time
  can be ensured for these operations.
\end{itemize} 

\noindent We now present the different elements that comprise the
\ip\ language  
and we will gradually introduce a number of restrictions on the kind
of programs which we admit as valid.  The main reason to impose these
restrictions is to achieve the central goal in this paper: generating
efficient emulators starting from a high-level language.

In a nutshell, the \ip\ language both restricts and extends Prolog.
The impure features (e.g., the dynamic database) of Prolog are not part of
\ip\ and only programs meeting some strict requirements
about determinism, modes in the unification, and others, are allowed.
In a somewhat circular fashion, the requirements we impose on these
programs are those which allow us to compile them into efficient code.
Therefore, implementation matters somewhat influence the design and
semantics of the language.
On the other hand, \ip\ also extends Prolog in order to provide a
single, well-defined, and amenable to analysis mechanism to implement
constant-time access to data: a specific kind of mutable variables.  
Thanks to the restrictions 
aforementioned and this mechanism \ip\ programs can be compiled into
very efficient low-level (e.g., C) code.

Since we are starting from Prolog, which is well understood, and the
restrictions on the admissible programs can be introduced painlessly
(they just reduce the set of programs which are deemed valid by the
compiler), we will start by showing, in the next section, how we
tackle efficient data handling, which as we mentioned departs
significantly, but in a controlled way, from the semantics of Prolog.

\paragraph{Notation.}
We will use lowercase math font for variables
(\textit{x}, \textit{y}, \textit{z}, $...$) in the rules that
describe the compilation and language semantics. Prolog variable names
will be written in math capital font (\textit{X}, \textit{Y},
\textit{Z}, $...$). Keywords and identifiers in the target C language
use bold text (\texttt{\textbf{return}}). Finally, sans serif text is used
for other names and identifiers (\textsf{f}, \textsf{g},
\textsf{h}, $...$).
The typography will make it possible to easily distinguish a compilation pattern
for \quoted{\textsf{f}(\textit{a})}, where \quoted{\textit{a}} may be
any valid term, and \quoted{\textsf{f}(\textit{A})}, where
\quoted{\textit{A}} is a Prolog variable with the concrete name
``A.''
Similarly, the expression \textit{f}(\textit{a}$_1$, \textsf{...}, \textit{a}$_n$) denotes any
structure with functor name  \textit{f} and \textit{n} arguments,
whatever they may be.
It differs from \textsf{f}(\textit{A}$_1$, \textsf{...}, \textit{A}$_n$), where the functor name is
fixed to \textsf{f} and the arguments are variables).  If
\textit{n} is 0, the previous expression is tantamount to just
\textit{f}.

\subsection{Efficient Mechanisms for Data Access and Update}
\label{sec:ip-mut}

In this section we will describe the formal semantics of \emph{typed
  mutable variables}, our proposal for providing efficient (in terms of
time and memory) data handling in \ip.  These variables feature
backtrackable destructive assignment and are accessible and updatable
in constant time through the use of a unique, associated identifier.
This is relevant for us as it is required to efficiently implement a
wide variety of algorithms, some of which appear in WAM-based abstract
machines,\footnote{One obvious example is the unification algorithm of
  logical variables, itself based on the Union-Find algorithm. An
  interesting discussion of this point is available
  in~\cite{DBLP:journals/tplp/SchrijversF06}, where a CHR version of
  the Union-Find algorithm is implemented and its complexity studied.}
which we want to describe in enough detail as to obtain efficient
executables.

There certainly exist a number options for implementing constant-time
data access in Prolog.
Dynamic predicates (\textsf{assert} / \textsf{retract}) can in some
cases (maybe with the help of type information) provide constant-time
operations; existing destructive update primitives (such as
\textsf{setarg}\textsf{/}\textsf{3}) can do the same.  However, it is difficult 
for the analyses normally used in the context of logic programming 
to deal with them in a precise way, in a significant part because 
their semantics was not devised with analysis in mind, and
therefore they are difficult to optimize as much as we need herein.

Therefore, we opted for a generalization of mutable variables with
typing constraints as mentioned before.
In our experience, this fits in a less intrusive way with the rest of
the logic framework, and at the same time allows us to generate
efficient code for the purpose of the work in this
paper.\footnote{Note that this differs
  from~\cite{morales06:improlog-lopstr}, where changes in mutable data
  were non-backtrackable side effects. Notwithstanding, performance is
  not affected in this work, since we restrict at compile-time the
  emulator instructions to be deterministic.}
Let us introduce some preliminary definitions before presenting the
semantics of mutable variables:
\begin{description}
\item[Type:] \ensuremath{\tau} is a unary predicate that defines a set
  of values (e.g., regular types as
  in~\cite{Dart-Zobel,gallagher-types-iclp94,eterms-sas02}).  If 
  \ensuremath{\tau}(\textit{v}) holds, then \textit{v} is said to
  contain values of type \ensuremath{\tau}. 
\item[Mutable identifier:] the identifier \textit{id} of a mutable
  is a unique atomic value that uniquely identifies a mutable and does
  not unify with any other non-variable except itself.
\item[Mutable store:]
  \ensuremath{{\varphi}} is a mapping
  \{\textit{id}$_1$\textsf{/}(\ensuremath{\tau}$_1$, \textit{val}$_1$), \textsf{...}, \textit{id}$_n$\textsf{/}(\ensuremath{\tau}$_n$, \textit{val}$_n$)\}, where \textit{id}$_i$ are mutable identifiers,
  \textit{val}$_i$ are terms, and \ensuremath{\tau}$_i$ type names.
  The expression \ensuremath{{\varphi}}(\textit{id}) stands for the pair
  (\ensuremath{\tau}, \textit{val}) associated with \textit{id} in
  \ensuremath{{\varphi}}, while \ensuremath{{\varphi}}[\textit{id}\textsf{/}(\ensuremath{\tau}, \textit{val})] denotes a mapping \ensuremath{{\varphi}'} such that
  \begin{center}
    \vspace{-2em}
    \begin{displaymath}
     \mbox{\ensuremath{{\varphi}'}(\textit{id}$_i$)} = \left\{
           \begin{array}{ll}
             \mbox{(\ensuremath{\tau}, \textit{val})} & \mbox{if } \mbox{\textit{id}$_i$ \textsf{=} \textit{id}} \\
             \mbox{\ensuremath{{\varphi}}(\textit{id}$_i$)}  & \mbox{otherwise}
           \end{array}
     \right.
   \end{displaymath}
\end{center}

  We assume the availability of a function
  \textsf{new\_id}(\ensuremath{{\varphi}}) that obtains a new unique identifier not
  present in \ensuremath{{\varphi}}.
\item[Mutable environment:] By \ensuremath{{\varphi}}$_0$\ensuremath{\leadsto} \ensuremath{{\varphi}}\ensuremath{~\vdash~}\textit{g} we denote a judgment of \textit{g} in the context of
  a mutable environment (\ensuremath{{\varphi}}$_0$, \ensuremath{{\varphi}}).
  The pair (\ensuremath{{\varphi}}$_0$, \ensuremath{{\varphi}}) relates
  the initial and final mutable stores,
  and the interpretation of the \textit{g}
  explicit.
\end{description}

\begin{figure}[t]
  \fbox{
\scalebox{0.85}{
\begin{minipage}{\textwidth}
  \begin{center}
    \rulebegin{M-New}
     \ensuremath{\frac{\begin{array}{l}\mbox{\ensuremath{\vdash~}\ensuremath{\tau}(\textit{v})}~~~\mbox{\textit{id} \textsf{=} \textsf{new\_id}(\ensuremath{{\varphi}}$_0$)}~~~\mbox{\ensuremath{{\varphi}} \textsf{=} \ensuremath{{\varphi}}$_0$[\textit{id}\textsf{/}(\ensuremath{\tau}, \textit{v})]}~~~\mbox{\ensuremath{\vdash~}\textit{m} \textsf{=} \textit{id}}~~~\end{array}}{\begin{array}{l}\mbox{\ensuremath{{\varphi}}$_0$\ensuremath{\leadsto}\ensuremath{{\varphi}}\ensuremath{~\vdash~}\textit{m} \textsf{=} \textsf{initmut}(\ensuremath{\tau}, \textit{v})}~~~\end{array}}}
  \end{center}
  \begin{center}
    \rulebegin{M-Read}
     \ensuremath{\frac{\begin{array}{l}\mbox{\ensuremath{\vdash~}\textit{m} \textsf{=} \textit{id}}~~~\mbox{(\textit{\_}, \textit{v}) \textsf{=} \ensuremath{{\varphi}}(\textit{id})}~~~\mbox{\ensuremath{\vdash~}\textit{x} \textsf{=} \textit{v}}~~~\end{array}}{\begin{array}{l}\mbox{\ensuremath{{\varphi}}\ensuremath{\leadsto}\ensuremath{{\varphi}}\ensuremath{~\vdash~}\textit{x} \textsf{=} \textit{m}\textsf{@}}~~~\end{array}}}
  \end{center}
  \begin{center}
    \rulebegin{M-Assign}
     \ensuremath{\frac{\begin{array}{l}\mbox{\ensuremath{\vdash~}\textit{m} \textsf{=} \textit{id}}~~~\mbox{(\ensuremath{\tau}, \textit{\_}) \textsf{=} \ensuremath{{\varphi}}$_0$(\textit{id})}~~~\mbox{\ensuremath{\vdash~}\ensuremath{\tau}(\textit{v})}~~~\mbox{\ensuremath{{\varphi}} \textsf{=} \ensuremath{{\varphi}}$_0$[\textit{id}\textsf{/}(\ensuremath{\tau}, \textit{v})]}~~~\end{array}}{\begin{array}{l}\mbox{\ensuremath{{\varphi}}$_0$\ensuremath{\leadsto}\ensuremath{{\varphi}}\ensuremath{~\vdash~}\textit{m} \ensuremath{\Leftarrow} \textit{v}}~~~\end{array}}}
  \end{center}
  \begin{center}
    \rulebegin{M-Type}
     \ensuremath{\frac{\begin{array}{l}\mbox{\ensuremath{\vdash~}\textit{m} \textsf{=} \textit{id}}~~~\mbox{(\ensuremath{\tau}, \textit{\_}) \textsf{=} \ensuremath{{\varphi}}(\textit{id})}~~~\end{array}}{\begin{array}{l}\mbox{\ensuremath{{\varphi}}\ensuremath{\leadsto}\ensuremath{{\varphi}}\ensuremath{~\vdash~}\textsf{mut}(\ensuremath{\tau}, \textit{m})}~~~\end{array}}}
  \end{center}
  \begin{center}
    \rulebegin{M-Weak}
     \ensuremath{\frac{\begin{array}{l}\mbox{\ensuremath{\vdash~}\textit{a}}~~~\end{array}}{\begin{array}{l}\mbox{\ensuremath{{\varphi}}\ensuremath{\leadsto}\ensuremath{{\varphi}}\ensuremath{~\vdash~}\textit{a}}~~~\end{array}}}
  \end{center}
  \begin{center}
    \rulebegin{M-Conj}
     \ensuremath{\frac{\begin{array}{l}\mbox{\ensuremath{{\varphi}}$_0$\ensuremath{\leadsto}\ensuremath{{\varphi}}$_1$\ensuremath{~\vdash~}\textit{a}}~~~\mbox{\ensuremath{{\varphi}}$_1$\ensuremath{\leadsto}\ensuremath{{\varphi}}\ensuremath{~\vdash~}\textit{b}}~~~\end{array}}{\begin{array}{l}\mbox{\ensuremath{{\varphi}}$_0$\ensuremath{\leadsto}\ensuremath{{\varphi}}\ensuremath{~\vdash~}(\textit{a} \ensuremath{\wedge} \textit{b})}~~~\end{array}}}
  \end{center}
  \begin{center}
    \rulebegin{M-Disj-1}
     \ensuremath{\frac{\begin{array}{l}\mbox{\ensuremath{{\varphi}}$_0$\ensuremath{\leadsto}\ensuremath{{\varphi}}\ensuremath{~\vdash~}\textit{a}}~~~\end{array}}{\begin{array}{l}\mbox{\ensuremath{{\varphi}}$_0$\ensuremath{\leadsto}\ensuremath{{\varphi}}\ensuremath{~\vdash~}(\textit{a} \ensuremath{\vee} \textit{b})}~~~\end{array}}}
    $~$$~$
    \rulebegin{M-Disj-2}
     \ensuremath{\frac{\begin{array}{l}\mbox{\ensuremath{{\varphi}}$_0$\ensuremath{\leadsto}\ensuremath{{\varphi}}\ensuremath{~\vdash~}\textit{b}}~~~\end{array}}{\begin{array}{l}\mbox{\ensuremath{{\varphi}}$_0$\ensuremath{\leadsto}\ensuremath{{\varphi}}\ensuremath{~\vdash~}(\textit{a} \ensuremath{\vee} \textit{b})}~~~\end{array}}}
  \end{center}
  \end{minipage}}
  }
  \caption{Rules for the implicit mutable store (operations and
    logical connectives).}
  \label{fig:mut-rules}
\end{figure}

We can now define the rules that manipulate mutable variables
(Figure~\ref{fig:mut-rules}).
For the sake of simplicity, they do not impose any evaluation
order. In practice, and in order to keep the computational cost low,
we will use the Prolog resolution strategy, and impose limitations on
the instantiation level of some particular terms; we postpone
discussing this issue until Section~\ref{sec:ip-boundaries}:

\begin{description}
\item[Creation:] The \rulename{M-New} rule defines the creation of new
  mutable placeholders. A goal \textit{m} \textsf{=} \textsf{initmut}(\ensuremath{\tau}, \textit{v})\footnote{To improve readability, we use the functional 
    notation of~\cite{functional-lazy-notation-flops2006} for the new
    \textsf{initmut}\textsf{/}3 and (\textsf{@})\textsf{/}2 \emph{built-in}
    predicates.  } checks that \ensuremath{\tau}(\textit{v}) holds (i.e.,
  \textit{v} has type \ensuremath{\tau}),
  creates a new mutable identifier \textit{id} that does not appear
  as a key in \ensuremath{{\varphi}}$_0$ and does not unify with any other
  term in the program, defines the updated \ensuremath{{\varphi}} as
  \ensuremath{{\varphi}}$_0$ where the value associated with \textit{id} is
  \textit{v}, and unifies \textit{m} with \textit{id}.  
  These restrictions ensure that \textit{m} 
  is an unbound variable before the mutable is created.

\item[Access:] Reading the contents of a mutable variable is defined
  in the \rulename{M-Read} rule. The goal \textit{x} \textsf{=} \textit{m}\textsf{@} holds if the
  variable \textit{m} is bound with a mutable identifier
  \textit{id}, for which an associated \textit{v} value exists in
  the mutable store \ensuremath{{\varphi}}, and the variable \textit{x}
  unifies with \textit{v}.\footnote{Since the variable in the
    mutable store is constrained to the type, it is not necessary to
    check that \textit{x} belongs to that type.}

\item[Assignment:] Assignment of values to mutables is described in
  the \rulename{M-Assign} rule. The goal \textit{m} \ensuremath{\Leftarrow} \textit{v}, which assigns
  a value to a mutable identifier, holds iff:
  \begin{itemize}
  \item \textit{m} is unified with a mutable identifier
    \textit{id}, for which a value is stored in \ensuremath{{\varphi}}$_0$
    with associated type \ensuremath{\tau}; 
  \item \textit{v} has type \ensuremath{\tau}, i.e., the value
    to be stored is compatible with the type associated with the
    mutable; 
  \item \ensuremath{{\varphi}}$_0$ is the result of replacing the associated
    type and value for \textit{id} by \ensuremath{\tau} and
    \textit{v}, respectively.
  \end{itemize}

\item[Typing:] The \rulename{M-Type} rule allows checking that a
  variable contains a mutable identifier of a given type. A goal
  \textsf{mut}(\ensuremath{\tau}, \textit{m}) is true if \textit{m} is unified with a
  mutable identifier \textit{id} that is associated with the type
  \ensuremath{\tau} in the mutable store \ensuremath{{\varphi}}.  

\end{description}

Note that although some of the rules above enforce typing constraints,
the compiler, as we will see, is actuallly able to statically remove
these checks and there is no dynamic typing involved in the execution
of admissible \ip\ programs.
The rest of the rules define how the previous operations on mutable
variables can be joined and combined together, and with other
predicates:

\begin{description}
\item[Weakening:] The weakening rule \rulename{M-Weak} states that if
  some goal can be solved without a mutable store, then it can also be
  solved in the context of a mutable store that is left
  untouched. This rule allows the integration of the new rules with
  predicates (and built-ins) that do not use mutables.

\item[Conjunction:] The \rulename{M-Conj} rule defines how to solve a
  goal \textit{a} \ensuremath{\wedge} \textit{b} (written as (\textit{a}, \textit{b}) in code) in an
  environment where the input mutable store \ensuremath{{\varphi}}$_0$ is
  transformed into \ensuremath{{\varphi}}, by solving \textit{a} and
  \textit{b} and connecting the output mutable store of the former
  (\ensuremath{{\varphi}}$_1$) with the input mutable store of the latter.
  This conjunction is not commutative, since the updates performed by
  \textit{a} may alter the values read in \textit{b}. If none of
  those goals modify the mutable store, then commutativity can be
  ensured. If none of them access the mutable store, then it is
  equivalent to the classic definition of conjunction (by applying the
  \rulename{M-Weak} rule).

\item[Disjunction:] The disjunction of two goals is defined in the
  \rulename{M-Disj} rule, where \textit{a} \ensuremath{\vee} \textit{b} (written as
  (\textit{a} \textsf{;} \textit{b}) in code) holds in a given environment if either
  \textit{a} or \textit{b} holds in such environment, with no
  relation between the mutable stores of both branches. That means
  that changes in the mutable store would be \emph{backtrackable} (e.g.,
  any change introduced by an attempt to solve one branch must be
  undone when trying another alternative). As with conjunction, if the
  goals in the disjunction do not update nor access the mutable store,
  then it is equivalent to the classic definition of disjunction (by
  applying the \rulename{M-Weak} rule).
\end{description}

Mutable terms, conceptually similar to the mutables we present here,
were introduced in SICStus Prolog as a replacement for
\textsf{setarg}\textsf{/}\textsf{3}, 
and also appear in proposals for global variables in logic programming
(such as ~\cite{global-vars-iclp97}, Bart Demoen's implementation of
(non)backtrackable global variables for
hProlog/SWI-Prolog~\cite{Demoen98globalvariables}), or imperative
assignment~\cite{giannesini85:prolog}.  In the latter case there was
no notion of types and the terms assigned to had to be (ground) atoms
at the time of assignment.

We will consider that two types unify if their names match.  Thus,
typing in mutables divide the mutable store into separate, independent
regions, which will facilitate program analysis.  
For the purpose of this work we will treat mutable variables as a
native part of the language.
It would however be possible to emulate the mutable store as a pair of
additional arguments, threaded from left to right in the goals of the
predicate bodies. A similar translation is commonly used to implement
DCGs or state variables in Mercury \cite{mercury-jlp}.

\subsection{Compilation Strategy}
\label{sec:ip-boundaries}
\label{sec:compilation}

In the previous section the operations on mutable data were presented
separately from the host language, and no commitment was made
regarding their implementation other than assuming that it could be
done efficiently.  However, when the host language \lsourcerefl\ has a
Prolog-like semantics (featuring unification and backtracking) and
even if backtrackable destructive assignment is used, the compiled
code can be unaffordably inefficient for deterministic computations
unless extensive analysis and optimization is performed.  On the other
hand, the same computations may be easy to optimize in lower-level
languages.

A way to overcome those problems is to specialize
the translated code for a relevant subset of the initial input data.
This subset can be \emph{abstractly} specified: let us consider a predicate
\textsf{bool}\textsf{/}\textsf{1} that defines truth values, and a call
\textsf{bool}(\textit{X}) where \textit{X} is known to be always bound to a
de-referenced atom.  The information about the dereferencing state of
\textit{X} is a 
partial specification of the initial conditions, and replacing
the call to the generic \textsf{bool}\textsf{/}\textsf{1} predicate by a call to
another predicate that implements a version of \textsf{bool}\textsf{/}\textsf{1}
that avoids unnecessary work (e.g., there is no need for \emph{choice
  points} or \emph{tag} testing on \textit{X}) produces
the same computation, but using code that is both shorter and faster.
To be usable in the compilation process, it is necessary to propagate
this knowledge about the program behavior as predicate-level
assertions or program-point annotations, usually by means of automatic
methods such as static analysis.
Such techniques has been tested
elsewhere~\cite{warren-phd,Taylor91,van-roy-computer,vanroy-survey,morales04:p-to-c-padl}.

This approach (as most automatic optimization techniques) has
obviously its own drawbacks when high performance is a requirement:
\emph{a}) the analysis is not always precise enough, which makes the
compiler require manual annotations or generate under-optimized
programs; \emph{b}) the program is not always optimized, even if
the analysis is able to infer interesting properties about it, since the
compiler may not be clever enough to improve the algorithm; and
\emph{c}) the final program performance is hard to predict, as we
leave more optimization decisions to the compiler, of which the 
programmer may not be aware.

For the purposes of this paper, cases \emph{a} and \emph{b} do not
represent a major problem. Firstly, if some of the annotations cannot
be obtained automatically and need to be provided by hand, the
programming style still encourages separation of optimization
annotations (as hints for the compiler) and the actual algorithm,
which we believe makes code easier to manage. Secondly, we adapted the
language \ip\ and compilation process to make the algorithms
implemented in the emulator easier to represent and compile.
For case \emph{c}, we took an approach  different from that in other systems.
Since our goal is to generate low-level code that \emph{ensures}
efficiency, we impose some constraints on the compilation output to
avoid generation of code known to be suboptimal.  This restricts the
admissible code and the compiler informs the user when the constraints
do not hold, by reporting efficiency errors.
This is obviously too drastic a solution for general programs, but we
found it a good compromise in our application.

\subsubsection{Preprocessing \ip\ programs}
\label{sec:prepro}

The compilation algorithm starts with the expansion of syntactic
extensions (such as, e.g., functional notation), followed by
normalization and analysis.
Normalization
is helpful to reduce programs to simpler building blocks for which
compilation schemes are described.

\begin{figure}
  \centering
  \begin{displaymath}
\begin{array}{rclr}
    \mbox{\textit{pred}} & ::= & \mbox{\textit{head} \textsf{:--} \textit{body}} & \mbox{(Predicates)}\\  
    \mbox{\textit{head}} & ::= & \mbox{\ensuremath{\lfloor}\ensuremath{\beta}\ensuremath{\rfloor}\textit{id}(\textit{var}, \textsf{...}, \textit{var})\ensuremath{\lfloor}\ensuremath{\beta}\ensuremath{\rfloor}} & \mbox{(Heads)}\\  
    \mbox{\textit{body}} & ::= & \mbox{\textit{goals} \textsf{$\mid$} (\textit{goals} \ensuremath{\rightarrow} \textit{goals} \textsf{;} \textit{body})} & \mbox{(Body)}\\
    \mbox{\textit{goals}} & ::= & \mbox{\ensuremath{\lfloor}\ensuremath{\beta}\ensuremath{\rfloor}\textit{goal}, \textsf{...}, \ensuremath{\lfloor}\ensuremath{\beta}\ensuremath{\rfloor}\textit{goal}} & \mbox{(Conjunction of goals)}\\
    \mbox{\textit{goal}} & ::= & \mbox{\textit{var} \textsf{=} \textit{var} \textsf{$\mid$} \textit{var} \textsf{=} \textit{cons}} ~| & \mbox{(Unifications)}\\
                           &     & \mbox{\textit{var} \textsf{=} \textit{var}\textsf{@} \textsf{$\mid$} \textit{var} \ensuremath{\Leftarrow} \textit{var}} ~| & \mbox{(Mutable ops.)} \\  
                           &     & \mbox{\textit{id}(\textit{var}, \textsf{...}, \textit{var})} & \mbox{(Built-In/User call)}\\  
    \mbox{\textit{var}} & ::= & \mbox{uppercase name} & \mbox{(Variables)}\\  
    \mbox{\textit{id}} & ::= & \mbox{lowercase name} & \mbox{(Atom names)}\\
    \mbox{\textit{cons}} & ::= & \mbox{atom names, integers, floats, characters, etc.} & \mbox{(Other constants)}\\
    \mbox{\ensuremath{\beta}} & ::= & \mbox{abstract substitution} & \mbox{(Program point anots.)}\\  
  \end{array}
      \end{displaymath}
    \caption{Syntax of normalized programs.}
  \label{fig:normalized-improlog}
\end{figure}

The syntax of the normalized programs is shown in
Figure~\ref{fig:normalized-improlog}, and is similar to that used in
\texttt{ciaocc}~\cite{morales04:p-to-c-padl}.  It focuses on
simplifying code generation rules and making analysis information
easily accessible. Additionally, operations on mutable variables are
considered built-ins.
Normalized predicates contain a single clause, composed of a head and
a body which contains a conjunction of goals or
\emph{if-then-else}s. Every goal and head are prefixed with
\emph{program point} information which contains the abstract
substitution inferred during analysis, relating every variable in the
goal / head with an abstraction of its value or state. However,
compilation needs that information to be also available for every
temporary variable that may appear during code generation and which is
not yet present in the normalized program.  In order to overcome this
problem, most auxiliary variables are already introduced before the
analysis.  The code reaches a homogeneous form by requiring that both
head and goals contain only syntactical variables as arguments, and
making unification and matching explicit in the body of the clauses.
Each of these basic data-handling steps can therefore be annotated
with the corresponding abstract state.  Additionally, unifications are
restricted to the \emph{variable-variable} and
\emph{variable-constant} cases.  As we will see later, this is enough
for our purposes.

\paragraph{Program Transformations.}
Normalization groups the bodies of the clauses of the
same predicate in a disjunction, sharing common new variables in the
head arguments, introducing unifications as explained before,
and taking care of renaming  variables local to each
clause.\footnote{This is required to make their information independent 
  in each branch during analysis and compilation.}
As a simplification for the purposes of this paper, we restrict
ourselves to treating
atomic, ground data types in the language.  Structured data is 
created by invoking built-in or predefined predicates.
Control structures such as disjunctions (\textit{\_} \textsf{;} \textit{\_}) and negation
\mbox{(\textsf{$\backslash$\texttt{+}} \textit{\_})} are only supported when they can be translated to
\emph{if-then-else}s, and a compilation error is emitted (and
compilation is aborted) if they
cannot. If cuts are not explicitly written in the program, mode analysis and
determinism analysis help in detecting the mutually exclusive
prefixes of the bodies, and delimit them with cuts.
Note that the restrictions that we impose on the accepted programs 
make it easier to treat some \emph{non-logical} Prolog features, such
as \emph{red} cuts, which make the language semantics more complex but
are widely used in practice.
We allow the use of \emph{red} cuts (explicitly or as
\mbox{(\textsf{...} \ensuremath{\rightarrow} \textsf{...} \textsf{;} \textsf{...})} constructs) as long as it is possible
to insert a mutually exclusive prefix in all the alternatives of the
disjunctions where they appear:  (\textit{b}$_1$, \textsf{!}, \textit{b}$_2$ \textsf{;} \textit{b}$_3$) is
treated as equivalent to \mbox{(\textit{b}$_1$, \textsf{!}, \textit{b}$_2$ \textsf{;} \textsf{$\backslash$\texttt{+}} \textit{b}$_1$, \textsf{!}, \textit{b}$_3$)} if
analysis (e.g.,~\cite{non-failure-iclp97})
is able to determine that \textit{b}$_1$ does not generate multiple solutions and
does not further instantiate any variables shared with the head or the
rest of the alternatives.

\paragraph{Predicate and Program Point Information.} 

The information that the analysis infers from (and is annotated in)
the normalized program is represented using the Ciao assertion language~\cite{prog-glob-an,assert-lang-disciplbook,ciaopp-sas03-journal-scp}
(with suitable property definitions and mode macros for our purposes).
This information 
can be divided into predicate-level assertions
and program point assertions.
Predicate-level assertions relate an abstract input state with an
output state (\ensuremath{\lfloor}\ensuremath{\beta}$_0$\ensuremath{\rfloor}\textit{f}(\textit{a}$_1$, \textsf{...}, \textit{a}$_n$)\ensuremath{\lfloor}\ensuremath{\beta}\ensuremath{\rfloor}), or state
facts about some properties (see below) of the predicate. Given a
predicate \textit{f}\textsf{/}\textit{n}, the properties needed for the compilation
rules used in this work are:
\begin{itemize}
\item \textsf{det}(\textit{f}\textsf{/}\textit{n}): The predicate \textit{f}\textsf{/}\textit{n} is
  deterministic (it has exactly one solution).
\item \textsf{semidet}(\textit{f}\textsf{/}\textit{n}): The predicate \textit{f}\textsf{/}\textit{n} is
  semideterministic (it has one or zero solutions).
\end{itemize}
We assume
that there is a single call pattern, or that all the possible call
patterns have been aggregated into a single one, i.e., the analysis we
perform does not take into account the different modes in which a
single predicate can be called.
Note that this does not prevent effectively supporting different
separate call patterns, as a previous specialization phase can
generate a different predicate version for each calling pattern.

The second kind of annotations keeps track of the abstract state of the
execution at each program point. For a goal \ensuremath{\lfloor}\ensuremath{\beta}\ensuremath{\rfloor}\textit{g},
the following judgments are defined on the abstract substitution
\ensuremath{\beta} on variables of \textit{g}:
\begin{itemize}
\item \ensuremath{\beta}~$\vdash$~\textsf{fresh}(\textit{x}): The variable \textit{x} is a fresh
  variable (not instantiated to any value, not sharing with any other
  variable).
\item \ensuremath{\beta}~$\vdash$~\textsf{ground}(\textit{x}): The variable \textit{x} contains a
  ground term (it does not contain any free variable).
\item \ensuremath{\beta}~$\vdash$~\textit{x}\ensuremath{:}\ensuremath{\tau}: The values that the variable
  \textit{x} can take at this program point are of type
  \ensuremath{\tau}.
\end{itemize}

\subsubsection{Overview of the Analysis of Mutable Variables}
\label{sec:imper-state-analysis}

The basic operations on mutables are restricted to some instantiation
state on input and have to obey to some typing rules.  In order to
make the analysis as parametric as possible to the concrete rules,
these are stated using the following assertions on the predicates
\textsf{@}/\textsf{2}, \textsf{initmut}/\textsf{3}, and
\ensuremath{\Leftarrow}/\textsf{2}: 

\begin{quotedcode}\begin{tabbing}
\\
\textsf{:--} \textbf{\textsf{pred}} \textsf{@}(\=\textsf{\texttt{+}}\textsf{mut}(\=\textit{T}), \textsf{$-$}\textit{T}).\\
\textsf{:--} \textbf{\textsf{pred}} \textsf{initmut}(\=\textsf{\texttt{+}}(\=\textsf{\^}\textit{T}), \textsf{\texttt{+}}\textit{T}, \textsf{$-$}\textsf{mut}(\=\textit{T})).\\
\textsf{:--} \textbf{\textsf{pred}} (\textsf{\texttt{+}}\textsf{mut}(\=\textit{T})) \ensuremath{\Leftarrow} (\textsf{\texttt{+}}\textit{T}).\\
\end{tabbing}
\end{quotedcode}

\noindent 
These state that:
\begin{itemize}
\item Reading the value associated with a mutable identifier of type
  \textsf{mut}(\textit{T}) (which must be ground) gives a value type
  \textit{T}.
\item Creating a mutable variable with type \textit{T} (escaped in
  the assertion to indicate that it is the type name that is provided
  as argument, not a value of that type) takes an initial value of
  type \textit{T} and gives a new mutable variable of type
  \textsf{mut}(\textit{T}).
\item  Assigning a mutable identifier (which must be ground
and of type \textsf{mut}(\textit{T})) a value requires the latter to be of
type \textit{T}.
\end{itemize}

Those assertions instruct the type analysis about the
meaning of the built-ins, requiring no further changes w.r.t.\
equivalent\footnote{In the sense that the behaviour of the built-ins 
  is not hard-wired into the analysis itself.}
type analyses for plain Prolog.
However, in our case more precision is needed. E.g., given
\textsf{mut}(\textsf{int}, \textit{A}) and 
(\textit{A} \ensuremath{\Leftarrow} \textsf{3}, \textsf{p}(\textit{A}\textsf{@})) we want to infer that \textsf{p}\textsf{/}\textsf{1} is
called with an integer value 3 and not with any integer (as inferred
using just the assertion).  With no information about the built-ins,
that code is equivalent to (\textit{T}$_0$ \textsf{=} \textsf{3}, \textit{A} \ensuremath{\Leftarrow} \textit{T}$_0$, \textit{T}$_1$ \textsf{=} \textit{A}\textsf{@}, \textsf{p}(\textit{T}$_1$)), and no relation between \textit{T}$_0$ and \textit{T}$_1$ is
established.

However, based on the semantics of mutables variables and their operations
(Figure~\ref{fig:mut-rules}), it is possible to define an analysis
based on abstract interpretation to infer properties of the values
stored in the mutable store.
To natively understand the built-ins, it is necessary to abstract the
mutable identifiers and the mutable store, and represent it in the
abstract domain, for which different options exist.

One is to explicitly keep the relation between the abstraction of the mutable
identifier and the variable containing its associated value. For
every newly created mutable or mutable assignment, the associated
value is changed, and the previous code would be
equivalent 
to (\textit{T} \textsf{=} \textsf{3}, \textit{A} \ensuremath{\Leftarrow} \textit{T}, \textit{T} \textsf{=} \textit{A}\textsf{@}, \textsf{p}(\textit{T})).
The analysis in this case will lose information when the value associated
with the mutable is unknown. That is, given \textsf{mut}(\textsf{int}, \textit{A}) and
\textsf{mut}(\textsf{int}, \textit{B}), it is not possible to prove that \textit{A} \ensuremath{\Leftarrow} \textsf{3}, \textsf{p}(\textit{B}\textsf{@}) will not call \textsf{p}\textsf{/}\textsf{1} with a value of 3.

Different abstractions of the mutable identifiers yield different
precision levels in the analysis. E.g., given an abstract domain for
mutable identifiers that distinguishes newly created mutables, the
chunk of code
(\textit{A} \textsf{=} \textsf{initmut}(\textsf{int}, \textsf{1}), \textit{B} \textsf{=} \textsf{initmut}(\textsf{int}, \textsf{2}), \textit{A} \ensuremath{\Leftarrow} \textsf{3}, \textsf{p}(\textit{B}\textsf{@}))
has enough information to ensure that \textit{B} is unaffected by the
assignment to \textit{A}.
In the current state, and for the purpose of the paper, the abstraction
of mutable identifiers is able to take into account newly created
mutables and mutables of a particular type.  When an assignment is
performed on an unknown mutable, it only needs to change the values of
mutables of exactly the same type, improving precision.\footnote{That was
  enough to specialize pieces of \ip\ code implementing the
  unification of tagged words, which was previously optimized by
  hand.}
If mutable identifiers are interpreted as pointers, that problem is
related to \emph{pointer aliasing} in imperative programming
(see~\cite{hind00which} for a tutorial overview).

\subsection{Data Representation and Operations}
\label{sec:comp-variables}

Data representation in most Prolog systems often chooses a 
general mapping from Prolog terms and variables to C data so that full
unification and backtracking can be implemented.
However, for the logical and mutable variables of \ip, we need the
least expensive mapping to C types and variables possible, since
anything else would bring an unacceptable overhead in critical code
(such as emulator instructions).
A general way to overcome this problem, which is taken in this work,
is to start from a general representation 
and replacing it by a more specific encoding.

Let us recall the general representation in WAM-based
implementations~\cite{Warren83,hassan-wamtutorial}.  The abtract
machine state is composed of a set of registers and stacks of memory
cells. The values stored in those registers and memory cells are
called \emph{tagged} words. Every tagged word has a \emph{tag} and a
\emph{value}, the tag indicating the kind of value stored in
it. Possible values are constants (such as integers up to some fixed
length, indexes for atoms, etc.), or pointers to larger data which
does not fit in the space for the value (such as larger integers,
multiple precision numbers, floating point numbers, arrays of
arguments for structures).\footnote{Those will be ignored in this
  paper, since all data can be described using atomic constants,
  mutable identifiers, and built-ins to control the emulator stacks.}
There exist a special \emph{reference} tag that indicates that the
value is a pointer to another cell.  That reference tag allows a cell
to point to itself (for unbound variables), or set of cells point to
the same value (for unified variables).  Given our assumption that
mutable variables can be efficiently implemented, we want to point out
that these representations can be extended for this case, using, for
example, an additional \emph{mutable} tag, to denote that the value is
a pointer to another cell which contains the associated value.  How
terms are created and unified using a tagged cell representation
is well described in the relevant literature.

When a Prolog-like language is (naively) translated to C, a large part of the
overhead comes from the use of tags
(including bits reserved for automatic memory management) and machine words
with fixed sizes  (e.g., \texttt{unsigned int}) for tagged cells.  If
we are able to enforce
a fixed tag for every variable (which we can in principle map to a
word) at compile time at every program point, 
those additional tag bits can be removed from the representation and
the whole machine word 
can be used for the value. This makes it possible to use different
C types for each kind of value (e.g., \texttt{char}, \texttt{float},
\texttt{double}, etc.). 
Moreover, the restrictions that we have imposed on program determinism
(Section~\ref{sec:prepro}), variable scoping, and visibility of
mutable identifiers make trailing unnecessary.

\subsubsection{C types for values}
The associated C type that stores the value for an \ip\ type
\ensuremath{\tau} is defined in a two-layered approach. First, the type
\ensuremath{\tau} is inferred by means of Prolog type analysis.  In our
case we are using the regular type analysis of~\cite{eterms-sas02},
which can type more programs than a Hindley-Damas-Milner type 
analysis~\cite{DamasMilner82}.  Then, for each variable a compatible
\emph{encoding type}, which contains all the values that the variable
can take, is assigned, and the corresponding C type for that encoding
type is used.  Encoding types are Prolog types annotated with an
assertion that indicates the associated C type.  A set of heuristics
is used to assign economic encodings so that memory usage and type
characteristics are adjusted as tightly as possible.  Consider,  for
example, the type \textsf{flag}/\textsf{1} defined as

\begin{quotedcode}\begin{tabbing}
\\
\textsf{:--} \textbf{\textsf{regtype}} \textsf{flag}\textsf{/}\textsf{1} \textsf{\texttt{+}} \textsf{low}(\=\textsf{int32}).\\
\textsf{flag} \textsf{:=} \textsf{off} \textsf{$\mid$} \textsf{on}.\\
\end{tabbing}
\end{quotedcode}

\noindent It specifies that the values in the declared type must be
represented using 
the \texttt{int32} C type. In this case, \textsf{off} will be encoded
as \texttt{0} and \textsf{on} encoded as \texttt{1}. The set of
available encoding types must be fixed at compile time, either defined
by the user or 
provided as libraries.  Although it is not always possible to
automatically provide a mapping,
we believe that this is a viable alternative to more 
restrictive typing options such as Hindley-Damas-Milner based typings.
A more detailed description of data types in \ip\ is available
in~\cite{tagschemes-ppdp08}.

\subsubsection{Mapping \ip\ variables to C variables}
\label{sec:mapping-c-vars}

The logical and mutable variables of \ip\ are mapped onto imperative,
low-level variables which can be global, local, or passed as
function arguments. Thus, pointers to the actual memory locations
where the value, mutable identifier, or mutable value are stored may
be necessary.  However, as stated before, we need to statically determine the
number of references required.
The \emph{reference mode}s of a variable will define the shape of the
memory cell (or C variable), indicating how the value or mutable value
is accessed:
\begin{itemize}
\item \textbf{0v}: the cell contains the value.
\item \textbf{1v}: the cell contains a pointer to the value.
\item \textbf{0m}: the cell contains the mutable value.
\item \textbf{1m}: the cell contains a pointer to the mutable cell, which contains the value.
\item \textbf{2m}: the cell contains a pointer to another cell, which contains a pointer to the mutable cell, which contains the mutable value.
\end{itemize}

For an \ip\ variable \textsf{x}, with associated C symbol
\texttt{x}, and given the C type for its value \ensuremath{c\tau},
Table~\ref{tab:rvallval} gives the full C type definition to be used
in the variable declaration, the \emph{r-value} (for the left part of
C assignments) and \emph{l-value} (as C expressions) for the reference
(or address) to the variable, the variable value, the reference to the
mutable value, and the mutable value itself.  These definitions relate
C and \ip\ variables and will be used later in the compilation rules.
Note that the translation for \textsf{ref\_rval} and
\textsf{val\_lval} is not defined for
\textbf{0m}. That indicates that it is
impossible to modify the mutable identifier itself for that mutable,
since it is fixed. This tight mapping to C types, avoiding when
possible unnecessary indirections, allows the C compiler to apply
optimizations such as using machine registers for mutable variables.

\begin{table}
\begin{tabular}{|l|r|r|r|r|r|}
  \cline{2-6}
  \multicolumn{1}{c|}{}     & \multicolumn{5}{c|}{\textsf{refmode}(\textit{x})} \\
  \cline{2-6}
  \multicolumn{1}{c|}{}     & \textbf{0v} & \textbf{1v} & \textbf{0m} & \textbf{1m}    & \textbf{2m} \\
  \cline{1-6}
  Final C type      & \ensuremath{c\tau} & \ensuremath{c\tau} \texttt{\textbf{*}} & \ensuremath{c\tau}   & \ensuremath{c\tau} \texttt{\textbf{*}} & \ensuremath{c\tau} \texttt{\textbf{*}} \texttt{\textbf{*}} \\
  \cline{1-6}
  \textsf{ref\_rval}\ensuremath{\llbracket}\textit{x}\ensuremath{\rrbracket}  & \texttt{\&x}     & \texttt{x}     & -             & \texttt{\&x}   & \texttt{x}   \\
  \textsf{val\_lval}\ensuremath{\llbracket}\textit{x}\ensuremath{\rrbracket}  & \texttt{x}       & \texttt{*x}    & -             & \texttt{x}     & \texttt{*x} \\
  \textsf{val\_rval}\ensuremath{\llbracket}\textit{x}\ensuremath{\rrbracket}  & \texttt{x}       & \texttt{*x}    & \texttt{\&x}   & \texttt{x}     & \texttt{*x} \\
  \cline{1-6}
  \textsf{mutval}\textsf{/}\textsf{val\_lval}\ensuremath{\llbracket}\textit{x}\ensuremath{\rrbracket} &              &                & \texttt{x}     & \texttt{*x}  & \texttt{**x}  \\
  \textsf{mutval}\textsf{/}\textsf{val\_rval}\ensuremath{\llbracket}\textit{x}\ensuremath{\rrbracket} &              &                & \texttt{x}     & \texttt{*x}  & \texttt{**x}  \\
  \cline{1-6}
\end{tabular}
  \caption{Operation and translation table for different mapping modes of \ip\ variables}
  \label{tab:rvallval}
\end{table}

The following algorithm infers the \emph{reference mode}
(\textsf{refmode}(\textit{\_})) of each predicate variable making use of type
and mode annotations:
\begin{itemize}
\item[1:] Given the head \ensuremath{\lfloor}\ensuremath{\beta}$_0$\ensuremath{\rfloor}\textit{f}(\textit{a}$_1$, \textsf{...}, \textit{a}$_n$)\ensuremath{\lfloor}\ensuremath{\beta}\ensuremath{\rfloor},
  the \emph{i}th-argument mode \mbox{\textsf{argmode}(\textit{f}\textsf{/}\textit{n}, \textit{i})} for a predicate
  argument \textit{a}$_i$, is defined as:
  \begin{displaymath}
    \mbox{\textsf{argmode}(\textit{f}\textsf{/}\textit{n}, \textit{i})} = \left\{
      \begin{array}{ll}
    \mbox{in} & \mbox{if \ensuremath{\beta}~$\vdash$~\textsf{ground}(\textit{a}$_i$)} \\
    \mbox{out} & \mbox{if \ensuremath{\beta}$_0$~$\vdash$~\textsf{fresh}(\textit{a}$_i$), \ensuremath{\beta}~$\vdash$~\textsf{ground}(\textit{a}$_i$)}
      \end{array}
    \right.
  \end{displaymath}

\item[2:] For each predicate argument \textit{a}$_i$, depending on
  \textsf{argmode}(\textit{f}\textsf{/}\textit{n}, \textit{a}$_i$): 
  \begin{itemize}
  \item If \textsf{argmode}(\textit{f}\textsf{/}\textit{n}, \textit{a}$_i$) = in, then
    \begin{itemize}
    \item if \ensuremath{\beta}~$\vdash$~\textit{a}$_i$\ensuremath{:}\textsf{mut}(\textit{t}) then
      \textsf{refmode}(\textit{a}$_i$) \textsf{=} \textbf{1m}, else  \textsf{refmode}(\textit{a}$_i$) \textsf{=} \textbf{0v}.
    \end{itemize}
  \item If \textsf{argmode}(\textit{f}\textsf{/}\textit{n}, \textit{a}$_i$) = out, then
    \begin{itemize}
    \item if \ensuremath{\beta}~$\vdash$~\textit{a}$_i$\ensuremath{:}\textsf{mut}(\textit{t}) then
        \textsf{refmode}(\textit{a}$_i$) \textsf{=} \textbf{2m}, else
        \textsf{refmode}(\textit{a}$_i$) \textsf{=} \textbf{1v}. 
    \end{itemize}
  \end{itemize}
\item[3:] For each unification \ensuremath{\lfloor}\ensuremath{\beta}\ensuremath{\rfloor}\textit{a} \textsf{=} \textit{b}:
  \begin{itemize}
  \item if \ensuremath{\beta}~$\vdash$~\textsf{fresh}(\textit{a}), \ensuremath{\beta}~$\vdash$~\textsf{ground}(\textit{b}), \ensuremath{\beta}~$\vdash$~\textit{b}\ensuremath{:}\textsf{mut}(\textit{t}), then \textsf{refmode}(\textit{a}) \textsf{=} \textbf{1m}.
  \item Otherwise, if \ensuremath{\beta}~$\vdash$~\textsf{fresh}(\textit{a}), then \textsf{refmode}(\textit{a}) \textsf{=} \textbf{0v}.
  \end{itemize}
\item[4:] For each mutable initialization \ensuremath{\lfloor}\ensuremath{\beta}\ensuremath{\rfloor}\textit{a} \textsf{=} \textsf{initmut}(\textit{t}, \textit{b}):
  \begin{itemize}
  \item if \ensuremath{\beta}~$\vdash$~\textsf{fresh}(\textit{a}), \ensuremath{\beta}~$\vdash$~\textsf{ground}(\textit{b}), \ensuremath{\beta}~$\vdash$~\textit{b}\ensuremath{:} \textsf{mut}(\textit{t}), then \textsf{refmode}(\textit{a}) \textsf{=} \textbf{0m}. 
  \end{itemize}
\item[5:] Any case not covered above is a compile-time error.
\end{itemize}

\paragraph{Escape analysis of mutable identifiers.}
According to the compilation scheme we follow, if a mutable variable
identifier cannot be reached outside the scope of a predicate, it can
be safely mapped to a (local) C variable. That requires the equivalent
of escape analysis for mutable identifiers.  A conservative
approximation to decide that mutables can be assigned to local C
variables is the following: the mutable variable identifier can be
read from, assigned to, and passed as argument to other predicates,
but it cannot be assigned to anything else than other local variables.
This is easy to check and has been precise enough for our purposes.

\subsection{Code Generation Rules} 
\label{sec:cgen}

Compilation processes a set of predicates, each one
composed of a \emph{head} and \emph{body} as defined in
Section~\ref{sec:prepro}. The body can contain 
control constructs, calls to user predicates,
calls to built-ins, and calls to external predicates written in C.
For each of these cases we will summarize the compilation as
translation rules, where \textit{p} stands for the
predicate compilation output that stores the C functions for the
compiled predicates.  The compilation state for a predicate is denoted
as \ensuremath{\theta}, and it is composed of a set of variable
declarations and a mapping from identifiers to \emph{basic
  blocks}.  Each basic block, identified by \ensuremath{\delta}, contains a
sequence of sentences and a terminal control
sentence. 

Basic blocks are finally translated to C code as labels, sequences of
sentences, and jumps or conditional branches generated as \emph{goto}s and
\emph{if-then-else}s.
Note that the use of labels and jumps in the generated code should not
make the C compiler generate suboptimal code, as simplification of
control logic to basic blocks and jumps is one of the first steps
performed by C compilers.  It was experimentally checked that using
\emph{if-then-else} constructs (when possible) does not necessarily
help mainstream C compilers in generating better code.  In any case,
doing so is a code generation option. 

For simplicity, in the following rules we will use the syntax
\ensuremath{\langle}\ensuremath{\theta}$_0$\ensuremath{\rangle} \ensuremath{\forall}\textit{i}\ensuremath{=}1\ensuremath{..}\textit{n} \textit{g} \ensuremath{\langle}\ensuremath{\theta}$_n$\ensuremath{\rangle} to denote
the evaluation of \textit{g} for every value of \textit{i} between
1 and \textit{n}, where the intermediate states
\ensuremath{\theta}$_j$ are adequately threaded to link every state with
the following one.

\subsubsection{Compilation of Goals}
\label{sec:comp-control}

The compilation of goals is described by the rule
$\langle$\ensuremath{\theta}$_0$$\rangle$~\textsf{gcomp}(\textit{goal}, \ensuremath{\eta}, \ensuremath{\delta})~$\Rightarrow$~$\langle$ \ensuremath{\theta}$\rangle$. \ensuremath{\eta} is a mapping which goes from
continuation identifiers (e.g., \textsf{s} for the success
continuation, \textsf{f} for the failure continuation, and possibly
more identifiers for other continuations, such as those needed for
exceptions) to basic blocks identifiers. Therefore
\ensuremath{\eta}(\textsf{s}) and \ensuremath{\eta}(\textsf{f}) denote the
continuation addresses in case of success (resp.\ failure) of
\textit{goal}.
The compilation state \ensuremath{\theta} is obtained from
\ensuremath{\theta}$_0$ by \emph{appending} the generated code for
\textit{goal} to the \ensuremath{\delta} basic block, and optionally
introducing more basic blocks connected by the continuations
associated to them.

\begin{ruleset}{Control compilation rules.}{fig:controlcomprules}
  \begin{center}
    \rulebegin{Conj}
     \ensuremath{\frac{\begin{array}{l}\mbox{$\langle$\ensuremath{\theta}$_0$$\rangle$~\textsf{bb\_new}~$\Rightarrow$~\ensuremath{\delta}$_b$~$\langle$\ensuremath{\theta}$_1$$\rangle$}\\
\mbox{$\langle$\ensuremath{\theta}$_1$$\rangle$~\textsf{gcomp}(\textit{a}, \ensuremath{\eta}[\textsf{s}\ensuremath{\mapsto}\ensuremath{\delta}$_b$], \ensuremath{\delta})~$\Rightarrow$~$\langle$\ensuremath{\theta}$_2$$\rangle$}\\
\mbox{$\langle$\ensuremath{\theta}$_2$$\rangle$~\textsf{gcomp}(\textit{b}, \ensuremath{\eta}, \ensuremath{\delta}$_b$)~$\Rightarrow$~$\langle$\ensuremath{\theta}$\rangle$}~~~\end{array}}{\begin{array}{l}\mbox{$\langle$\ensuremath{\theta}$_0$$\rangle$~\textsf{gcomp}((\textit{a}, \textit{b}), \ensuremath{\eta}, \ensuremath{\delta})~$\Rightarrow$~$\langle$\ensuremath{\theta}$\rangle$}~~~\end{array}}}
  \end{center}

  \begin{center}
    \rulebegin{IfThenElse}
     \ensuremath{\frac{\begin{array}{l}\mbox{$\langle$\ensuremath{\theta}$_0$$\rangle$~\textsf{bb\_newn}(2)~$\Rightarrow$~[\ensuremath{\delta}$_t$, \ensuremath{\delta}$_e$]~$\langle$\ensuremath{\theta}$_1$$\rangle$}\\
\mbox{$\langle$\ensuremath{\theta}$_1$$\rangle$~\textsf{gcomp}(\textit{a}, \ensuremath{\eta}[\textsf{s}\ensuremath{\mapsto}\ensuremath{\delta}$_t$, \textsf{f}\ensuremath{\mapsto}\ensuremath{\delta}$_e$], \ensuremath{\delta})~$\Rightarrow$~$\langle$\ensuremath{\theta}$_2$$\rangle$}\\
\mbox{$\langle$\ensuremath{\theta}$_2$$\rangle$~\textsf{gcomp}(\textit{then}, \ensuremath{\eta}, \ensuremath{\delta}$_t$)~$\Rightarrow$~$\langle$\ensuremath{\theta}$_3$$\rangle$}~~~\mbox{$\langle$\ensuremath{\theta}$_3$$\rangle$~\textsf{gcomp}(\textit{else}, \ensuremath{\eta}, \ensuremath{\delta}$_e$)~$\Rightarrow$~$\langle$\ensuremath{\theta}$\rangle$}~~~\end{array}}{\begin{array}{l}\mbox{$\langle$\ensuremath{\theta}$_0$$\rangle$~\textsf{gcomp}((\textit{a} \ensuremath{\rightarrow} \textit{then} \textsf{;} \textit{else}), \ensuremath{\eta}, \ensuremath{\delta})~$\Rightarrow$~$\langle$\ensuremath{\theta}$\rangle$}~~~\end{array}}}
  \end{center}

  \begin{center}
    \rulebegin{True}
     \ensuremath{\frac{\begin{array}{l}\mbox{$\langle$\ensuremath{\theta}$_0$$\rangle$~\textsf{emit}(\texttt{\textbf{goto}} \ensuremath{\eta}(\textsf{s}), \ensuremath{\delta})~$\Rightarrow$~$\langle$\ensuremath{\theta}$\rangle$}~~~\end{array}}{\begin{array}{l}\mbox{$\langle$\ensuremath{\theta}$_0$$\rangle$~\textsf{gcomp}(\textsf{true}, \ensuremath{\eta}, \ensuremath{\delta})~$\Rightarrow$~$\langle$\ensuremath{\theta}$\rangle$}~~~\end{array}}}
    $~$$~$
    \rulebegin{Fail}
     \ensuremath{\frac{\begin{array}{l}\mbox{$\langle$\ensuremath{\theta}$_0$$\rangle$~\textsf{emit}(\texttt{\textbf{goto}} \ensuremath{\eta}(\textsf{f}), \ensuremath{\delta})~$\Rightarrow$~$\langle$\ensuremath{\theta}$\rangle$}~~~\end{array}}{\begin{array}{l}\mbox{$\langle$\ensuremath{\theta}$_0$$\rangle$~\textsf{gcomp}(\textsf{fail}, \ensuremath{\eta}, \ensuremath{\delta})~$\Rightarrow$~$\langle$\ensuremath{\theta}$\rangle$}~~~\end{array}}}
  \end{center}

\end{ruleset}

The rules for the compilation of control are presented in
Figure~\ref{fig:controlcomprules}.  We will assume some predefined
operations to request a new basic block identifier (\textsf{bb\_new}) and
a list of new identifiers (\textsf{bb\_newn}), and to
add a C sentence to a given basic block (\textsf{emit}).
The conjunction (\textit{a}, \textit{b}) is translated by rule
\rulename{Conj} by reclaiming a new new basic block identifier
\ensuremath{\delta}$_b$ for the subgoal \textit{b}, generating code for
\textit{a} in the target \ensuremath{\delta}, using as success
continuation \ensuremath{\delta}$_b$, and then generating code for
\textit{b} in \ensuremath{\delta}$_b$.
The construct \mbox{(\textit{a} \ensuremath{\rightarrow} \textit{b} \textsf{;} \textit{c})} is similarly compiled by
the \rulename{IfThenElse} rule. The compilation of \textit{a} 
takes place using as success and failure continuations the basic
block identifiers where \textit{b} and \textit{c} are emitted,
respectively. Then, the process continues by compiling both
\textit{b} and \textit{c} using the original continuations. 
The goals \textsf{true} and \textsf{fail} are compiled by emitting
a jump statement (\texttt{\textbf{goto}} \textit{\_}) that goes
directly to the success and failure continuation (rules
\rulename{True} and \rulename{Fail}).

As stated in Section~\ref{sec:prepro}, there is no compilation rule
for disjunctions (\textit{a} \textsf{;} \textit{b}).  Nevertheless, program
transformations can change them into \emph{if-then-else}
structures, following the constraints on the input language.  E.g.,
(\textit{X} \textsf{=} \textsf{1} \textsf{;} \textit{X} \textsf{=} \textsf{2}) is accepted if \textit{X} is 
ground on entry, since the code can be translated into the equivalent
(\textit{X} \textsf{=} \textsf{1} \ensuremath{\rightarrow} \textsf{true} \textsf{;} \textit{X} \textsf{=} \textsf{2} \ensuremath{\rightarrow} \textsf{true}).  It will not be accepted if
\textit{X} is unbound, since the \emph{if-then-else} code and the
initial disjunction are not equivalent.

Note that since continuations are taken using C \texttt{\textbf{goto}} \textit{\_}
statements, there is a great deal of freedom in the physical ordering of
the basic blocks in the program.  The current implementation
emits code in an order roughly corresponding to the source program, but it has
internal data structures which make it easy to change this order.
Note that different orderings can impact performance, by, for example,
changing code locality, affecting how the processor speculative
execution units perform, and changing which \texttt{\textbf{goto}} \textit{\_}
statements which jump to an immediate label can be simplified by the compiler.

\subsubsection{Compilation of Goal Calls}
\label{sec:comp-calls}

External predicates explicitly defined in C and user predicates
compiled to C code have both the same external interface. Thus we use
the same call compilation rules for them.

Predicates that may fail are mapped to functions with \emph{boolean}
return types (indicating success / failure), and those which cannot
fail are mapped to procedures (with no return result -- as explained
later in Section~\ref{sec:compilation-heads}).
Figure~\ref{fig:callcomprules} shows the rules to compile calls to
external or user predicates.  Function \textsf{argpass}(\textit{f}\textsf{/}\textit{n})
returns the list \mbox{[\textit{r}$_1$, \textsf{...}, \textit{r}$_n$]} of argument passing
modes for predicate \textit{f}\textsf{/}\textit{n}. Depending on
\textsf{argmode}(\textit{f}\textsf{/}\textit{n}, \textit{i}) (see Section~\ref{sec:mapping-c-vars}) ,
\textit{r}$_i$ is \textsf{val\_rval} for \textsf{in} or
\textsf{ref\_rval} for \textsf{out}. Using the translation in
Table~\ref{tab:rvallval}, the C expression for each variable is given
as \textit{r}$_i$\ensuremath{\llbracket}\textit{a}$_i$\ensuremath{\rrbracket}. Taking the C identifier assigned to 
predicate (\textsf{c\_id}(\textit{f}\textsf{/}\textit{n})), we have all the pieces to perform
the call.
If the predicate is semi-deterministic (i.e., it either fails or gives
a single solution), the \rulename{Call-S} rule emits code that checks
the return value and jumps to the success or failure continuation.
If the predicate is deterministic, the \rulename{Call-D} rule emits
code that continues at the success continuation.  To reuse those code
generation patterns, rules \rulename{Emit-S} and \rulename{Emit-D} are
defined.

\begin{ruleset}{Compilation of calls.}{fig:callcomprules}
  \begin{center}
    \rulebegin{Call-S}
     \ensuremath{\frac{\begin{array}{l}\mbox{\textsf{semidet}(\textit{f}\textsf{/}\textit{n})}\\
\mbox{[\textit{r}$_1$, \textsf{...}, \textit{r}$_n$] \textsf{=} \textsf{argpass}(\textit{f}\textsf{/}\textit{n})}\\
\mbox{\ensuremath{\forall}\textit{i}\ensuremath{=}1\ensuremath{..}\textit{n} \textit{c}$_i$ \textsf{=} \textit{r}$_i$\ensuremath{\llbracket}\textit{a}$_i$\ensuremath{\rrbracket}}\\
\mbox{\textit{c}$_f$ \textsf{=} \textsf{c\_id}(\textit{f}\textsf{/}\textit{n})}~~~\mbox{$\langle$\ensuremath{\theta}$_0$$\rangle$~\textsf{emit\_s}(\textit{c}$_f$\texttt{\textbf{(}}\textit{c}$_1$\textbf{,} \textsf{...}\textbf{,} \textit{c}$_n$\texttt{\textbf{)}}, \ensuremath{\eta}, \ensuremath{\delta})~$\Rightarrow$~$\langle$\ensuremath{\theta}$\rangle$}~~~\end{array}}{\begin{array}{l}\mbox{$\langle$\ensuremath{\theta}$_0$$\rangle$~\textsf{gcomp}(\textit{f}(\textit{a}$_1$, \textsf{...}, \textit{a}$_n$), \ensuremath{\eta}, \ensuremath{\delta})~$\Rightarrow$~$\langle$\ensuremath{\theta}$\rangle$}~~~\end{array}}}
  \end{center}

  \begin{center}
    \rulebegin{Emit-S}
     \ensuremath{\frac{\begin{array}{l}\mbox{$\langle$\ensuremath{\theta}$_0$$\rangle$~\textsf{emit}(\texttt{\textbf{if (}}\textit{expr}\texttt{\textbf{) }}\texttt{\textbf{goto}} \ensuremath{\eta}(\textsf{s})\textbf{;} \texttt{\textbf{else}} \texttt{\textbf{goto}} \ensuremath{\eta}(\textsf{f})\textbf{;} , \ensuremath{\delta})~$\Rightarrow$~$\langle$\ensuremath{\theta}$\rangle$}~~~\end{array}}{\begin{array}{l}\mbox{$\langle$\ensuremath{\theta}$_0$$\rangle$~\textsf{emit\_s}(\textit{expr}, \ensuremath{\eta}, \ensuremath{\delta})~$\Rightarrow$~$\langle$\ensuremath{\theta}$\rangle$}~~~\end{array}}}
  \end{center}

  \begin{center}
    \rulebegin{Call-D}
     \ensuremath{\frac{\begin{array}{l}\mbox{\textsf{det}(\textit{f}\textsf{/}\textit{n})}\\
\mbox{[\textit{r}$_1$, \textsf{...}, \textit{r}$_n$] \textsf{=} \textsf{argpass}(\textit{f}\textsf{/}\textit{n})}\\
\mbox{\ensuremath{\forall}\textit{i}\ensuremath{=}1\ensuremath{..}\textit{n} \textit{c}$_i$ \textsf{=} \textit{r}$_i$\ensuremath{\llbracket}\textit{a}$_i$\ensuremath{\rrbracket}}\\
\mbox{\textit{c}$_f$ \textsf{=} \textsf{c\_id}(\textit{f}\textsf{/}\textit{n})}~~~\mbox{$\langle$\ensuremath{\theta}$_0$$\rangle$~\textsf{emit\_d}(\textit{c}$_f$\texttt{\textbf{(}}\textit{c}$_1$\textbf{,} \textsf{...}\textbf{,} \textit{c}$_n$\texttt{\textbf{)}}, \ensuremath{\eta}, \ensuremath{\delta})~$\Rightarrow$~$\langle$\ensuremath{\theta}$\rangle$}~~~\end{array}}{\begin{array}{l}\mbox{$\langle$\ensuremath{\theta}$_0$$\rangle$~\textsf{gcomp}(\textit{f}(\textit{a}$_1$, \textsf{...}, \textit{a}$_n$), \ensuremath{\eta}, \ensuremath{\delta})~$\Rightarrow$~$\langle$\ensuremath{\theta}$\rangle$}~~~\end{array}}}
  \end{center}

  \begin{center}
    \rulebegin{Emit-D}
     \ensuremath{\frac{\begin{array}{l}\mbox{$\langle$\ensuremath{\theta}$_0$$\rangle$~\textsf{emit}(\textit{stat}, \ensuremath{\delta})~$\Rightarrow$~$\langle$\ensuremath{\theta}$_1$$\rangle$}\\
\mbox{$\langle$\ensuremath{\theta}$_1$$\rangle$~\textsf{emit}(\texttt{\textbf{goto}} \ensuremath{\eta}(\textsf{s}), \ensuremath{\delta})~$\Rightarrow$~$\langle$\ensuremath{\theta}$\rangle$}~~~\end{array}}{\begin{array}{l}\mbox{$\langle$\ensuremath{\theta}$_0$$\rangle$~\textsf{emit\_d}(\textit{stat}, \ensuremath{\eta}, \ensuremath{\delta})~$\Rightarrow$~$\langle$\ensuremath{\theta}$\rangle$}~~~\end{array}}}
  \end{center}
\end{ruleset}

\subsubsection{Compilation of Built-in Calls}
\label{sec:comp-builtin}

\begin{ruleset}{Unification compilation rules.}{fig:unifcomprules}
  \begin{center}
    \rulebegin{Unify-FG}
     \ensuremath{\frac{\begin{array}{l}\mbox{\textsf{var}(\textit{a})}~~~\mbox{\textsf{var}(\textit{b})}~~~\mbox{\ensuremath{\beta}~$\vdash$~\textsf{fresh}(\textit{a})}~~~\mbox{\ensuremath{\beta}~$\vdash$~\textsf{ground}(\textit{b})}\\
\mbox{\textit{c}$_a$ \textsf{=} \textsf{val\_lval}\ensuremath{\llbracket}\textit{a}\ensuremath{\rrbracket}}~~~\mbox{\textit{c}$_b$ \textsf{=} \textsf{val\_rval}\ensuremath{\llbracket}\textit{b}\ensuremath{\rrbracket}}\\
\mbox{$\langle$\ensuremath{\theta}$_0$$\rangle$~\textsf{emit\_d}(\textit{c}$_a$\texttt{\textbf{=}}\textit{c}$_b$, \ensuremath{\eta}, \ensuremath{\delta})~$\Rightarrow$~$\langle$\ensuremath{\theta}$\rangle$}~~~\end{array}}{\begin{array}{l}\mbox{$\langle$\ensuremath{\theta}$_0$$\rangle$~\textsf{gcomp}(\ensuremath{\lfloor}\ensuremath{\beta}\ensuremath{\rfloor}\textit{a} \textsf{=} \textit{b}, \ensuremath{\eta}, \ensuremath{\delta})~$\Rightarrow$~$\langle$\ensuremath{\theta}$\rangle$}~~~\end{array}}}
  \end{center}

  \begin{center}
    \rulebegin{Unify-GG}
     \ensuremath{\frac{\begin{array}{l}\mbox{\textsf{var}(\textit{a})}~~~\mbox{\textsf{var}(\textit{b})}~~~\mbox{\ensuremath{\beta}~$\vdash$~\textsf{ground}(\textit{a})}~~~\mbox{\ensuremath{\beta}~$\vdash$~\textsf{ground}(\textit{b})}\\
\mbox{\textit{c}$_a$ \textsf{=} \textsf{val\_rval}\ensuremath{\llbracket}\textit{a}\ensuremath{\rrbracket}}~~~\mbox{\textit{c}$_b$ \textsf{=} \textsf{val\_rval}\ensuremath{\llbracket}\textit{b}\ensuremath{\rrbracket}}\\
\mbox{$\langle$\ensuremath{\theta}$_0$$\rangle$~\textsf{emit\_s}(\textit{c}$_a$\texttt{\textbf{==}}\textit{c}$_b$, \ensuremath{\eta}, \ensuremath{\delta})~$\Rightarrow$~$\langle$\ensuremath{\theta}$\rangle$}~~~\end{array}}{\begin{array}{l}\mbox{$\langle$\ensuremath{\theta}$_0$$\rangle$~\textsf{gcomp}(\ensuremath{\lfloor}\ensuremath{\beta}\ensuremath{\rfloor}\textit{a} \textsf{=} \textit{b}, \ensuremath{\eta}, \ensuremath{\delta})~$\Rightarrow$~$\langle$\ensuremath{\theta}$\rangle$}~~~\end{array}}}
  \end{center}

  \begin{center}
    \rulebegin{Instance-FC}
     \ensuremath{\frac{\begin{array}{l}\mbox{\textsf{var}(\textit{a})}~~~\mbox{\textsf{cons}(\textit{b})}~~~\mbox{\ensuremath{\beta}~$\vdash$~\textsf{fresh}(\textit{a})}\\
\mbox{\textit{c}$_a$ \textsf{=} \textsf{val\_lval}\ensuremath{\llbracket}\textit{a}\ensuremath{\rrbracket}}~~~\mbox{\textit{c}$_b$ \textsf{=} \textsf{encodecons}(\textit{b}, \textsf{encodingtype}(\textit{a}))}\\
\mbox{$\langle$\ensuremath{\theta}$_0$$\rangle$~\textsf{emit\_d}(\textit{c}$_a$\texttt{\textbf{=}}\textit{c}$_b$, \ensuremath{\eta}, \ensuremath{\delta})~$\Rightarrow$~$\langle$\ensuremath{\theta}$\rangle$}~~~\end{array}}{\begin{array}{l}\mbox{$\langle$\ensuremath{\theta}$_0$$\rangle$~\textsf{gcomp}(\ensuremath{\lfloor}\ensuremath{\beta}\ensuremath{\rfloor}\textit{a} \textsf{=} \textit{b}, \ensuremath{\eta}, \ensuremath{\delta})~$\Rightarrow$~$\langle$\ensuremath{\theta}$\rangle$}~~~\end{array}}}
  \end{center}
\end{ruleset}

When compiling goal calls, we distinguish the special case of
built-ins, which are natively understood by the \ip\ compiler and
which treats them especially.
The unification \textit{a} \textsf{=} \textit{b} is handled as shown in
Figure~\ref{fig:unifcomprules}.
If \textit{a} is a fresh variable and \textit{b} is ground (resp.\
for the symmetrical case), the \rulename{Unify-FG} rule specifies a
translation that generates an assignment statement that copies the
value stored in \textit{b} into \textit{a} (using the translation
for their \emph{r-value} and \emph{l-value}, respectively).
When \textit{a} and \textit{b} are both ground, the built-in is
translated into a comparison of their values (rule \rulename{Unify-GG}).
When \textit{a} is a variable and \textit{b} is a constant, 
the built-in is
translated into an assignment statement that copies the C value
encoded from \textit{b}, using the encoding type required by
\textit{a}, into \textit{a} (rule \rulename{Instance-FC}).
Note that although full unification may be assumed during program
transformations and analysis, it must be ultimately reduced to one of
the cases above. Limiting to the simpler cases is expected, in order
to avoid bootstrapping problems when defining the full unification in
\ip\ as part of the emulator definition.

\begin{ruleset}{Compilation rules for mutable operations.}{fig:mutcomprules}
  \begin{center}
    \rulebegin{InitMut}
     \ensuremath{\frac{\begin{array}{l}\mbox{\textsf{var}(\textit{a})}~~~\mbox{\ensuremath{\beta}~$\vdash$~\textsf{fresh}(\textit{a})}~~~\mbox{\textsf{refmode}(\textit{a}) \textsf{=} \textbf{0m}}\\
\mbox{$\langle$\ensuremath{\theta}$_0$$\rangle$~\textsf{gcomp}(\textit{a} \ensuremath{\Leftarrow} \textit{b}, \ensuremath{\eta}, \ensuremath{\delta})~$\Rightarrow$~$\langle$\ensuremath{\theta}$\rangle$}~~~\end{array}}{\begin{array}{l}\mbox{$\langle$\ensuremath{\theta}$_0$$\rangle$~\textsf{gcomp}(\textit{a} \textsf{=} \textsf{initmut}(\ensuremath{\tau}, \textit{b}), \ensuremath{\eta}, \ensuremath{\delta})~$\Rightarrow$~$\langle$\ensuremath{\theta}$\rangle$}~~~\end{array}}}
  \end{center}

  \begin{center}
    \rulebegin{AssignMut}
     \ensuremath{\frac{\begin{array}{l}\mbox{\textsf{var}(\textit{a})}~~~\mbox{\ensuremath{\beta}~$\vdash$~\textsf{ground}(\textit{a})}~~~\mbox{\ensuremath{\beta}~$\vdash$~\textit{a}\ensuremath{:}\textsf{mut}(\textit{\_})}\\
\mbox{\textsf{var}(\textit{b})}~~~\mbox{\ensuremath{\beta}~$\vdash$~\textsf{ground}(\textit{b})}\\
\mbox{\textit{c}$_a$ \textsf{=} \textsf{mutval}\textsf{/}\textsf{val\_lval}\ensuremath{\llbracket}\textit{a}\ensuremath{\rrbracket}}\\
\mbox{\textit{c}$_b$ \textsf{=} \textsf{val\_rval}\ensuremath{\llbracket}\textit{b}\ensuremath{\rrbracket}}\\
\mbox{$\langle$\ensuremath{\theta}$_0$$\rangle$~\textsf{emit\_d}(\textit{c}$_a$\texttt{\textbf{=}}\textit{c}$_b$, \ensuremath{\eta}, \ensuremath{\delta})~$\Rightarrow$~$\langle$\ensuremath{\theta}$\rangle$}~~~\end{array}}{\begin{array}{l}\mbox{$\langle$\ensuremath{\theta}$_0$$\rangle$~\textsf{gcomp}(\textit{a} \ensuremath{\Leftarrow} \textit{b}, \ensuremath{\eta}, \ensuremath{\delta})~$\Rightarrow$~$\langle$\ensuremath{\theta}$\rangle$}~~~\end{array}}}
  \end{center}

  \begin{center}
    \rulebegin{ReadMut}
     \ensuremath{\frac{\begin{array}{l}\mbox{\textsf{var}(\textit{a})}~~~\mbox{\ensuremath{\beta}~$\vdash$~\textsf{ground}(\textit{a})}~~~\mbox{\ensuremath{\beta}~$\vdash$~\textit{a}\ensuremath{:}\textsf{mut}(\textit{\_})}~~~\mbox{\ensuremath{\beta}~$\vdash$~\textsf{ground}(\textit{a}\textsf{@})}\\
\mbox{\textsf{var}(\textit{b})}~~~\mbox{\ensuremath{\beta}~$\vdash$~\textsf{fresh}(\textit{b})}\\
\mbox{\textit{c}$_a$ \textsf{=} \textsf{mutval}\textsf{/}\textsf{val\_rval}\ensuremath{\llbracket}\textit{a}\ensuremath{\rrbracket}}\\
\mbox{\textit{c}$_b$ \textsf{=} \textsf{val\_lval}\ensuremath{\llbracket}\textit{b}\ensuremath{\rrbracket}}\\
\mbox{$\langle$\ensuremath{\theta}$_0$$\rangle$~\textsf{emit\_d}(\textit{c}$_b$\texttt{\textbf{=}}\textit{c}$_a$, \ensuremath{\eta}, \ensuremath{\delta})~$\Rightarrow$~$\langle$\ensuremath{\theta}$\rangle$}~~~\end{array}}{\begin{array}{l}\mbox{$\langle$\ensuremath{\theta}$_0$$\rangle$~\textsf{gcomp}(\textit{b} \textsf{=} \textit{a}\textsf{@}, \ensuremath{\eta}, \ensuremath{\delta})~$\Rightarrow$~$\langle$\ensuremath{\theta}$\rangle$}~~~\end{array}}}
  \end{center}
\end{ruleset}

The compilation rules for operations on mutable variables are defined
in Figure~\ref{fig:mutcomprules}. The initialization of a mutable
\textit{a} \textsf{=} \textsf{initmut}(\ensuremath{\tau}, \textit{b}) (rule \rulename{InitMut}) is compiled
as a mutable assignment, but limited to the case where the reference
mode of \textit{a} is \textbf{0m} (that is, it has been
inferred that it will be a local mutable variable).
The built-in \textit{a} \ensuremath{\Leftarrow} \textit{b} is translated into an assignment
statement (rule \rulename{AssignMut}), that copies the value of
\textit{b} as the mutable value of \textit{a}.
The \rulename{ReadMut} rule defines the translation of \textit{b} \textsf{=}  \textit{a}\textsf{@}, an assignment statement that copies the value stored in the
mutable value of \textit{a} into \textit{b}, which must be a fresh
variable.    
Note that the case \textit{x} \textsf{=} \textit{a}\textsf{@} where \textit{x} is not fresh
can be reduced to (\textit{t} \textsf{=} \textit{a}\textsf{@}, \textit{x} \textsf{=} \textit{t}), with \textit{t} a new
variable, for which compilation rules exist.

\begin{ruleset}{Predicate compilation rules.}{fig:predcomprules}
  \begin{center}
    \rulebegin{Pred-D}
     \ensuremath{\frac{\begin{array}{l}\mbox{\textsf{det}(\textit{name})}~~~\mbox{([\textit{a}$_1$, \textsf{...}, \textit{a}$_n$], \textit{body}) \textsf{=} \textsf{lookup}(\textit{name})}\\
\mbox{\ensuremath{\theta}$_0$ \textsf{=} \textsf{bb\_empty}}\\
\mbox{$\langle$\ensuremath{\theta}$_3$$\rangle$~\textsf{bb\_newn}(2)~$\Rightarrow$~[\ensuremath{\delta}, \ensuremath{\delta}$_s$]~$\langle$\ensuremath{\theta}$_4$$\rangle$}\\
\mbox{$\langle$\ensuremath{\theta}$_4$$\rangle$~\textsf{gcomp}(\textit{body}, [\textsf{s}\ensuremath{\mapsto}\ensuremath{\delta}$_s$], \ensuremath{\delta})~$\Rightarrow$~$\langle$\ensuremath{\theta}$_5$$\rangle$}\\
\mbox{$\langle$\ensuremath{\theta}$_5$$\rangle$~\textsf{emit}(\texttt{\textbf{return}}, \ensuremath{\delta}$_s$)~$\Rightarrow$~$\langle$\ensuremath{\theta}$\rangle$}\\
\mbox{\textit{c}$_f$ \textsf{=} \textsf{c\_id}(\textit{name})}\\
\mbox{\textit{argdecls} \textsf{=} \textsf{argdecls}([\textit{a}$_1$, \textsf{...}, \textit{a}$_n$])}~~~\mbox{\textit{vardecls} \textsf{=} \textsf{vardecls}(\textit{body})}~~~\mbox{\textit{code} \textsf{=} \textsf{bb\_code}(\ensuremath{\delta}, \ensuremath{\theta})}\\
\mbox{$\langle$\textit{p}$_0$$\rangle$~\textsf{emitdecl}(\texttt{\textbf{void}} \textit{c}$_f$\texttt{\textbf{(}}\textit{argdecls}\texttt{\textbf{)}} \texttt{\textbf{\{}} \textit{vardecls}\texttt{\textbf{; }}\textit{code} \texttt{\textbf{\}}})~$\Rightarrow$~$\langle$\textit{p}$\rangle$}~~~\end{array}}{\begin{array}{l}\mbox{$\langle$\textit{p}$_0$$\rangle$~\textsf{pcomp}(\textit{name})~$\Rightarrow$~$\langle$\textit{p}$\rangle$}~~~\end{array}}}
  \end{center}
  \begin{center}
    \rulebegin{Pred-S}
     \ensuremath{\frac{\begin{array}{l}\mbox{\textsf{semidet}(\textit{name})}~~~\mbox{([\textit{a}$_1$, \textsf{...}, \textit{a}$_n$], \textit{body}) \textsf{=} \textsf{lookup}(\textit{name})}\\
\mbox{\ensuremath{\theta}$_0$ \textsf{=} \textsf{bb\_empty}}\\
\mbox{$\langle$\ensuremath{\theta}$_3$$\rangle$~\textsf{bb\_newn}(3)~$\Rightarrow$~[\ensuremath{\delta}, \ensuremath{\delta}$_s$, \ensuremath{\delta}$_f$]~$\langle$\ensuremath{\theta}$_4$$\rangle$}\\
\mbox{$\langle$\ensuremath{\theta}$_4$$\rangle$~\textsf{gcomp}(\textit{body}, [\textsf{s}\ensuremath{\mapsto}\ensuremath{\delta}$_s$, \textsf{f}\ensuremath{\mapsto}\ensuremath{\delta}$_f$], \ensuremath{\delta})~$\Rightarrow$~$\langle$\ensuremath{\theta}$_5$$\rangle$}\\
\mbox{$\langle$\ensuremath{\theta}$_5$$\rangle$~\textsf{emit}(\texttt{\textbf{return}} \texttt{\textbf{TRUE}}, \ensuremath{\delta}$_s$)~$\Rightarrow$~$\langle$\ensuremath{\theta}$_6$$\rangle$}\\
\mbox{$\langle$\ensuremath{\theta}$_6$$\rangle$~\textsf{emit}(\texttt{\textbf{return}} \texttt{\textbf{FALSE}}, \ensuremath{\delta}$_f$)~$\Rightarrow$~$\langle$\ensuremath{\theta}$\rangle$}\\
\mbox{\textit{c}$_f$ \textsf{=} \textsf{c\_id}(\textit{name})}\\
\mbox{\textit{argdecls} \textsf{=} \textsf{argdecls}([\textit{a}$_1$, \textsf{...}, \textit{a}$_n$])}~~~\mbox{\textit{vardecls} \textsf{=} \textsf{vardecls}(\textit{body})}~~~\mbox{\textit{code} \textsf{=} \textsf{bb\_code}(\ensuremath{\delta}, \ensuremath{\theta})}\\
\mbox{$\langle$\textit{p}$_0$$\rangle$~\textsf{emitdecl}(\texttt{\textbf{bool}} \textit{c}$_f$\texttt{\textbf{(}}\textit{argdecls}\texttt{\textbf{)}} \texttt{\textbf{\{}} \textit{vardecls}\texttt{\textbf{; }}\textit{code} \texttt{\textbf{\}}})~$\Rightarrow$~$\langle$\textit{p}$\rangle$}~~~\end{array}}{\begin{array}{l}\mbox{$\langle$\textit{p}$_0$$\rangle$~\textsf{pcomp}(\textit{name})~$\Rightarrow$~$\langle$\textit{p}$\rangle$}~~~\end{array}}}
  \end{center}
\end{ruleset}

\subsubsection{Compilation of Predicates}
\label{sec:compilation-heads}

The rules in the previous sections defined how goals are compiled.
In this section we will use those rules to compile predicates as C
functions. 
Figure~\ref{fig:predcomprules} provides rules that distinguish between
deterministic and semi-deterministic predicates. For a predicate with
\textit{name} \textsf{=} \textit{f}\textsf{/}\textit{n}, the \textsf{lookup}(\textit{name}) function returns
its arguments and body. The information from analysis of encoding
types and reference modes (Section~\ref{sec:comp-variables}) is used
by \textsf{argdecls} and \textsf{vardecls} to obtain the list of
argument and variable declarations for the program.  On the other
hand, \textsf{bb\_code} is a predefined operation that flattens the
basic blocks in its second argument \ensuremath{\theta} as a C block
composed of labels and statements. Finally, the \textsf{emitdecl}
operation is responsible for inserting the function declarations in
the compilation output \textit{p}.
Those definitions are used in the \rulename{Pred-D} and
\rulename{Pred-S} rules. 
The former compiles deterministic predicates by
binding a single success to a \texttt{\textbf{return}} statement, and
emits a C function returning no value. The latter compiles
semi-deterministic predicates by binding the continuations to code
that returns a \emph{true} or \emph{false} value depending on the
success and failure status.
Note that this code matches exactly the scheme needed in
Section~\ref{sec:comp-calls} to perform calls to \ip\ predicates
compiled as C functions.

\subsubsection{A Compilation Example}
\label{sec:compilation-example}

In order to clarify how the previous rules generate code, we include
here a code snippet (Figure~\ref{fig:compexample}) with several
types of variables accessed both from the scope of their first
appearance, and from outside that frame. We show also how this code is
compiled into two C functions. Note that redundant jumps and labels
have been simplified. It is composed of an encoding type definition
\textsf{flag}\textsf{/}\textsf{1}, two predicates that are compiled to C functions
(\textsf{p}\textsf{/}\textsf{1} semi-deterministic, \textsf{swmflag}\textsf{/}\textsf{1} deterministic),
and two predicates with annotations to unfold the code during
preprocessing (\textsf{mflag}\textsf{/}\textsf{2} and \textsf{swflag}\textsf{/}\textsf{2}). Note that by
unfolding the \textsf{mflag}\textsf{/}\textsf{2} predicate, a piece of illegal code (passing a
reference to a local mutable) becomes legal.  Indeed, this kind of
predicate unfolding has proved to be a good, manageable replacement
for the macros which usually appear in emulators written in
lower-level languages and which are often a source of mistakes.

\noindent
\begin{figure}[t]
  \begin{minipage}[t]{0.40\linewidth}
    \textbf{Source:}
    \vspace{-1em}
    \begin{tabbing}
\\
\textsf{:--} \textbf{\textsf{regtype}} \textsf{flag}\textsf{/}\textsf{1} \textsf{\texttt{+}} \textsf{low}(\=\textsf{int32}).\\
\textsf{flag} \textsf{:=} \textsf{off} \textsf{$\mid$} \textsf{on}.\\
\textsf{:--} \textbf{\textsf{pred}} \textsf{p}(\=\textsf{\texttt{+}}\textit{I}) \textsf{::} \textsf{flag}.\\
\textsf{p}(\=\textit{I}) \textsf{:--}\\
\hspace{2em}\=\textsf{mflag}(\=\textit{I}, \textit{A}),\\
\>\=\textit{A} \textsf{=} \=\textit{B},\\
\>\textsf{swmflag}(\=\textit{B}),\\
\>\=\textit{A}\textsf{@} \textsf{=} \=\textsf{on}.\\
\textsf{:--} \textbf{\textsf{pred}} \textsf{mflag}\textsf{/}\textsf{2} \textsf{\texttt{+}} \textsf{unfold}.\\
\textsf{mflag}(\=\textit{I}, \textit{X}) \textsf{:--}\\
\hspace{2em}\=\=\textit{X} \textsf{=} \=\textsf{initmut}(\=\textsf{flag}, \textit{I}).\\
\textsf{:--} \textbf{\textsf{pred}} \textsf{swmflag}(\=\textsf{\texttt{+}}\textit{I}) \textsf{::} \textsf{mut}(\=\textsf{flag}).\\
\textsf{swmflag}(\=\textit{X}) \textsf{:--}\\
\hspace{2em}\=\textsf{swflag}(\=\textit{X}\textsf{@}, \textit{X}$_2$),\\
\>\textit{X} \ensuremath{\Leftarrow} \textit{X}$_2$.\\
\textsf{:--} \textbf{\textsf{pred}} \textsf{swflag}\textsf{/}\textsf{2} \textsf{\texttt{+}} \textsf{unfold}.\\
\textsf{swflag}(\=\textsf{on}, \textsf{off}).\\
\textsf{swflag}(\=\textsf{off}, \textsf{on}).\\
\end{tabbing}

  \end{minipage}
  \hspace{2em}
  \begin{minipage}[t]{0.45\linewidth}
    \textbf{Output:}
\begin{lstlisting}[basicstyle=\footnotesize\ttfamily\selectfont, numbers=left, numberstyle=\tiny, frame=single]{}
bool p(int32 i) {
    int32 a;
    int32 *b;
    int32 t;
    b = &a;
    swmflag(b);
    t = a;
    if (t == 1) goto l1; else goto l2;
l1: return TRUE;
l2: return FALSE;
}
void smwflag(int32 *x) {
    int32 t;
    int32 x2;
    t = *x;
    if (t == 1) goto l1; else goto l2;
l1: x2 = 0;
    goto l3;
l2: x2 = 1;
l3: *x = x2;
    return;
}
\end{lstlisting}
  \end{minipage}
  \caption{\ip\ compilation example}
  \label{fig:compexample}
\end{figure}

\subsubsection{Related compilation schemes}
\label{sec:relat-comp-schem}
Another compilation scheme
which produces similar code is described in~\cite{fergus-mercury-hlc}.
There are, however, significant differences, of which  we will mention just a
few.  One of them is the source language and the constraints imposed
on it.
In our case we aim at writing a WAM emulator in \ip\ from which C code
is generated with the constraint that it has to be identical (or, at
least, very close) to a hand-written and hand-optimized emulator,
including the implementation of the internal data structures.  This
has forced us to pay special attention to the compilation and
placement of data, use mutable variables, and ignore for now
non-deterministic control.
Also, in this work we use an intermediate representation based
on basic blocks, which makes it easier to interface with internal
back-ends for compilers other than GCC, such as LLVM~\cite{LLVM:CGO04}
(which enables JIT compilation from the same representation).

\section{Extensions for Emulator Generation in \ip}
\label{sec:generating-emul-main}

The dialect and compilation process that has been described so far is
general enough to express the instructions in a typical WAM emulator,
given some basic built-ins about operations on data types, memory
stacks, and O.S. interface. However, combining those pieces of code
together to build an \emph{efficient} emulator requires a compact
encoding of the bytecode language, and a bytecode fetching and
dispatching loop that usually needs a tight control on low-level data
and operations that we have not included in the \ip\ language.  In
~\cite{morales05:generic_eff_AM_implem_iclp} we showed that it is
possible to automate the generation of the emulator from generic
instruction definitions and annotations stating how the bytecode is
encoded and decoded.  Moreover, this process was found to be highly
mechanizable, while making instruction code easier to manage and
other optimizations (such as instruction merging) easier to perform.
In this section we show how this approach is integrated in the
compilation process, by including the emulator generation as part of
it.

\subsection{Defining  WAM instructions in \ip}
\label{sec:reflecting}

The definition of every WAM instruction in \ip\ looks just like a
regular 
predicate, and
the types, modes, etc.\ of each of their arguments have to be declared
using (Ciao) assertions.
As an example, Figure~\ref{fig:unif-const} shows \ip\ code corresponding to the
definition of an instruction
which tries to unify a term and a constant. The \textbf{\textsf{pred}}
declaration states that the first argument is a mutable variable
and that the second is a tagged word containing a constant.
It includes a sample implementation of the WAM dereference
operation,
which follows a reference chain and
stops when the value pointed to is the same as the pointing term, or
when the chain cannot be followed any more. Note the use of the native
type \textsf{tagged}\textsf{/}\textsf{2} and the operations \textsf{tagof}\textsf{/}\textsf{2} and
\textsf{tagval}\textsf{/}\textsf{2} which access the tag and the associated value of a
tagged word, respectively. Also note that the \textsf{tagval}\textsf{/}\textsf{2} of a
tagged word with \textsf{ref} results in a mutable variable, as can
be recognized in the code. Other native operations include
\textsf{trail\_cond}\textsf{/}\textsf{1}, \textsf{trail\_push}\textsf{/}\textsf{1}, and operations to
manage the emulator stacks.
Note the special predicates \textsf{next\_ins} and
\textsf{fail\_ins}. They execute the next instruction or the failure
instruction, respectively. The purpose of the next instruction is to
continue the emulation of the next bytecode instruction (which can be
considered as a recursive call to the emulator itself, but which will be
defined as a built-in).
The failure instruction must take care of unwinding the stacks at the
WAM 
level and selecting the next bytecode instruction to execute (to
implement the failure in the emulator). As a usual instruction, it can
be defined by calling built-ins or other \ip\ code, and it should
finally 
include a call to \textsf{next\_ins} to continue the emulation. Since
this instruction is often invoked from other instructions, a special
treatment is given to share its code, which will be described later.

\begin{figure}
\centering\small
\begin{minipage}[t]{0.80\linewidth}
\begin{tabbing}
\\
\textsf{:--} \textbf{\textsf{pred}} \textsf{u\_cons}(\=\textsf{\texttt{+}}, \textsf{\texttt{+}}) \textsf{::} \textsf{mut}(\=\textsf{tagged}) \textsf{*} \textsf{constagged}.\\
\textsf{u\_cons}(\=\textit{A}, \textit{Cons}) \textsf{:--}\\
\hspace{2em}\=\textsf{deref}(\=\textit{A}\textsf{@}, \textit{T}$_d$),\\
\>( \=\textsf{tagof}(\=\textit{T}$_d$, \textsf{ref}) \ensuremath{\rightarrow} \textsf{bind\_cons}(\=\textit{T}$_d$, \textit{Cons}), \textsf{next\_ins}\\
\hspace{2em}\=\textsf{;} \=\=\textit{T}$_d$ \textsf{=} \=\textit{Cons} \ensuremath{\rightarrow} \textsf{next\_ins}\\
\hspace{2em}\=\textsf{;} \=\textsf{fail\_ins}\\
\>).\\
\end{tabbing}

\vspace{-8ex}
\begin{tabbing}
\\
\textsf{:--} \textbf{\textsf{pred}} \textsf{deref}\textsf{/}\textsf{2}.\\
\textsf{deref}(\=\textit{T}, \textit{T}$_d$) \textsf{:--}\\
\hspace{2em}\=( \=\textsf{tagof}(\=\textit{T}, \textsf{ref}) \ensuremath{\rightarrow}\\
\> \> \hspace{2em}\=\=\textit{T}$_1$ \textsf{=} \=\ensuremath{\sim}\textsf{tagval}(\=\textit{T})\textsf{@},\\
\> \> \>(\=\=\textit{T} \textsf{=} \=\textit{T}$_1$ \ensuremath{\rightarrow} \=\textit{T}$_d$ \textsf{=} \=\textit{T}$_1$ \textsf{;} \=\textsf{deref}(\=\textit{T}$_1$, \textit{T}$_d$))\\
\hspace{2em}\=\textsf{;} \=\=\textit{T}$_d$ \textsf{=} \=\textit{T}\\
\>).\\
\end{tabbing}

\vspace{-8ex}
\begin{tabbing}
\\
\textsf{:--} \textbf{\textsf{pred}} \textsf{bind}\textsf{/}\textsf{2}.\\
\textsf{bind\_cons}(\=\textit{Var}, \textit{Cons}) \textsf{:--}\\
\hspace{2em}\=(\=\textsf{trail\_cond}(\=\textit{Var}) \ensuremath{\rightarrow} \textsf{trail\_push}(\=\textit{Var}) \textsf{;} \=\textsf{true}),\\
\hspace{2em}\=\ensuremath{\sim}\textsf{tagval}(\=\textit{Var}) \ensuremath{\Leftarrow} \textit{Cons}.\\
\end{tabbing}

\end{minipage}
  \caption{Unification with a constant and auxiliary definitions.}
  \label{fig:unif-const}
\end{figure}

The compilation process is able to unfold (if so desired) the
definition of the predicates called by \textsf{u\_cons}\textsf{/}\textsf{2}
and to propagate information inside the instruction, in
order to optimize the resulting piece of the emulator. After the set
of transformations that instruction definitions are subject to, and
other optimizations on the output (such as transformation of some
recursions into loops) the generated C code is of high quality (see,
for example, Figure~\ref{fig:code-gen-ux_cons}, for the code
corresponding to a specialization of this instruction).

Our approach has been to define a reduced number of instructions (50
is a ballpark figure) and let the merging and specialization process
(see Section~\ref{sec:automatic-spec-merg}) generate all instructions
needed to have a competitive emulator. Note that efficient emulators
tend to have a large number of instructions (hundreds, or even
thousands, in the case of Quintus Prolog)
and many of them are variations (obtained through specialization,
merging, etc., normally done manually) on ``common blocks.''  These common
blocks are the simple
instructions we aim at representing explicitly in \ip.

In the experiments we performed (Section~\ref{sec:evaluation}) the
emulator with a largest number of instructions had 199 different
opcodes (not counting those which result from padding some other
instruction with zeroes to ensure a correct alignment in memory).  A
simple instruction set is easier to maintain and its consistency is easier
to ensure.  Complex instructions are generated automatically in a (by
construction) sound way from this initial ``seed.''

\subsection{An Emulator Specification in \ip}
\label{sec:emul-spec}

Although \ip\ could be powerful enough to describe the emulation loop,
as mentioned before we leverage on previous
work~\cite{morales05:generic_eff_AM_implem_iclp} in which \lc\
emulators were automatically built from definitions of instructions
written in \la\ and their corresponding code written in \lc.  Bytecode
representation, compiler back-end, and an emulator (including the
emulator loop) able to understand \lb\ code can be automatically
generated from those components.  In our current setting, definitions for
\la\ instructions are written in \lsourcerefl\ (recall
Figure~\ref{fig:lang-levels-new}) and these definitions can be
automatically translated into \lc\ by the \ip\ compiler.  We are thus
spared of making this compiler more complex than needed.  More details
on this process will be given in the following section.

\subsection{Assembling the Emulator}
\label{sec:generating-emul}

\begin{figure}[t]
\begin{center}

\pgfdeclarelayer{background}
\pgfdeclarelayer{foreground}
\pgfsetlayers{background,main,foreground}

\tikzstyle{part} = [draw, dashed, text width=5em, fill=white, text centered, minimum height=2.5em]
\tikzstyle{epart} = [draw, densely dashed, text width=7em, fill=white, text centered, minimum height=2.5em]
\tikzstyle{boldline} = [line width=0.05cm]
\tikzstyle{ann} = [above, text width=5em]
\tikzstyle{annbelow} = [below, text width=5em]
\tikzstyle{programpart} = [fill=blue!10, draw, drop shadow]
\tikzstyle{mpart} = [draw, dashed, text width=3em, fill=white, text centered, minimum height=2.5em]
\tikzstyle{missingpart} = [draw, dashed, text width=3em, fill=white, text centered, minimum height=1em]
\def\blockdist{2.3}
\def\edgedist{2.5}
\def\cmargin{0.2}
\def\bigmargin{0.4}

\scalebox{0.85}{
  \begin{tikzpicture}
    \node (EmuDef) {\progtext{\proglang{E$^r$}{\lsourcerefl}}{\la\ emulator}{source code}};
    \path (EmuDef)+(0,-2em) node (bcdefs) [epart] {Bytecode definitions};
    \path (bcdefs)+(0,-3em) node (insdefs) [epart] {Instruction definitions in \lsourcerefl};
    \path (EmuDef)+(4,2em) node (machname) {Machine \machall};
    \path (machname)+(0,-3em) node (machbc) [mpart] {\mach{enc} \\ \mach{dec} \\ \mach{arg_I}};
    \path (machbc)+(0,-4em) node (machins) [mpart] {\mach{ins'} \\ \mach{def_I}};
    \path [draw, ->] (bcdefs) -- node [ann] {} (machbc.west |- bcdefs) ;
    \path [draw, ->] (insdefs) -- node [ann] {} (machins.west |- insdefs);
    \draw [->] (machbc.east -| machname.north east)+(\cmargin,0) -- node [ann] {\mach{enc} \mach{ins'}} +(\edgedist,0) 
        node[right] (backend) [part] {\la\ to \lb\ back-end};
    \path (backend)+(0,2.5em) node (compdefs) [missingpart] {...};
    \path (compdefs)+(0,1.5em) node (compname) {\lsource\ \textbf{to} \lb\ \textbf{compiler}};
    \draw [->, boldline] (machins.east -| machname.north east)+(\cmargin,-\bigmargin) 
        -- node [annbelow] {\textbf{emucomp} \mach{dec} \mach{arg_I} \mach{ins'} \mach{def_I}} +(\edgedist,-\bigmargin) 
        node[right] [programpart, minimum height=3em] {\progtext{\proglang{E$^r$}{\lc}}{\lb\ emulator}{low-level code}};
    \begin{pgfonlayer}{background}
        \path (EmuDef.north west)+(-\cmargin,\cmargin) node (a) {};
        \path (insdefs.south east -| EmuDef.south east)+(+\cmargin,-\cmargin) node (b) {};
        \path[programpart, boldline] (a) rectangle (b);
        \path (machname.north west)+(-\cmargin,\cmargin) node (a) {};
        \path (machins.south east -| machname.south east)+(+\cmargin,-\cmargin) node (b) {};
        \path[programpart, fill=white] (a) rectangle (b);
        \path (compname.north west)+(-\cmargin,\cmargin) node (a) {};
        \path (backend.south east -| compname.south east)+(+\cmargin,-\cmargin) node (b) {};
        \path[programpart] (a) rectangle (b);
        \coordinate (c) at ($ (insdefs.south) + (0,-\cmargin) $);
        \coordinate (d) at ($ (machins.south) + (0,-\cmargin) $);
        \draw[->, boldline] (c) -- +(0, -\bigmargin) -| node[near start, below] {\textbf{mgen}} (d);
    \end{pgfonlayer}
\end{tikzpicture}}
\end{center}
  \caption{From \ip\ definitions to \lb\ emulator in \lc.}
  \label{fig:ip-emucomp}
\end{figure}

We will describe now the process that takes an \ip\ representation of
an abstract machine and obtains a full-fledged implementation of this
machine.  The overall process is sketched in
Figure~\ref{fig:ip-emucomp}, and can be divided into two stages, which
we have termed \mgen\ and \emucomp.  The emulator definition, \emudef,
is a set of predicates and assertions written in \ip, and \mgen\ is
basically a normalization process where the source \emudef\ is
processed to obtain a machine definition \machall. This definition
contains components describing the instruction semantics written in
\ip\ and a set of \emph{hints} about the bytecode representation
(e.g., numbers for the bytecode instructions).  \machall\ is then
processed by an emulator compiler \emucomp\ which generates a bytecode
emulator for the language \lb, written in the language \lc. The
machinery to encode \la\ programs into the bytecode representation
\lb\ is also given by definitions in \machall.

Using the terminology in~\cite{morales05:generic_eff_AM_implem_iclp},
we denote the components of \machall\ as follows:

\begin{center}
   \ensuremath{\machall} \textsf{=} (\ensuremath{\machall_{enc}}, \ensuremath{\machall_{dec}}, \ensuremath{\machall_{arg_I}}, \ensuremath{\machall_{def_I}}, \ensuremath{\machall_{ins'}})
\end{center}

First, the relation between
\la\ and \lb\ is given by means of several components:\footnote{The
  complete description
  includes all elements for a WAM: \texttt{X} and
  \texttt{Y} registers, atoms, numbers, functors, etc.}

\begin{description}
\item[\underline{\ensuremath{\machall_{enc}}}] declares how the bytecode
  encodes \la\ instructions and
  data: e.g.,  \texttt{X(0)} is encoded as the number \textsf{0} for 
  an instruction which needs access to some \texttt{X} register.
\item[\underline{\ensuremath{\machall_{dec}}}] declares how the bytecode should
  be decoded to give back the initial instruction format in \la: e.g.,
  for an instruction which uses as argument an \texttt{X} register, a
  \textsf{0} means \texttt{X(0)}.
\end{description}

The remaining of the components of \machall\ capture the meaning of the
(rather low level) constituents of \la, providing a description of
each instruction.  Those components do not make bytecode
representation issues explicit, as they have already been specified in
\ensuremath{\machall_{enc}} and \ensuremath{\machall_{dec}}. In the present
work definitions for \la\ instructions are given in \lsourcerefl,
instead of in \lc\ as was done in the formalization presented 
in~\cite{morales05:generic_eff_AM_implem_iclp}. The reason for this
change is that in~\cite{morales05:generic_eff_AM_implem_iclp} the
final implementation language (\lc, in which emulators were generated)
was also the language in which each basic instruction was assumed to
be written.  However, in our case, instructions are obviously
written in \lsourcerefl\ (i.e., \ip, which is more amenable to automatic program
transformations) and it makes more sense to use it directly in the
definition of \machall.
Using \lsourcerefl\ requires, however, extending and/or modifying the remaining
parts of \machall\ with respect to the original definition as follows:

\begin{description}
\item[\underline{\ensuremath{\machall_{arg_I}}}] which assigns a pair $(T, mem)$ to
  every expression in \la, where $T$ is the type of the expression and
  $mem$ is the translation of the expression into \lc. For example,
  the type of \texttt{X(0)} is \textsf{mut}(\textsf{tagged}) and its memory
  location is \texttt{\textbf{\&}}(\texttt{\textbf{x[}}0\texttt{\textbf{]}}), assuming
  \texttt{X} registers end up in an array.%
  \footnote{This definition has been expanded with respect to its
    original \mach{arg} definition in order to include the \ip\ type
    in addition to the memory location.}

\item[\underline{\mach{def_I}}] which contains the definition of each
  instruction in \lsourcerefl.%

\item[\underline{\ensuremath{\machall_{ins'}}}] which describes the instruction set
  with opcode numbers and the format of each instruction, i.e., the
  type in \la\ for each instruction argument: e.g., \texttt{X}
  registers, \texttt{Y} registers, integers, atoms, functors.
\end{description}

The rest of the components and \ensuremath{\machall_{ins'}} are used by
the emulator compiler to generate an \lb\ emulator written in \lc.
A summarized definition of the emulator compiler and how it uses the
different pieces in \machall\ can be found in
Figure~\ref{fig:emucomp}. The \rulename{Emu} rule defines a function
that contains the emulator loop. It is similar to the 
\rulename{Pred-D} rule already presented, but takes parts of the source
code from \machall. It generates a list of basic block identifiers for
each instruction, and a basic block identifier for the emulator loop
entry. The \rulename{Swr} rule is used to insert a \emph{switch}
statement that implements the opcode fetching, and jumps to the code
of each instruction. The \rulename{Ins} rule is used to generate the
code for each instruction. To implement the built-ins
\textsf{next\_ins} and \textsf{fail\_ins}, two special continuations
\textsf{ni} and \textsf{fi} are stored in the continuation
mapping. The continuation to the failure instruction is bound to the
\ensuremath{\delta}$_f$ basic block identifier (assuming that the 
\textit{op}$_f$ opcode is that of the failure instruction). The 
\rulename{FailIns} rule 
includes a special case in \textsf{gcomp} that
implements this call. The continuation to the next instruction is a
pair of the basic block that begins the emulator switch, and a piece
of C code that moves the bytecode pointer to the next instruction
(that is particular to each instruction, and is returned by
\textsf{insdef} alongside with its code). The 
\rulename{NextIns} rule 
emits code that executes that code and jumps to 
opcode fetching.

\begin{ruleset}{Emulator compiler.}{fig:emucomp}
  \begin{center}
    \rulebegin{Emu}
     \ensuremath{\frac{\begin{array}{l}\mbox{\ensuremath{\machall_{ops}} \textsf{=} [\textit{op}$_1$, \textsf{...}, \textit{op}$_n$]}\\
\mbox{\ensuremath{\theta}$_0$ \textsf{=} \textsf{bb\_empty}}\\
\mbox{$\langle$\ensuremath{\theta}$_0$$\rangle$~\textsf{bb\_newn}(\textit{n} \textsf{\texttt{+}} 1)~$\Rightarrow$~[\ensuremath{\delta}, \ensuremath{\delta}$_1$, \textsf{...}, \ensuremath{\delta}$_n$]~$\langle$\ensuremath{\theta}$_1$$\rangle$}\\
\mbox{$\langle$\ensuremath{\theta}$_1$$\rangle$~\textsf{emit\_switch}(\texttt{\textbf{get\_opcode}}\texttt{\textbf{(}}\texttt{\textbf{)}}, \ensuremath{\machall_{ops}}, [\ensuremath{\delta}$_1$, \textsf{...}, \ensuremath{\delta}$_n$], \ensuremath{\delta})~$\Rightarrow$~$\langle$\ensuremath{\theta}$_2$$\rangle$}\\
\mbox{\ensuremath{\langle}\ensuremath{\theta}$_2$\ensuremath{\rangle} \ensuremath{\forall}\textit{i}\ensuremath{=}1\ensuremath{..}\textit{n} \textsf{inscomp}(\textit{op}$_i$, \ensuremath{\delta}, \ensuremath{\delta}$_i$, [\textsf{fi}\ensuremath{\mapsto}\ensuremath{\delta}$_f$]) \ensuremath{\langle}\ensuremath{\theta}\ensuremath{\rangle}}\\
\mbox{\textit{code} \textsf{=} \textsf{bb\_code}(\ensuremath{\delta}, \ensuremath{\theta})}\\
\mbox{$\langle$\textit{p}$_0$$\rangle$~\textsf{emitdecl}(\texttt{\textbf{void}} \texttt{\textbf{emu}}\texttt{\textbf{(}}\texttt{\textbf{)}} \texttt{\textbf{\{}} \textit{code} \texttt{\textbf{\}}})~$\Rightarrow$~$\langle$\textit{p}$\rangle$}~~~\end{array}}{\begin{array}{l}\mbox{$\langle$\textit{p}$_0$$\rangle$~\textsf{emucomp}~$\Rightarrow$~$\langle$\textit{p}$\rangle$}~~~\end{array}}}
  \end{center}
  \begin{center}
    \rulebegin{Swr}
     \ensuremath{\frac{\begin{array}{l}\mbox{$\langle$\ensuremath{\theta}$_0$$\rangle$~\textsf{emit}(\texttt{\textbf{switch (}}\textit{x}\texttt{\textbf{) \{}}\texttt{\textbf{case}} \textit{v}$_1$\texttt{\textbf{:}} \texttt{\textbf{goto}} \ensuremath{\delta}$_1$\textbf{;} \textsf{...}\textbf{;} \texttt{\textbf{case}} \textit{v}$_n$\texttt{\textbf{:}} \texttt{\textbf{goto}} \ensuremath{\delta}$_n$\textbf{;} \texttt{\textbf{\}}}, \ensuremath{\delta})~$\Rightarrow$~$\langle$\ensuremath{\theta}$\rangle$}~~~\end{array}}{\begin{array}{l}\mbox{$\langle$\ensuremath{\theta}$_0$$\rangle$~\textsf{emit\_switch}(\textit{x}, [\textit{v}$_1$, \textsf{...}, \textit{v}$_n$], [\ensuremath{\delta}$_1$, \textsf{...}, \ensuremath{\delta}$_n$], \ensuremath{\delta})~$\Rightarrow$~$\langle$\ensuremath{\theta}$\rangle$}~~~\end{array}}}
  \end{center}
  \begin{center}
    \rulebegin{Ins}
     \ensuremath{\frac{\begin{array}{l}\mbox{(\textit{body}, \textit{nextp}) \textsf{=} \textsf{insdef}(\textit{opcode})}\\
\mbox{$\langle$\ensuremath{\theta}$_2$$\rangle$~\textsf{gcomp}(\textit{body}, \ensuremath{\eta}[\textsf{ni}\ensuremath{\mapsto}(\ensuremath{\delta}$_0$, \textit{nextp})], \ensuremath{\delta})~$\Rightarrow$~$\langle$\ensuremath{\theta}$\rangle$}~~~\end{array}}{\begin{array}{l}\mbox{$\langle$\ensuremath{\theta}$_0$$\rangle$~\textsf{inscomp}(\textit{opcode}, \ensuremath{\delta}$_0$, \ensuremath{\delta}, \ensuremath{\eta})~$\Rightarrow$~$\langle$\ensuremath{\theta}$\rangle$}~~~\end{array}}}
  \end{center}
  \begin{center}
    \hspace{-1em}
    \rulebegin{NextIns}
     \ensuremath{\frac{\begin{array}{l}\mbox{(\ensuremath{\delta}$_s$, \textit{nextp}) \textsf{=} \ensuremath{\eta}(\textsf{ni})}\\
\mbox{$\langle$\ensuremath{\theta}$_0$$\rangle$~\textsf{emit}(\textit{nextp}, \ensuremath{\delta})~$\Rightarrow$~$\langle$\ensuremath{\theta}$_1$$\rangle$}\\
\mbox{$\langle$\ensuremath{\theta}$_1$$\rangle$~\textsf{emit}(\texttt{\textbf{goto}} \ensuremath{\delta}$_s$, \ensuremath{\delta})~$\Rightarrow$~$\langle$\ensuremath{\theta}$\rangle$}~~~\end{array}}{\begin{array}{l}\mbox{$\langle$\ensuremath{\theta}$_0$$\rangle$~\textsf{gcomp}(\textsf{next\_ins}, \ensuremath{\eta}, \ensuremath{\delta})~$\Rightarrow$~$\langle$\ensuremath{\theta}$\rangle$}~~~\end{array}}}
    $~$
    \rulebegin{FailIns}
     \ensuremath{\frac{\begin{array}{l}\mbox{\ensuremath{\delta}$_s$ \textsf{=} \ensuremath{\eta}(\textsf{fi})}\\
\mbox{$\langle$\ensuremath{\theta}$_0$$\rangle$~\textsf{emit}(\texttt{\textbf{goto}} \ensuremath{\delta}$_s$, \ensuremath{\delta})~$\Rightarrow$~$\langle$\ensuremath{\theta}$\rangle$}~~~\end{array}}{\begin{array}{l}\mbox{$\langle$\ensuremath{\theta}$_0$$\rangle$~\textsf{gcomp}(\textsf{fail\_ins}, \ensuremath{\eta}, \ensuremath{\delta})~$\Rightarrow$~$\langle$\ensuremath{\theta}$\rangle$}~~~\end{array}}}
  \end{center}
\end{ruleset}

The abstract machine component \ensuremath{\machall_{ins'}} is used to
obtain the \textit{name} and data \textit{format} of the
instruction identified by a given \textit{opcode}. From the format
and \ensuremath{\machall_{arg_I}} definition, a type, an encoding type, and a
custom \emph{r-value} for each instruction argument are filled. In
this way, the compilation process can transparently work with
variables whose value is defined from the operand information in a
bytecode stream (e.g., an integer, a machine register, etc.).

\paragraph{Relation with Other Compilation Schemes.}
The scheme of the generated emulator code is somewhat similar to what
the Janus compilation scheme~\cite{jc-janus} produces for general
programs. In Janus, addresses for continuations are either known
statically (e.g., for calls, and therefore a direct, static jump to a
label can be performed) or are popped from the stack when returning.
Since labels cannot be directly assigned to variables in standard C,
an implementation workaround is made by assigning a number to each possible
return address (and it is this number which is pushed onto / popped
from the stack) and using a \emph{switch} to relate these numbers with
the code executing them.
In our case we have a similar \emph{switch}, but it relates each
opcode with its corresponding instruction code, and it is executed
every time a new instruction is dispatched.

We want to note that we deliberately stay within standard C in this
presentation: taking advantage of C extensions, such as storing labels
in variables, which are provided by \texttt{gcc} and used, for
example, in~\cite{henderson95:compiling_using_gnu_c,codognet95:wamcc},
is out of the scope of this paper.  These optimizations are not
difficult to add as code generation options, and therefore they should
not be part of a basic scheme.  Besides, that would make it difficult
to use compilers other than \texttt{gcc}.

\begin{example}
  
As an example, from the instruction in Figure~\ref{fig:unif-const},
which unifies a term living in some variable with a constant, we can
derive a specialized version in which the term is assumed to live in
an \texttt{X} register. The declaration:
\begin{quotedcode}\begin{tabbing}
\\
\textsf{:--} \textbf{\textsf{ins\_alias}}(\=\textsf{ux\_cons}, \textsf{u\_cons}(\=\textsf{xreg\_mutable}, \textsf{constagged})).\\
\end{tabbing}
\end{quotedcode}
\noindent
assigns the (symbolic) name \textsf{ux\_cons} to the new instruction,
and specifies that the first argument lives in an \texttt{X}
register. The declaration: 
\begin{quotedcode}\begin{tabbing}
\\
\textsf{:--} \textbf{\textsf{ins\_entry}}(\=\textsf{ux\_cons}).\\
\end{tabbing}
\end{quotedcode}
\noindent
indicates that the emulator has an entry for that
instruction.\footnote{We optionally allow a pre-assignment of an
  opcode number to each instruction entry. Different assignments of
  instruction numbers to opcodes can impact the final performance, as
  they dictate how the code is laid out in the emulator switch which
  affects, for example, the behavior of the cache.}
Figure~\ref{fig:code-gen-ux_cons} shows the code generated for the
instruction (right) and a fragment of the emulator generated by the
emulator compiler in Figure~\ref{fig:emucomp}.
\end{example}

\begin{figure}[t]
  \centering
\hspace*{1em}
  \begin{minipage}[t]{0.45\linewidth}
  \begin{lstlisting}[basicstyle=\footnotesize\ttfamily\selectfont, numbers=left, numberstyle=\tiny, frame=single]{}
loop:
  switch(Op(short,P,0)) {
  ...
  case 97: goto ux_cons;
  ...
  }
  ...
ux_cons:
  tagged t;
  t = X(Op(short,P,2));
  deref(&t);
  if (tagged_tag(t) != REF)
    goto ux_cons__0;
  bind_cons(t, Op(tagged,P,4));
  goto ux_cons__1;
ux_cons__0:
  if (t != Op(tagged,P,4))
    goto fail_ins;
ux_cons__1:
  P = Skip(P,8);
  goto loop;
  ...
\end{lstlisting}
  \end{minipage}
\hfill
  \begin{minipage}[t]{0.4\linewidth}
  \begin{lstlisting}[basicstyle=\footnotesize\ttfamily\selectfont, numbers=left, numberstyle=\tiny, frame=single]{}
void deref(tagged_t *a0) {
  tagged_t t0;
deref:
  if (tagged_tag(*a0) == REF)
    goto deref__0;
    else goto deref__1;
deref__0:
  t0 = *(tagged_val(*a0));
  if ((*a0) != t0)
    goto deref__2;
    else goto deref__1;
deref__2:
  *a0 = t0;
  goto deref;
deref__1:
  return;
}
\end{lstlisting}
  \end{minipage}
\hspace*{1em}
  \caption{Code generated for a simple instruction.}
  \label{fig:code-gen-ux_cons}
\end{figure}

\section{Automatic Generation of Abstract Machine Variations}
\label{sec:automatic-spec-merg}

Using the techniques described in the previous section we now address
how abstract machine variations can be generated automatically.
Substantial work has been devoted to abstract machine generation
strategies
such as,
e.g.,~\cite{demoen00:wam-variations,insmerging-sicstus,Cs07aregister-free},
which explore different design variations with the objective of
putting together highly optimized emulators.  However, as shown
previously, by making the semantics of the abstract machine
instructions explicit in a language like \ip, which is easily amenable
to automatic processing, such variations can be formulated in a
straightforward way mostly as automatic transformations.  Adding new
transformation rules and testing them together with the existing ones
becomes then a relatively easy task.

We will briefly describe some of these transformations, which will be
experimentally evaluated in Section~\ref{sec:evaluation}.  Each
transformation is identified by a two-letter code.
We make a distinction between transformations which change the
instruction set (by creating new instructions) and those which only
affect the way code is generated.

\subsection{Instruction Set Transformations}
\label{sec:gen-instructions}

\begin{figure}[t]
  \centering
    \includegraphics[width=0.90\linewidth]{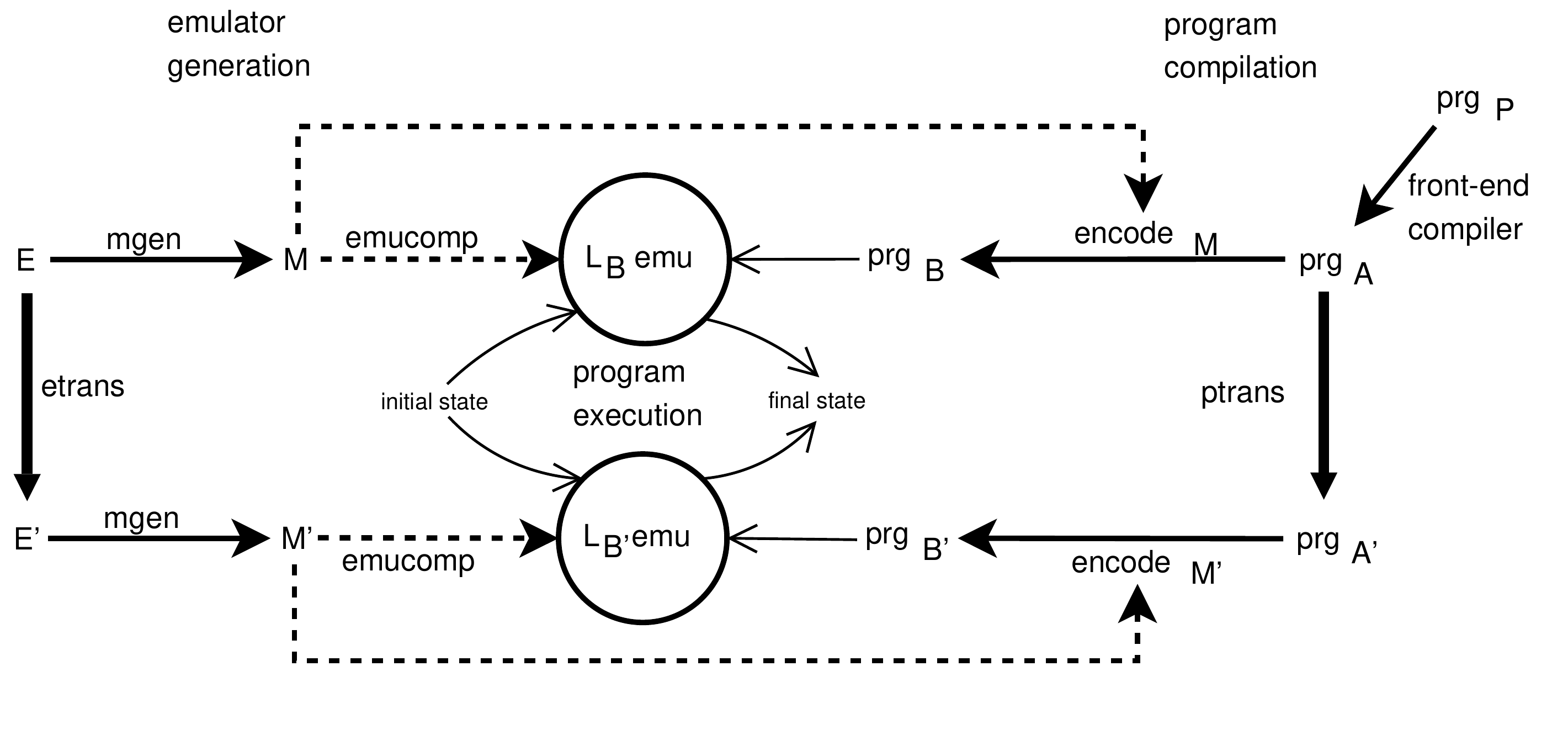}
  \caption{Application of an instruction set transformation ($ptrans$, $etrans$).}
  \label{fig:iset-trans}
\end{figure}

Let us define an instruction set transformation as a pair
($ptrans$,$etrans$), so that $ptrans$ transforms programs from two
symbolic bytecode languages \la\ and (a possibly different) \la'
\footnote{Those languages can be different, for example, if the
  transformation adds or removes some instructions}
and $etrans$ transforms the abstract machine definition within \lsourcerefl.
Figure~\ref{fig:iset-trans} depicts the relation between emulator
generation, program compilation, program execution, and instruction
set transformations. The full emulator generation includes $etrans$ as
preprocessing before \mgen\ is performed. The resulting emulator is
able to interpret transformed programs after $ptrans$ is applied
(before bytecode encoding), that is, the new compiler is obtained by
including $ptrans$ as a new compilation phase.

Note that both $ptrans$ and $etrans$ are working at the symbolic
bytecode level. It is easier to work with a symbolic \la\ program
than with the stream of bytes that represents \lb\ code, and it is easier 
to transform instructions written in \lsourcerefl\ and specified at the \la\ 
level, than those already written in \lc\ code (where references to
\lb\ code and implementation details obscure the actual
semantics). Reasoning about the correctness of the global
transformation that affects the \la\ program and the instruction code
is also easier in the \lsourcerefl\ specification of the emulator
instructions
than in a low-level \lc\ emulator (assuming the correctness of the
emulator generation process).

In the following sections we will review the instruction set
transformations currently available.  Although more transformations can of
course be applied, the current set is designed with the aim of
generating, from simple \ip\ definitions, an emulator which is as
efficient as a hand-crafted, carefully tuned one.

\subsubsection{Instruction Merging [\textbf{im}]}
\label{sec:instr-merging}

\emph{Instruction Merging} generates larger instructions from
sequences of smaller ones, and is aimed at saving fetch cycles at the
expense of a larger instruction set and, therefore, an increased
\texttt{switch} size.  This technique has been used extensively in
high-performance systems (e.g., Quintus Prolog, SICStus, Yap, 
etc.).  The performance of different combinations has been studied
empirically~\cite{insmerging-sicstus}, but in that work new
instructions were generated by hand, although deciding which
instructions had to be created was done by means of profiling. In our
framework only a single declaration is needed to emit code for a new,
merged instruction.  Merging is done automatically through code
unfolding and based on the definitions of the component instructions.
This makes it possible, in principle, to define
a set of (experimentally) optimal merging rules.  However, finding
exactly this set of rules is actually not straightforward.

Merging rules are specified separately from the instruction code
itself, and these rules state how basic instructions have to be
combined.  To start with, we will need to show how instructions are
defined based on their abstract versions.  For example, definition:

\begin{quotedcode}\begin{tabbing}
\\
\textsf{move}(\=\textit{A}, \textit{B}) \textsf{:--} \textit{B} \ensuremath{\Leftarrow} \textit{A}\textsf{@} .\\
\end{tabbing}
\end{quotedcode}

\noindent
moves data between two locations, i.e., the contents of the \textit{a}
mutable into the \textit{b} mutable. In order to specify precisely
the source and destination of the data, it is necessary to specify the
instruction \emph{format}, with a declaration such as:

\begin{quotedcode}\begin{tabbing}
\\
\textsf{:--} \textbf{\textsf{ins\_alias}}(\=\textsf{movexy}, \textsf{move}(\=\textsf{xreg\_mutable}, \textsf{yreg\_mutable})).\\
\end{tabbing}
\end{quotedcode}

\noindent which defines a \emph{virtual} instruction named
\textsf{movexy}, that corresponds to the instantiation of the code
for \textsf{move}\textsf{/}\textsf{2} for the case in which the first argument
corresponds to an \emph{X} register and the second one corresponds to
a \emph{Y} register. Both registers are seen from \ip\ as mutable
variables of type \emph{mut(tagged)}.
Then, and based on this more concrete instruction, 
the declaration: 

\begin{quotedcode}\begin{tabbing}
\\
\textsf{:--} \textbf{\textsf{ins\_entry}}(\=\textsf{movexy} \textsf{\texttt{+}} \textsf{movexy}).\\
\end{tabbing}
\end{quotedcode}

\noindent
forces the compiler to actually use during compilation an
instruction composed of two virtual ones and to emit bytecode
containing it (thanks to the \emph{ptrans} transformation in
Figure~\ref{fig:iset-trans}, which processes the program instruction
sequence to replace occurrences of the collapsed pattern by the new
instruction).  Emulator code will be generated implementing an
instruction which merges two \textsf{movexy} instructions (thanks to
the \emph{etrans} transformation).
The corresponding code is equivalent to:

\begin{quotedcode}\begin{tabbing}
\\
\textsf{:--} \textbf{\textsf{ins\_entry}}(\=\textsf{movexy\_movexy}).\\
\textsf{:--} \textbf{\textsf{pred}} \textsf{movexy\_movexy}(\=\textsf{xreg\_mutable}, \textsf{yreg\_mutable},\\
\>\textsf{xreg\_mutable}, \textsf{yreg\_mutable}).\\
\textsf{movexy\_movexy}(\=\textit{A}, \textit{B}, \textit{C}, \textit{D}) \textsf{:--} \textit{B} \ensuremath{\Leftarrow} \textit{A}\textsf{@}, \textit{D} \ensuremath{\Leftarrow} \textit{C}\textsf{@} .\\
\end{tabbing}
\end{quotedcode}

This can later be subject to other transformations and used to
generate emulator code as any other \ip\ instruction.

\subsubsection{Single-Instruction Encoding of Sequences of the Same
  Instruction [\textbf{ie}]}

In some cases a series of similar instructions (e.g.,
\textsf{unify\_with\_void}) with different arguments can be collapsed
into a single instruction with a series of operands which correspond
to the arguments of each of the initial instructions.  For example, a
bytecode sequence such as:

\begin{quotedterm} \textsf{unify\_with\_void}(\textsf{x}(\textsf{1})), \textsf{unify\_with\_void}(\textsf{x}(\textsf{2})), \textsf{unify\_with\_void}(\textsf{x}(\textsf{5}))\end{quotedterm}

\noindent can be compiled into:

\begin{quotedterm} \textsf{unify\_with\_void\_n}([\textsf{x}(\textsf{1}), \textsf{x}(\textsf{2}), \textsf{x}(\textsf{5})])\end{quotedterm}

\noindent which would perform exactly as in the initial instruction series, but
taking less space and needing fewer fetch cycles.  Such an instruction
can be created, emitted, and the corresponding emulator code generated
automatically based on the definition of \textsf{unify\_with\_void}.

In order to express this \emph{composite} instruction within \ip\ using a
single predicate, \textsf{unify\_with\_void\_n} needs to receive a
fixed number of arguments.  A different predicate for each of the
possible lengths of the array would have to be generated otherwise.
A single argument actually suffices; hence the square brackets, which
are meant to denote an array.

The \ip\ code which corresponds to the newly generated instruction is,
conceptually, as follows:

\begin{quotedcode}\begin{tabbing}
\\
\textsf{unify\_with\_void\_n}(\=\textit{Array}) \textsf{:--}\\
\hspace{2em}\=\textsf{array\_len}(\=\textit{Array}, \textit{L}),\\
\>\textsf{unify\_with\_void\_n\_2}(\=\textsf{0}, \textit{L}, \textit{Array}).\\
\textsf{unify\_with\_void\_n\_2}(\=\textit{I}, \textit{L}, \textit{Array}) \textsf{:--}\\
\hspace{2em}\=( \=\=\textit{I} \textsf{=} \=\textit{L} \ensuremath{\rightarrow} \textsf{true}\\
\hspace{2em}\=\textsf{;} \=\textsf{elem}(\=\textit{I}, \textit{Array}, \textit{E}),\\
\> \>\textsf{unify\_with\_void}(\=\textit{E}),\\
\> \>\textit{I}$_1$ \textsf{is} \textit{I} \textsf{\texttt{+}} \textsf{1},\\
\> \>\textsf{unify\_with\_void\_n\_2}(\=\textit{I}$_1$, \textit{L}, \textit{Array})\\
\>).\\
\end{tabbing}
\end{quotedcode}

It should be clear here why a fixed number of arguments is needed: a
series of \textsf{unify\_with\_void\_n}\textsf{/}\textsf{1},
\textsf{unify\_with\_void\_n}\textsf{/}\textsf{2}, etc. would have to be generated
otherwise.
Note that the loop code ultimately calls \textsf{unify\_with\_void}\textsf{/}\textsf{1},
the \lsourcerefl\ reflection of the initial instruction.

In this particular case the compiler to \lc\ performs some
optimizations not captured in the previous code.  For example, instead
of traversing explicitly the array with an index, this array is
expanded and inlined in the bytecode and the program counter is used to
retrieve the indexes of the registers by incrementing it after every
access.  As the length of the array is known when the bytecode is
generated, it is actually explicitly encoded in the final bytecode.
Therefore, all of the newly introduced operations
(\textsf{array\_len}\textsf{/}\textsf{2}, \textsf{elem}\textsf{/}\textsf{3}, etc.) need constant time and
are compiled efficiently.

\subsubsection{Instructions for Special Built-Ins [\textbf{ib}]}

As mentioned before, 
calling external library code or internal predicates (classically
termed ``built-ins'') requires following a protocol, to pass the
arguments, to check for failure, etc. Although the protocol can be the
same as for normal predicates (e.g., passing arguments as \texttt{X}
registers), some built-ins
require a special (more efficient) protocol (e.g., 
passing arguments as \lc\ arguments, avoiding movements in \texttt{X}
registers).  Calling those special built-ins is, by default,
taken care of by a generic family of instructions, one per arity.
This is represented as the instructions:

\begin{quotedcode}\begin{tabbing}
\\
\textsf{:--} \textbf{\textsf{ins\_alias}}(\=\textsf{bltin1d}, \textsf{bltin1}(\=\textsf{bltindet}(\=\textsf{tagged}), \textsf{xreg})).\\
\textsf{bltin1}(\=\textit{BltName}, \textit{A}) \textsf{:--} \textit{BltName}(\=\textit{A}\textsf{@}).\\
\textsf{:--} \textbf{\textsf{ins\_alias}}(\=\textsf{bltin2d}, \textsf{bltin2}(\=\textsf{bltindet}(\=\textsf{tagged}, \textsf{tagged}), \textsf{xreg}, \textsf{xreg})).\\
\textsf{bltin2}(\=\textit{BltName}, \textit{A}, \textit{B}) \textsf{:--} \textit{BltName}(\=\textit{A}\textsf{@}, \textit{B}\textsf{@}).\\
\end{tabbing}
\end{quotedcode}

\noindent where each \textsf{bltinI}\textsf{/}\textsf{i} \textsf{\texttt{+}} \textsf{1} acts as a bridge to call the external
code expecting $i$ parameters. The \textit{BltName} argument
represents a predicate abstraction that will contain a reference to
the actual code of the built-in during the execution. The type of the
\textit{BltName} argument reflects the accepted calling pattern of the
predicate abstraction. When compiling those instructions to \lc\ code,
that predicate abstraction is efficiently translated as an unboxed
pointer to an \lc\ procedure.%
\footnote{In the bytecode, the argument that corresponds to the
  predicate abstraction is stored as a number that uniquely
  identifies the built-in. When the bytecode is actually loaded, this number is
  used to look up the actual address of the built-in in a table
  maintained at runtime. This is needed since, in general, there is no
  way to know which address will be assigned to the entry point of a
  given built-in in different program executions.}
With the definition shown above, the \ip\ compiler can generate, for
different arities, an instruction which calls the built-in passed as
first argument.

However, by specifying at compile time a predefined set of built-ins or
predicates written in \lc\ (that is, a static value for
\textit{BltName} instead of a dynamic value),
the corresponding instructions can be statically specialized and an
instruction set which performs direct calls to the corresponding
built-ins can be generated.  This saves
an operand, generating slightly smaller code, and replaces an
indirection by a direct call, which saves memory accesses and helps
the processor pipeline, producing faster code.

\subsection{Transformations of Instruction Code} 
\label{sec:changes-emul-struct}

Some transformations do not create new instructions; they perform
instead a number of optimizations on already existing instructions by
manipulating the code or by applying selectively alternative
translation schemes.

\subsubsection{Unfolding Rules [\textbf{ur}]}

Simple predicates can be unfolded before compilation. In the case of
instruction merging, unfolding is used to merge two (or more)
instructions into a single piece of code, in order to avoid fetch
cycles (Section~\ref{sec:instr-merging}). However, uncontrolled
unfolding is not always an advantage, because an increased emulator
size can affect negatively the cache behavior.  Therefore the
\emph{\textbf{ur}} option turns on or off a predefined set of rules to
control which instruction mergings are actually performed.
Unfolding rules follow the scheme:

\begin{quotedcode}\begin{tabbing}
\\
\textsf{:--} \textbf{\textsf{ins\_entry}}(\=\textit{Ins}$_1$ \textsf{\texttt{+}} \textit{Ins}$_2$ \textsf{\texttt{+}} \textsf{...} \textsf{\texttt{+}} \textit{Ins}$_n$, \textit{WhatToUnfold}).\\
\end{tabbing}
\end{quotedcode}

\noindent where \textit{Ins}$_1$ to \textit{Ins}$_n$ are the basic instructions to be
merged, and \textit{WhatToUnfold} is a rule specifying exactly which
instruction(s) has to be unfolded when \textbf{\emph{ur}} is
activated.
As a concrete example, the unfolding rule:

\begin{quotedcode}\begin{tabbing}
\\
\textsf{:--} \textbf{\textsf{ins\_entry}}(\=\textsf{alloc} \textsf{\texttt{+}} \textsf{movexy} \textsf{\texttt{+}} \textsf{movexy}, \textsf{1}).\\
\end{tabbing}
\end{quotedcode}

\noindent means that in the instruction to be generated by combining one
\textsf{alloc} and two \textsf{movexy}, the code for \textsf{alloc} is
inlined (the value of the last argument \textsf{1} refers to the
\emph{first} instruction in the sequence), and the (shared) code for
\textsf{movexy} \textsf{\texttt{+}} \textsf{movexy} is invoked afterwards.  A companion instruction
merging rule for \textsf{movexy} \textsf{\texttt{+}} \textsf{movexy} exists:

\begin{quotedcode}\begin{tabbing}
\\
\textsf{:--} \textbf{\textsf{ins\_entry}}(\=\textsf{movexy} \textsf{\texttt{+}} \textsf{movexy}, \textsf{all}).\\
\end{tabbing}
\end{quotedcode}

\noindent which states that the code for both \textsf{movexy} has to be unfolded
in a combined instruction.  The instruction
\textsf{alloc} \textsf{\texttt{+}} \textsf{movexy} \textsf{\texttt{+}} \textsf{movexy} would generate code for \textsf{alloc}
plus a call to \textsf{movexy} \textsf{\texttt{+}} \textsf{movexy}.  The compiler eventually
replaces this call by an explicit jump to the location of
\textsf{movexy} \textsf{\texttt{+}} \textsf{movexy} in the emulator.  The program counter is
updated accordingly to access the arguments correctly.

\subsubsection{Alternative Tag Switching Schemes [\textbf{ts}]}
Tags are used to determine dynamically the type of basic data (atoms,
structures, numbers, variables, etc.) contained in a (tagged) memory
word. Many instructions and built-ins (like unification) take different
actions depending on the type (or tag) of the input arguments.  This is
called \emph{tag switching}, and it is a heavily-used operation which
is therefore worth optimizing as much as possible.  The tag is
identified (and the corresponding action taken) using tag switching
such as:

\begin{center}
  (\textit{tagtest}$_1$ \ensuremath{\rightarrow} \textit{tagcode}$_1$ \textsf{;} \textsf{...} \textsf{;} \textit{tagtest}$_n$ \ensuremath{\rightarrow} \textit{tagcode}$_n$)
\end{center}

\noindent
where every \textit{tagtest}$_i$ has the form \textsf{tagof}(\textit{v}, \textit{tag}$_i$)
(i.e., code that performs a different action depending on the tag
value of a tagged \textit{v}).
The \emph{\textbf{ts}} option chooses between either a \loexpr{switch}
control structure (when enabled) or a set of predefined test
patterns based on tag encodings and assertions on the possible tags
(when disabled).

Both possibilities are studied in more detail in~\cite{tagschemes-ppdp08}. 
Since the numbers that encode the tags are usually small, it is easy
for a modern C compiler (e.g., \texttt{gcc}) to generate an indirection
table and jump to the right code using it (that is, it does not
require a linear search).  It is difficult, however, to make the C
compiler aware that checks to ensure that the tag number will actually
be one of the cases in the \emph{switch} are, by construction,
unnecessary (i.e., there is no need for a \emph{default} case).  This
information could be propagated to the compiler with a type system
which not all low-level languages have.
The alternative compilation scheme (rule \rulename{Tif}) makes explicit use of
tag-checking primitives, where the sequence of \textit{ctest}$_i$ and
the code of each branch depends on the particular case.

  The latter approach is somewhat longer (and more complex as the
  number of allowed tags grows) than the former.  However, in some
  cases there are several advantages to the latter, besides the
  already mentioned avoidance of boundary checks:
\begin{itemize}
\item Tags with higher runtime probability can be checked before, in
  order to select the right branch as soon as possible.
\item Since the evaluation order is completely defined, tests can be
  specialized to determine as fast as possible which alternative
  holds.
  For example, if by initial assumption \textit{v} can only be either
  a heap variable, a stack variable, or a structure (having a
  different tag for each case), then the tests can check if it is a
  heap variable or a stack variable and assume that it is a structure
  in the last branch.
\end{itemize}

Deciding on the best option has to be based on experimentation, the
results of which we summarize in Section~\ref{sec:overall-best} and in
tables~\ref{tab:performance-x86} and~\ref{tab:performance-ppc}.

\subsubsection{Connected Continuations [\textbf{cc}]}

Some actions can be repeated unnecessarily because they appear at the
end of an operation and at the beginning of the next one.  Often they
have no effect the second time they are called (because they are,
e.g., tests which do not change the tested data, or data movements).
In the case of tests, for example, they are bound to fail or succeed
depending on what happened in the previous invocation.

As an example, in the fragment \textsf{deref}(\textit{T}), (\textsf{tagof}(\textit{T}, \textsf{ref}) \ensuremath{\rightarrow} \textit{A} \textsf{;} \textit{B})
the test \textsf{tagof}(\textit{R}, \textsf{ref}) is performed just before exiting
\textsf{deref}\textsf{/}\textsf{1} (see Figure~\ref{fig:unif-const}).  Code generation for
instructions which include similar patterns is able to insert a jump
either to \textit{A} or \textit{B} from the code generated for
\textsf{deref}\textsf{/}\textsf{1}.
This option enables or disables this optimization for a series of
preselected cases, by means of code annotations similar to the ones
already shown.

\subsubsection{Read/Write Mode Specialization [\textbf{rw}]}

WAM-based implementations use a flag to test whether heap structures
are being read (matched against) or written (created).  According to
the value of this flag, which is set by code executed immediately
before, several instructions adapt their behavior with an internal,
local \emph{if-then-else}.

A common optimization is to partially evaluate the 
\emph{switch}
statement which implements the fetch-and-execute
cycle inside the emulator loop.  Two different switches can be
generated, with the same structure, but with heap-related instructions
specialized to perform either reads or
writes~\cite{Carlsson96thesicstus}.  Enabling or disabling the
\emph{\textbf{rw}} optimization makes it possible to generate
instruction sets (and emulators) where this transformation has been
turned on or off.

This is conceptually performed by generating different versions of the
code for the instructions,  
depending on the value of a mutable variable \textsf{mode}, which can
only take the values \textsf{read} or \textsf{write}.
Deciding whether to generate different code versions or to generate
\emph{if-then-else}s to be checked at run-time is done based on a
series of heuristics which try to forecast the complexity and size of
the resulting code.

\subsection{Experimental Evaluation}
\label{sec:evaluation}

We present in this section experimental data regarding the performance
achieved on a set of benchmarks by a collection of
emulators, all of which were automatically generated by selecting different
combinations of the options presented in previous sections.  In
particular, by using all 
\textbf{compatible} possibilities for the transformation and
generation options given in Section~\ref{sec:automatic-spec-merg}
we generated 96 different emulators (instead of $2^7 = 128$,
as not all options are independent; for example, \textbf{ie} needs
\textbf{im} to be performed).
This bears a close relationship
with~\cite{demoen00:wam-variations}, but here we are not changing the
internal data structure representation (and of course our instructions
are all initially coded in \ip).  
It is also related to the experiment reported
in~\cite{insmerging-sicstus}, but the tests we perform are more
extensive and cover more variations on the kind of changes that the
abstract machine is subject to.  Also,~\cite{insmerging-sicstus}
starts off by being selective about the instructions to merge, which
may seem a good idea but, given the very intricate dependencies among
different optimizations, can also result in a loss of optimization
opportunities.  In any case, this is certainly a point we want to
address in the future by using instruction-level profiling.

Although most of the benchmarks we used are relatively well known, a
brief description follows:

\newcommand{\mdes}[1]{\parbox{4em}{\textbf{#1}}}

\begin{description}
\item[\mdes{boyer}] Simplified Boyer-Moore theorem prover kernel.
\item[\mdes{crypt}] Cryptoarithmetic puzzle involving multiplication.
\item[\mdes{deriv}] Symbolic derivation of polynomials.
\item[\mdes{factorial}] Compute the factorial of a number.
\item[\mdes{fib}] Simply recursive computation of the
  n$^{\mathrm{th}}$ Fibonacci number.
\item[\mdes{knights}] Chess knight tour, visiting only once every board cell.
\item[\mdes{nreverse}] Naive reversal of a list using append.
\item[\mdes{poly}]  Raises symbolically the expression
  \textsf{1} \textsf{\texttt{+}} \textsf{x} \textsf{\texttt{+}} \textsf{y} \textsf{\texttt{+}} \textsf{z} to the n$^{\mathrm{th}}$ power.
\item[\mdes{primes}] Sieve of Eratosthenes.
\item[\mdes{qsort}] Implementation of QuickSort.
\item[\mdes{queens11}] $N$-Queens with $N = 11$.
\item[\mdes{query}] \parbox[t]{0.87\linewidth}{Makes a natural language
    query to a knowledge database with in\-for\-ma\-tion about country
    names, population, and area.}\smallskip
\item[\mdes{tak}] Computation of the Takeuchi function.
\end{description}

Our starting point was a ``bare'' instruction set comprising the
common basic blocks of a relatively efficient abstract machine (the
``optimized'' abstract machine of Ciao 1.13, in the `optim\_comp'
directory of the Ciao 1.13 repository).\footnote{Changes in the
  optimized version include tweaks to clause jumping, arithmetic
  operations and built-ins and some code clean-ups that reduce the
  size of the emulator loop. 
} The Ciao abstract machines have their remote roots in the emulator of
SICStus Prolog 0.5/0.7 (1986-89), but have evolved over the years
quite independently and been the object of many optimizations and code
rewrites resulting in performance improvements and much added
functionality.\footnote{This includes modules, attributed variables,
  support for higher order, multiprocessing, parallelism, tabling,
  modern memory management, etc., etc.}  The performance of this, our
\emph{baseline} engine matches that of modern Prolog
implementations. Table~\ref{tab:comparison-speed} helps evaluating the
speed of this baseline optimized Ciao emulator w.r.t.\ to the
relatively unoptimized Ciao 1.13 emulator compiled by default in the
Ciao distribution, some high-performance Prolog implementations (Yap
5.1.2~\cite{costa:yap-design-tplp} and hProlog
2.7~\cite{demoen:hprolog}), and the popular SWI-Prolog system (version
5.6.55~\cite{wielemaker10:swi_prolog}).

\begin{table}[t]
  \centering
  \begin{tabular}{|l|rrrrr|}
    \cline{1-6}
    \textbf{Benchmark} & \textbf{Yap} & \textbf{hProlog} & \textbf{SWI} &
    \textbf{Ciao-std} & \textbf{Ciao-opt}\\
     & \textbf{5.1.2} & \textbf{2.7} & \textbf{5.6.55} &
    \textbf{1.13} & \textbf{(baseline)}    \\\cline{1-6}
    boyer&     1392&    1532&        11169& 2560&     1604\\
    crypt&     3208&    2108&        36159& 6308&     3460\\
    deriv&     3924&    3824&        12610& 6676&     3860\\
    exp&       1308&    1740&        2599& 1400&      1624\\
    factorial& 4928&    2368&        16979& 3404&     2736\\
    fft&       1020&    1652&        14351& 2236&     1548\\
    fib&       2424&    1180&        8159& 1416&      1332\\
    knights&   2116&    1968&        11980& 3432&     2352\\
    nreverse&  1820&     908&        18950& 3900&     2216\\
    poly&      1328&    1104&        6850& 1896&      1160\\
    primes&    4060&    2004&        28050& 3936&     2520\\
    qsort&     1604&    1528&        8810& 2600&      1704\\
    queens11&  1408&    1308&        24669& 3200&     1676\\
    query&     632&      676&        6180& 1448&       968\\
    tak&       3068&    1816&        27500& 5124&     2964\\
    \cline{1-6}
  \end{tabular}
  \caption{Speed comparison of baseline with other Prolog systems.}
  \label{tab:comparison-speed}
\end{table}

Figures~\ref{fig:geom-mean-x86} (in page~\pageref{fig:geom-mean-x86})
to~\ref{fig:tak-ppc} (in page~\pageref{fig:tak-ppc}) summarize
graphically the results of the experiments, as the data gathered
\mbox{---96} emulators $\times$ 13 benchmarks = 1248 performance
\mbox{figures---} is too large to be comfortably presented in regular
tables.

Each figure presents the speedup obtained by different emulators for a
given benchmark (or all benchmarks in the case of the summary
tables).  Such speedups are
relative to some
``default'' code generation options, which we have set to be those
which were active in the 
Ciao emulator we started with (our baseline), and which therefore
receive speedup 1.0.  Every point 
in each graph corresponds to the relative speed of a different
emulator obtained with a different combination of the options
presented in Sections~\ref{sec:gen-instructions}
and~\ref{sec:changes-emul-struct}.

The options are related to the points in the graphs as follows:
each option
is assigned a bit in a binary number, where `1' means activating the
option and `0' means deactivating it.  Every value in the $y$ axis of
the figures corresponds to a combination of the three options in
Section~\ref{sec:gen-instructions}.  Note that only 6 combinations
(out of the $2^3 = 8$ possible ones) are plotted due to dependencies
among options (for example, ``Instruction encoding'' always implies
``Instruction merging'').  The options in
Section~\ref{sec:changes-emul-struct}, which correspond to
transformations in the way code is generated and which need four bits,
are encoded using $2^4 = 16$ different dot shapes.  Every combination
of emulator generation options is thus assigned a different 7-bit
number encoded as a dot shape and a $y$ coordinate.
The $x$ coordinate represents the speedup as presented before (i.e.,
relative to the hand-coded emulator currently in Ciao 1.13).
The level of agressiveness of the instruction set transformations used
in the paper was selected to automatically generate an emulator
identical to the hand-crafted one.  We have experimentally confirmed
that it is difficult to outperform this engine without changing other
parameters, such as the overall architecture.

\begin{table}
  \centering
  \begin{tabular}{|c|c|c||c|c|c|c|}
    \cline{1-7}
    \multicolumn{3}{|c||}{Instruction} &
    \multicolumn{4}{c|}{Instruction}
    \\    
    \multicolumn{3}{|c||}{Generation} &
    \multicolumn{4}{c|}{Transformations}
    \\    \cline{1-7}
Instruction&Special &Instruction&Tag  &Connected&Unfolding&R/W  \\
Encoding   &Builtins&Merging&Switching&Conts.   &Rules    & Mode\\

    \textbf{(ie)} & \textbf{(ib)} & \textbf{(im)} & \textbf{(ts)} &
    \textbf{(cc)} & \textbf{(ur)} & \textbf{(rw)} \\\cline{1-7}
  \end{tabular}
  \caption{Meaning of the bits in the plots.}
  \label{tab:bit-meaning}
\end{table}

Different selections for the bits assigned to the $y$ coordinate and
to the dot shapes would of course yield different plot configurations.
However, our selection seems intuitively appropriate, as it uses two
different encodings for two different families of transformations, one
which affects the bytecode language itself, and another one which
changes the way these bytecode operands are interpreted.
Table~\ref{tab:bit-meaning} relates the bits in these two groups,
using the same order as in the plots.

Every benchmark was run several times on each emulator to make timings
stable.  The hosts used were an x86 machine with a Pentium 4
processor running Linux
and an iMac with a PowerPC 7450 running Mac OS X.
Arithmetic and geometric\footnote{The geometric average is known to be
  less influenced by extremely good or bad cases.} averages of all
benchmarks were calculated and are shown in
Figures~\ref{fig:geom-mean-x86}, \ref{fig:arith-mean-x86},
\ref{fig:geom-mean-ppc}, and \ref{fig:arith-mean-ppc}.  Their
similarity seems to indicate that there are no ``odd'' behaviors off
the average.  Additionally, we are including detailed plots for every
benchmark and all the engine generation variants, following the
aforementioned codification, first for the x86 architecture
(Figures~\ref{fig:factorial-x86} to~\ref{fig:fib-x86}) and then 
for the PowerPC architecture (Figures~\ref{fig:factorial-ppc}
to~\ref{fig:fib-ppc}), in the same order in both cases.  Plots for
specially relevant cases are shown first, followed by the rest of the
figures sorted following an approximate (subjective) ``more sparse'' to
``less sparse'' order.

\intelpict{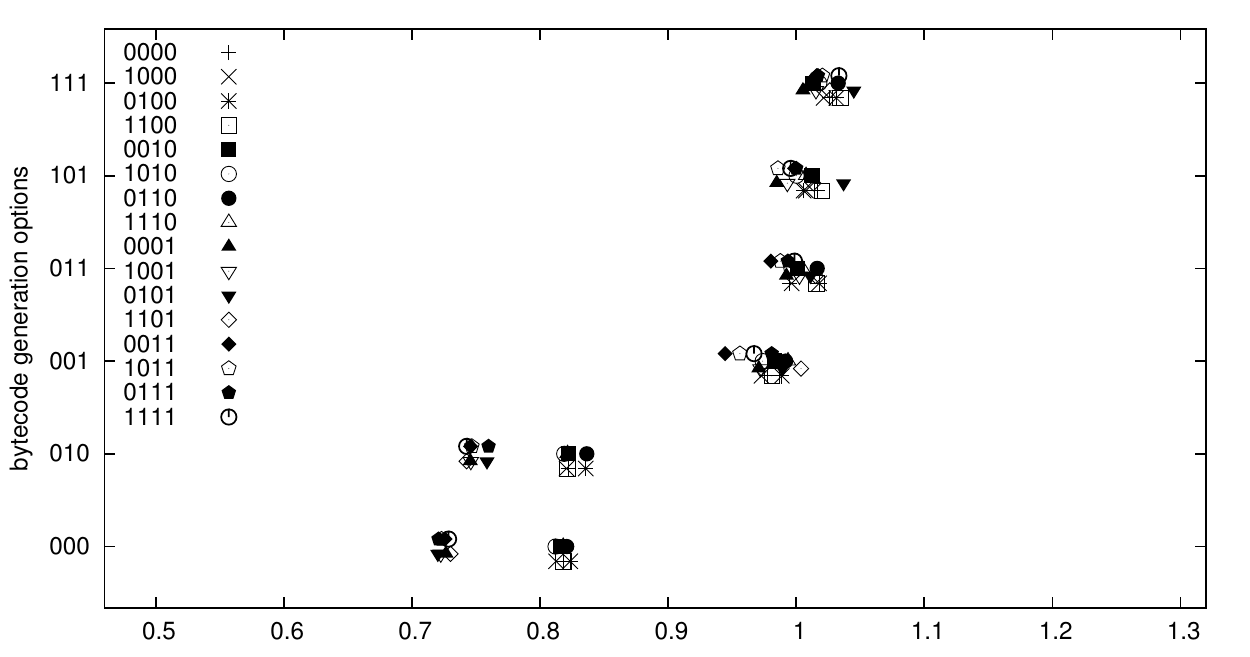}
{Geometric average of all benchmarks (with a dot per emulator)}
{fig:geom-mean-x86}
\intelpict{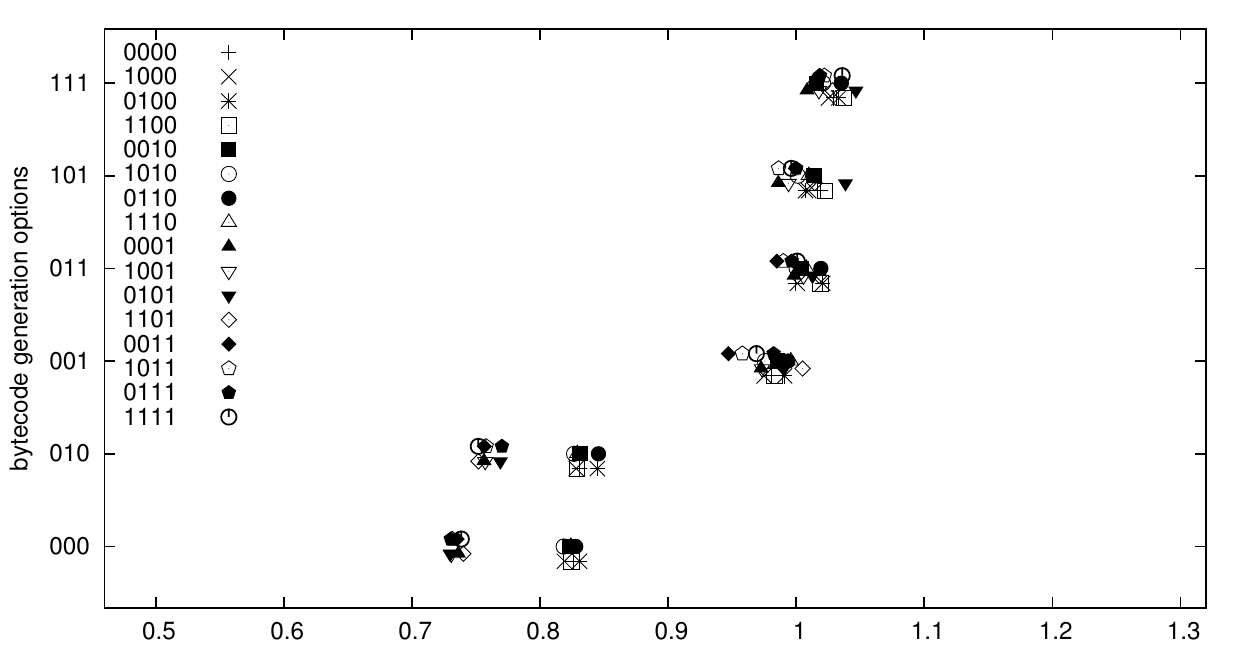}
{Arithmetic average of all benchmarks (with a dot per emulator)}
{fig:arith-mean-x86}

\subsubsection{General Analysis}
\label{sec:general-analysis}

The best speedup among all tried options, averaged across the
exercised benchmarks and with respect to the baseline Ciao 1.13
emulator, is 1.05$\times$ for the x86 processor
(Table~\ref{tab:performance-x86}, top section, under the column
\emph{w.r.t.\ def.}) and 1.01$\times$ for the PowerPC
(Table~\ref{tab:performance-ppc}, top section, same column heading).
While this is a modest average gain, some benchmarks achieve much
better speedups.  An alternative interpretation of this result is that
by starting with a relatively simple instruction set (coded directly
in \ip) and applying automatically and systematically a set of
transformation and code generation options which can be trusted to be
correct, we have managed to match (and even exceed) the time
performance of an emulator which was hand-coded by very proficient
programmers, and in which decisions were thoroughly tested along
several years.  Memory usage was unaltered.
Note (in the same tables) that the speedup obtained with respect to
the basic instruction set (under the column labeled \emph{w.r.t.\
  base}) is significantly higher.

Figure~\ref{fig:geom-mean-x86} depicts the geometric average of the
executions of all benchmarks in an Intel platform.  It aims at giving
an intuitive feedback of the overall performance of the option sets,
and indeed a well defined clustering around eight centers is clear.
Figure~\ref{fig:arith-mean-x86}, which uses the arithmetic average, is
very similar (but not identical --- it is very slightly biased towards
higher speedups), and it shows eight well-defined clusters as well.

From these pictures we can infer that bytecode transformation options
can be divided into two different sets: one which is barely
affected by options of the generation of code for the emulator
(corresponding to the upper four clusters), and another set (the
bottom four clusters) in which changes to the generation of the
emulator code does have an effect in the performance.

In general, the results obtained in the PowerPC show fewer variations
than those obtained in an x86 processor.  We attribute this behavior
to differences between these two architectures, as they greatly affect
the optimization opportunities and the way the C compiler can generate
code.  For example, the larger number of general-purpose registers
available in a PowerPC seems to make the job of the C compiler less
dependent on local variations of the code (as the transformations
shown in Section~\ref{sec:changes-emul-struct} produce).  Additionally,
internal differences between both processors (e.g., how branch
prediction is performed, whether there is register renaming, shadow
registers, etc.)  can also contribute to the differences we observed.

As a side note, while Figures~\ref{fig:geom-mean-x86}
and~\ref{fig:arith-mean-x86} portray an average behavior, there were
benchmarks whose performance depiction actually match this average
behavior very faithfully ---e.g., the simply recursive Factorial
(Figure~\ref{fig:factorial-x86}), which is often disregarded as an
unrealistic benchmark but which, for this particular experiment, turns
out to predict quite well the (geometric) average behavior of all
benchmarks.  Experiments in the PowerPC
(Figures~\ref{fig:geom-mean-ppc} and~\ref{fig:factorial-ppc}) generate
similar results.

\begin{figure}
  \centering
  \includegraphics[width=0.8\linewidth]{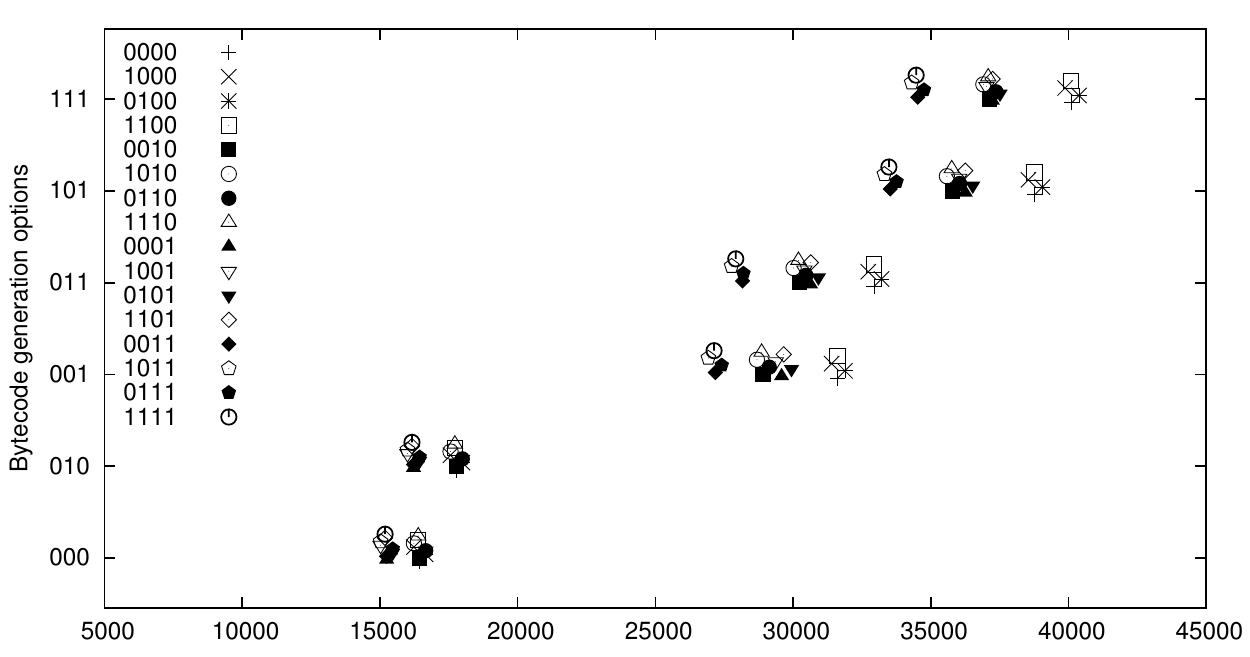}
  \caption{Size (in bytes) of WAM emulator with respect to the generation options (i86).}
  \label{fig:wam-size}
\end{figure}

Figure~\ref{fig:wam-size} 
presents the size of the WAM loop (using actual i86
object code size measured in bytes) for each bytecode and \lc\
code generation option. 
This size is independent from the benchmarks, and therefore only one
plot is shown.  It resembles notably the depictions of the speedup
graphs.  In fact, a detailed inspection of the distribution of
low-level code generation options (where each of them corresponds to
one of the 16 different dot shapes) inside each
bytecode language option shows some correlation among larger and
faster emulators.  This is not surprising as some code generation
schemes which tend to increase the size do so because they generate
additional, specialized code.

As in the case for speed, \lb\ generation options are the ones
which influence most heavily the code size.  This is understandable
because some options (for example, the \emph{\textbf{im}} switch for
instruction merging, corresponding to the leftmost bit of the
``bytecode generation options'') increment notably the size of the
emulator loop.
On the other hand, code generation options have a less significant
effect, as they do not necessarily affect all the instructions.

It is to be noted that the generation of specialized switches for the
write and read modes of the WAM (the \emph{\textbf{rw}} option) does
not increase the size of the emulator.  The reason is that when the
\emph{\textbf{rw}} flag is checked by all the instructions which need
to do so (and many instructions need it), a large number of
\emph{if-then-else} constructions with their associated code are
generated.  In the case of the specialized switches, only an
\emph{if-then-else} is needed and the savings from generating less
branching code make the emulator smaller.

\intelpict{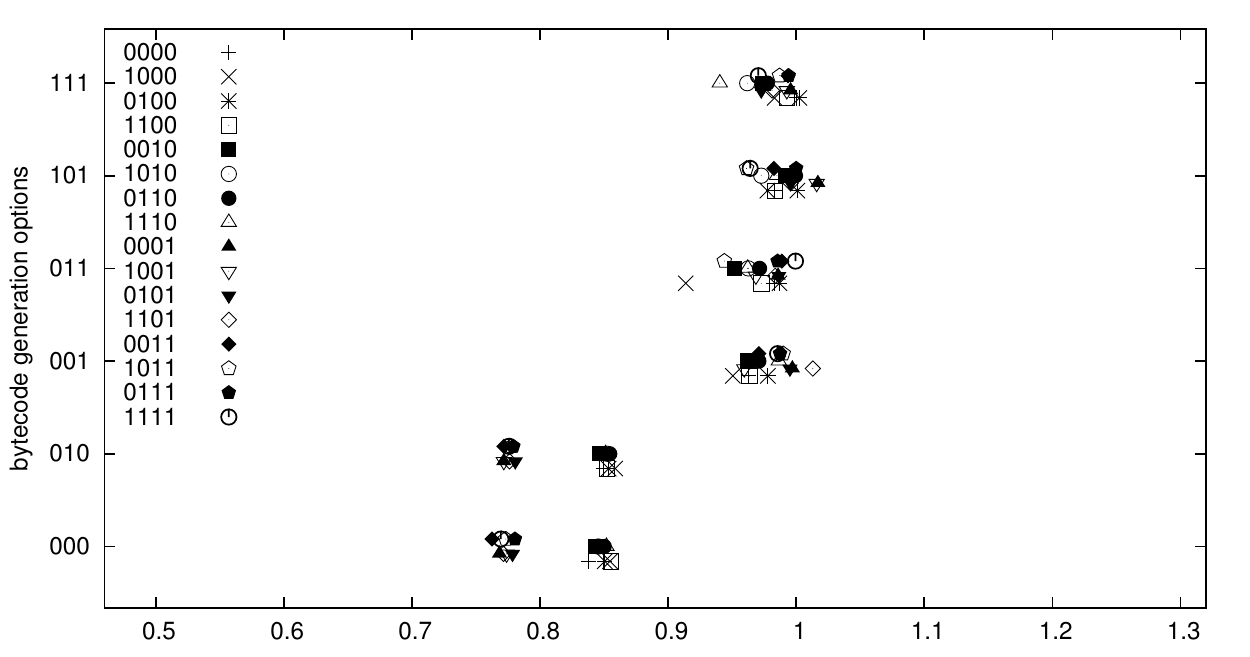}{Factorial involving large numbers}{fig:factorial-x86}

\intelpict{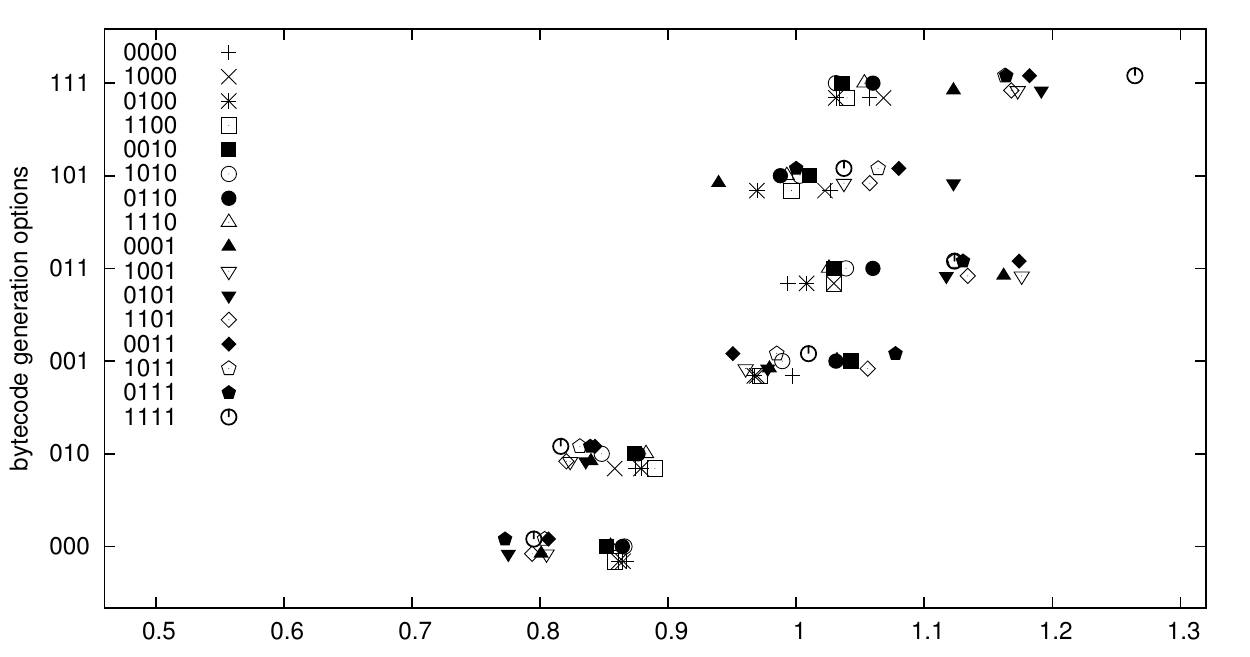}{Queens (with 11 queens to place)}{fig:queens11-x86}

\intelpict{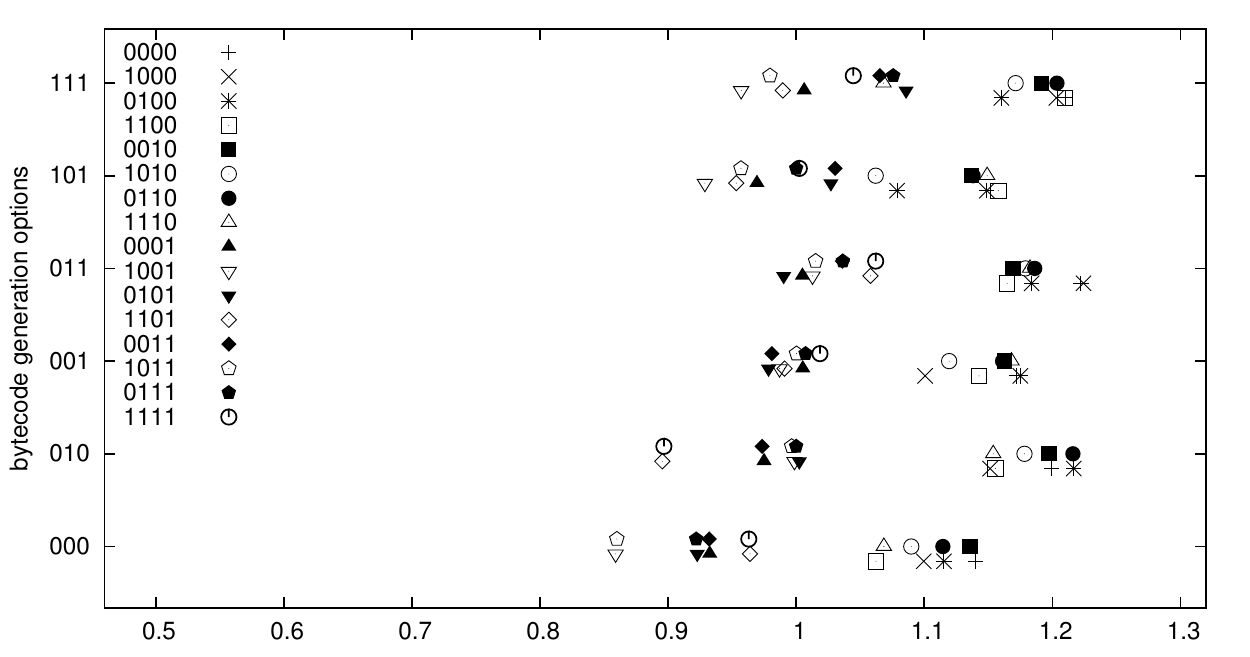}{Cryptoarithmetic puzzle}{fig:crypt-x86}

\intelpict{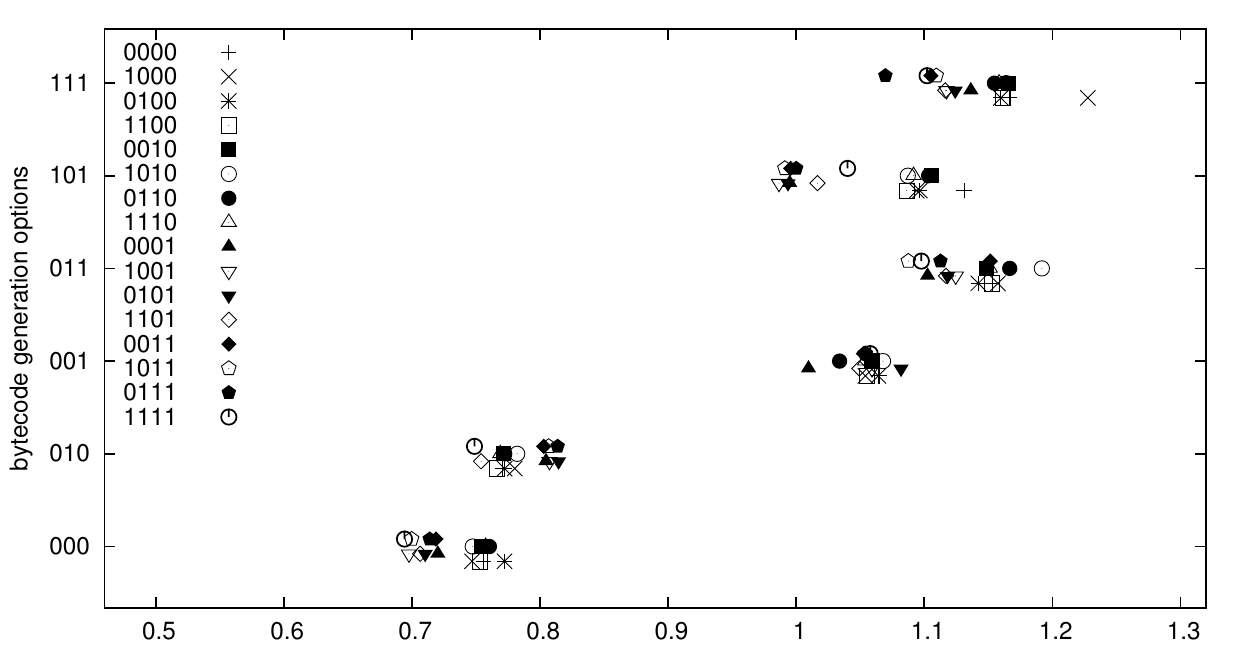}{Computation of the Takeuchi function}{fig:tak-x86}
\intelpict{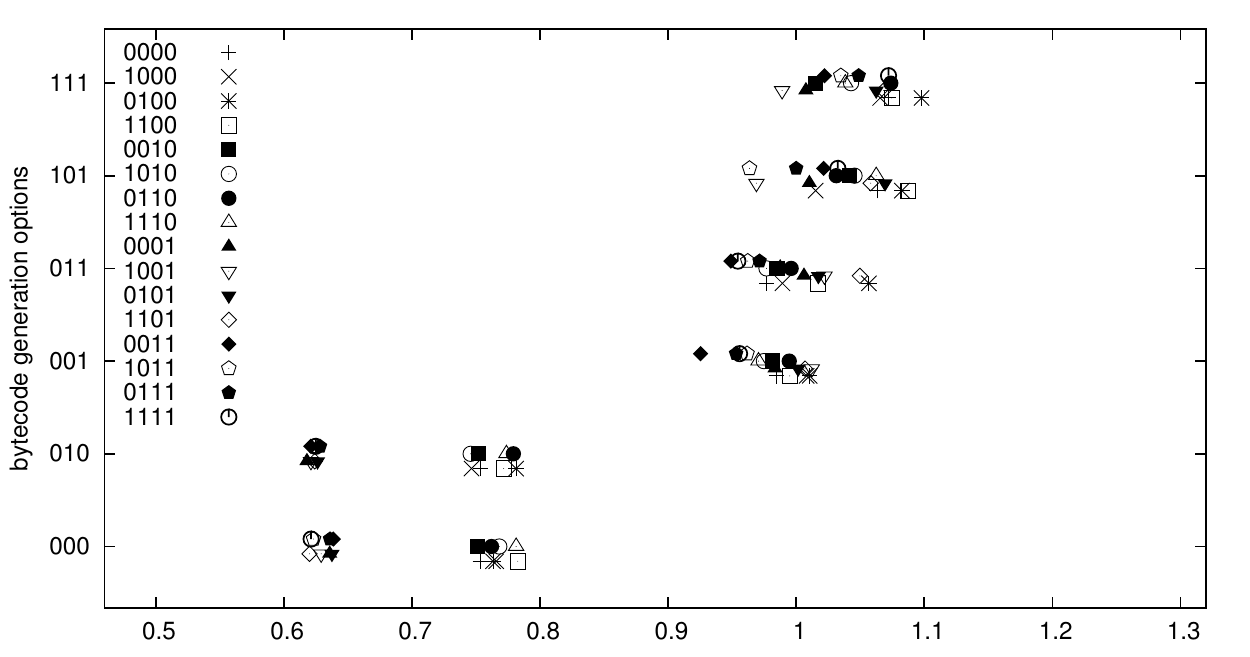}{Symbolic derivation of polynomials}{fig:deriv-x86}
\intelpict{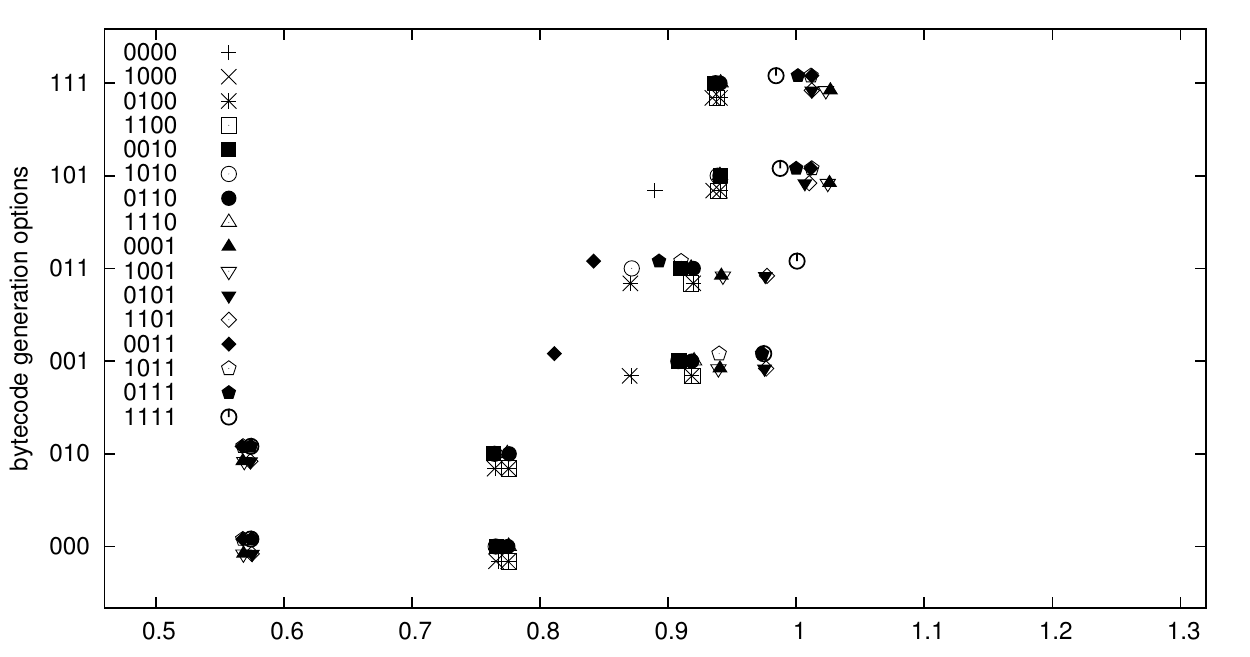}{Naive reverse}{fig:nrev-x86}
\intelpict{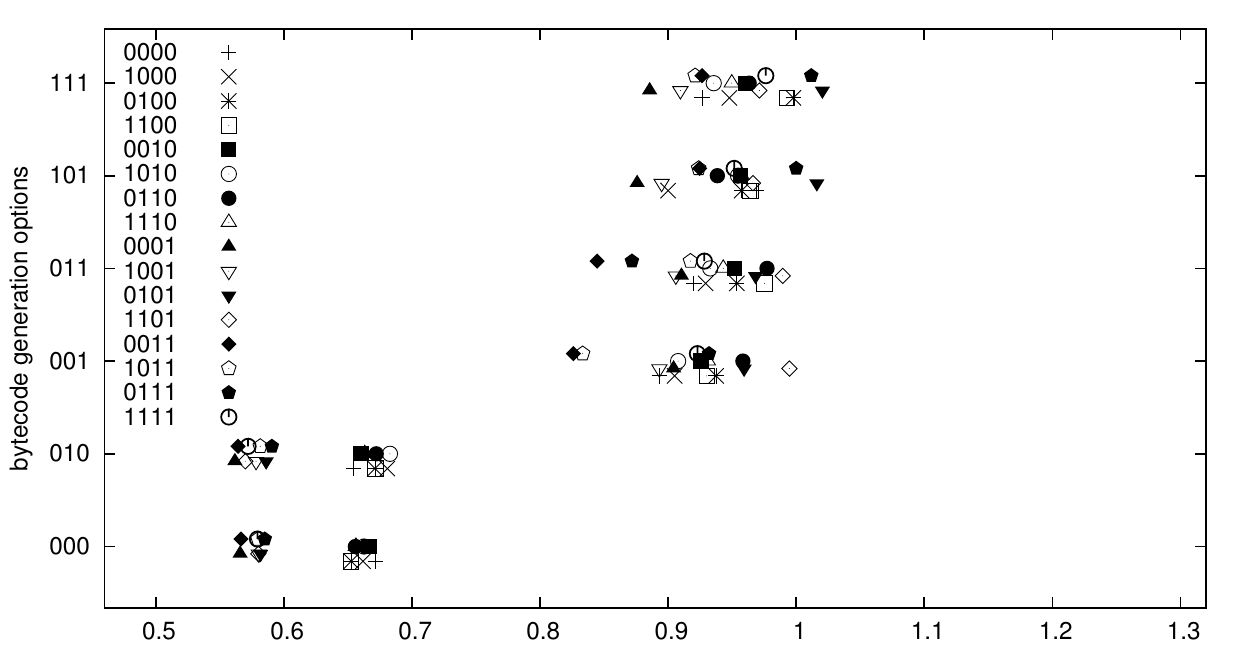}{Symbolic exponentiation of a polynomial}{fig:poly-x86}
\intelpict{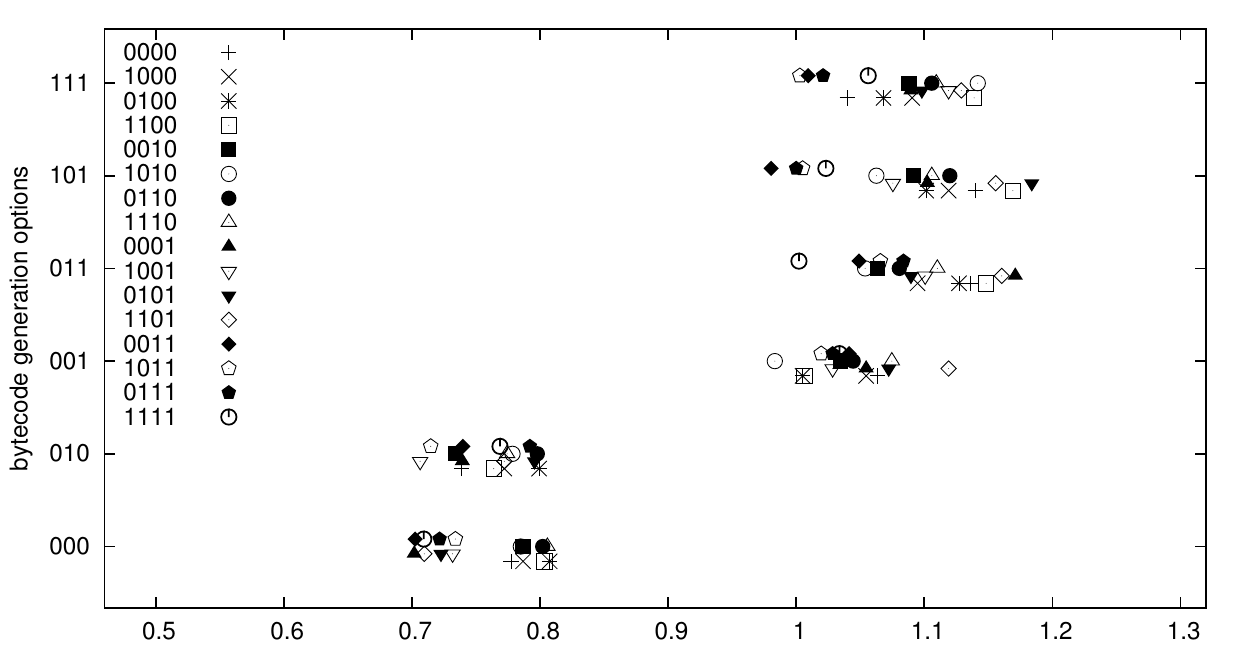}{Version of Boyer-Moore theorem prover}{fig:boyer-x86}
\intelpict{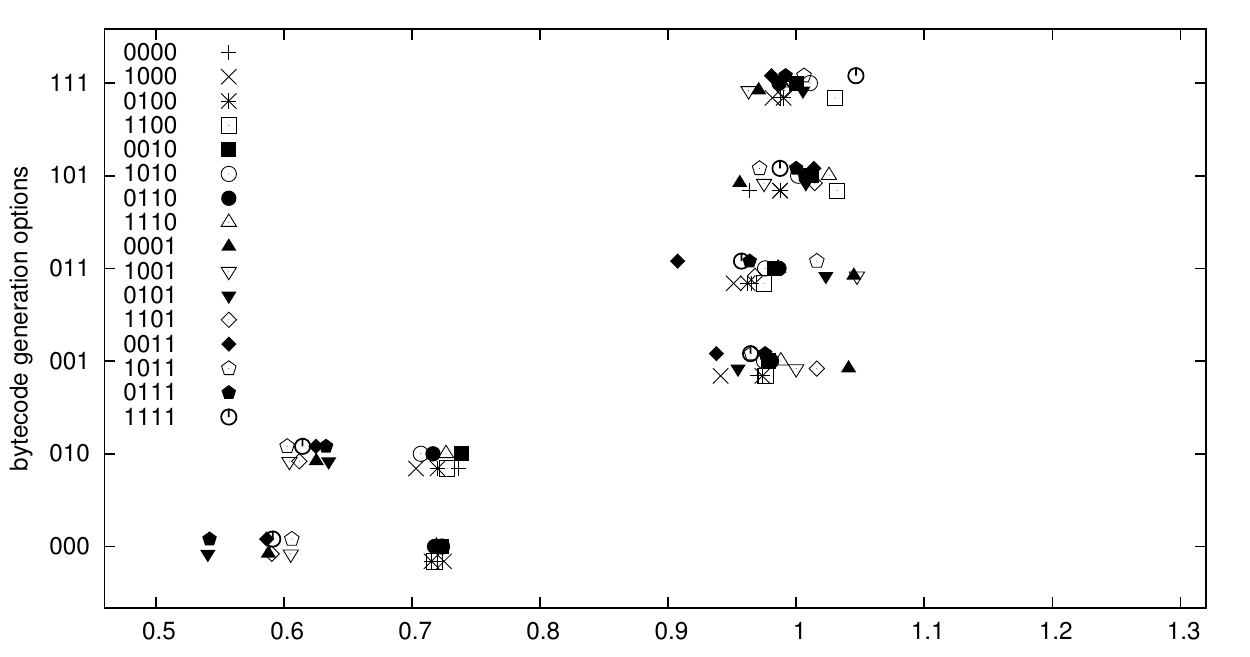}{QuickSort}{fig:qsort-x86}
\intelpict{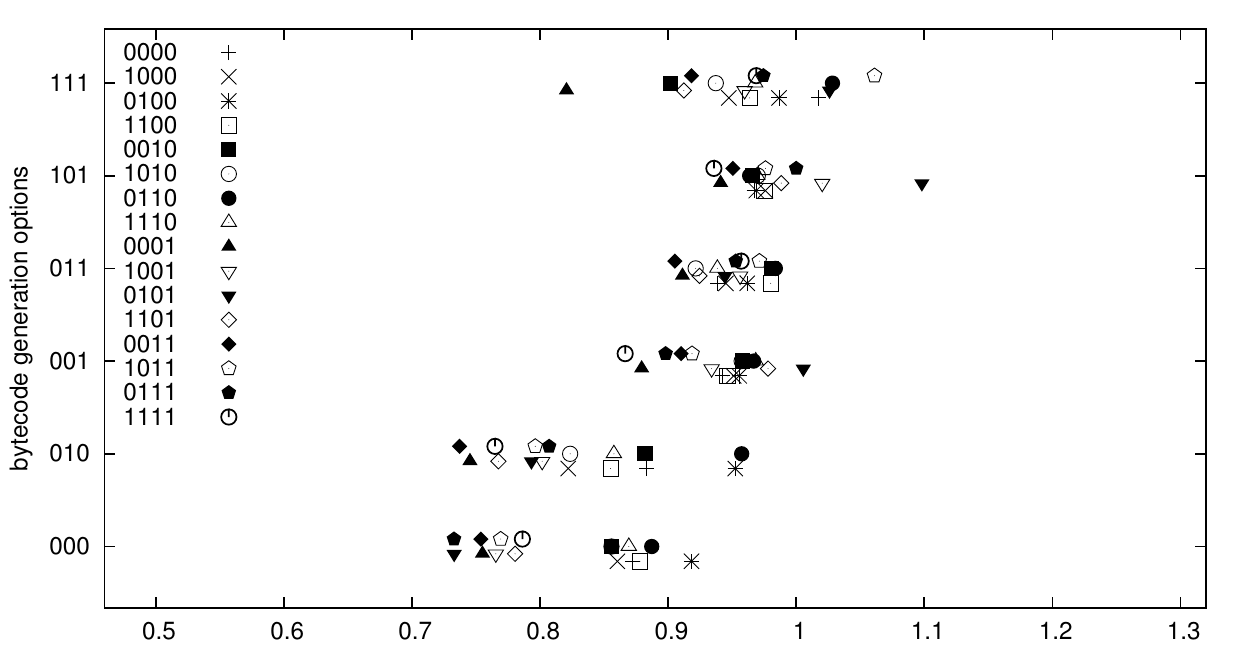}{Calculate primes using the sieve of Eratosthenes}{fig:primes-x86}
\intelpict{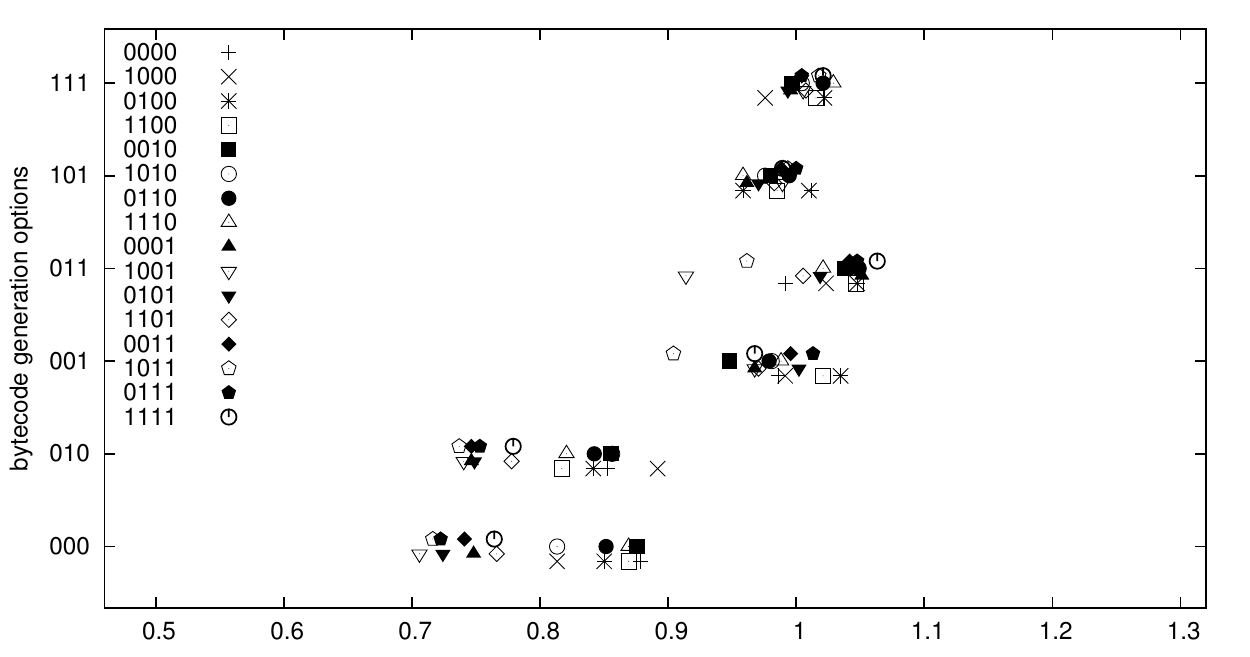}{Natural language query to a geographical database}{fig:query-x86}
\intelpict{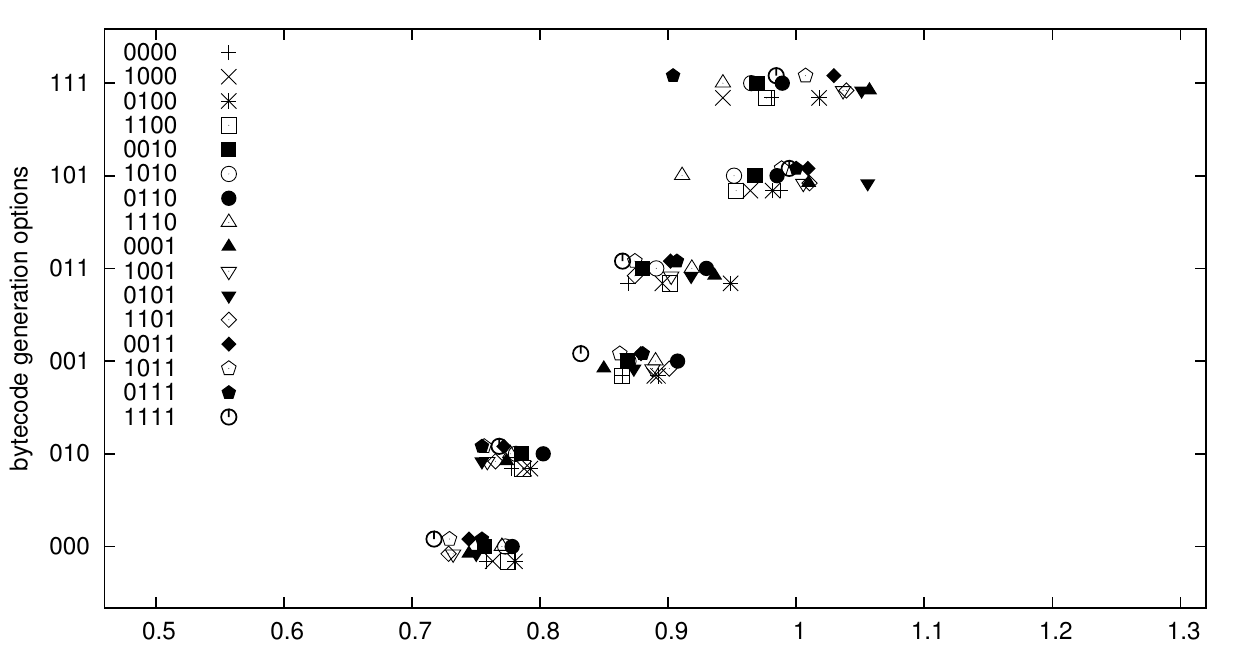}{Chess knights tour}{fig:knights-x86}
\intelpict{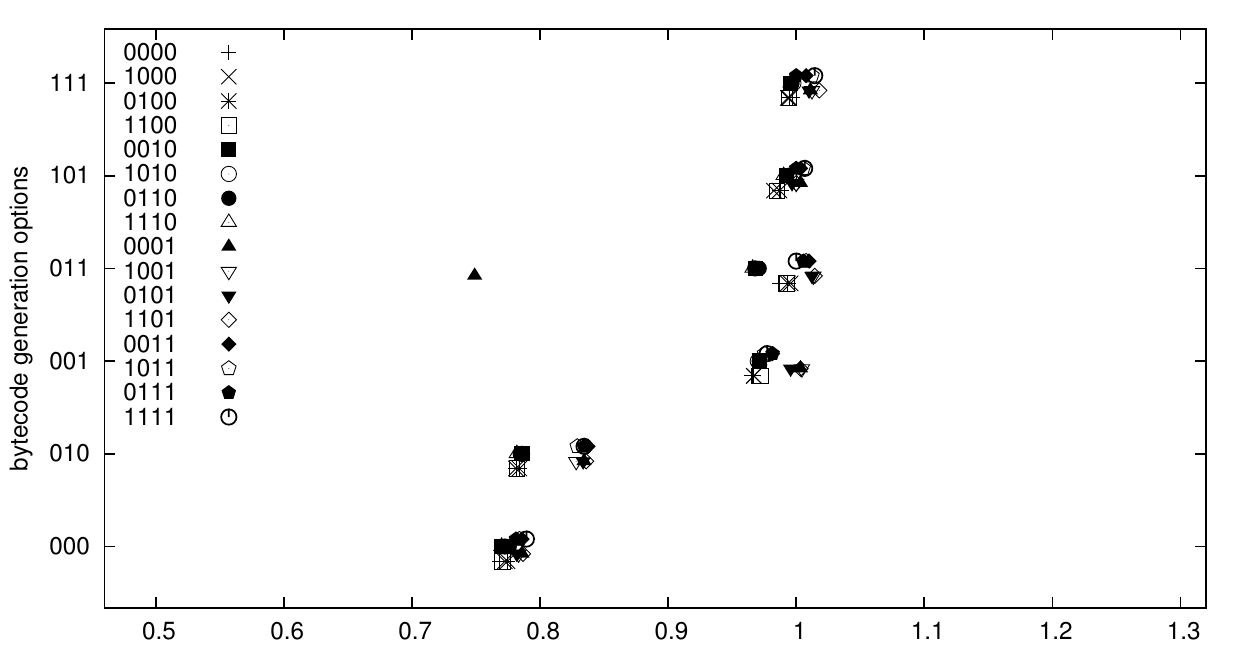}{Simply recursive Fibonacci}{fig:fib-x86}

\subsubsection{A More Detailed Inspection of Selected Cases}
\label{sec:detailed-inspection}

Figures~\ref{fig:queens11-x86} (Queens 11) and~\ref{fig:crypt-x86}
(Cryptoarithmetic puzzle) show two cases of interest.
The former corresponds to results which, while departing from the
average behavior, still resemble it in its structure,
although there is a combination of options which achieves a speedup
(around 1.25) that is significantly higher than average.
Figure~\ref{fig:crypt-x86} shows a different landscape where variations on
the code generation scheme appear to be as relevant as those on the
bytecode itself.  Both benchmarks are, however, search-based programs
which perform mainly arithmetic operations (with the addition of
some data structure management in the case of the Queens program), and
could in principle be grouped in the same class of programs.
This points to the need to perform a finer grain analysis to
determine, instruction by instruction, how \mbox{every}
engine/bytecode generation option affects execution time, and also how
these different options affect each other.

Studying which options are active inside each cluster sheds some light
about their contribution to the overall speedup.  For example,
the upper four clusters of Figures~\ref{fig:geom-mean-x86}
and~\ref{fig:arith-mean-x86} have in common the use of the
\emph{\textbf{ib}} option, which generates specialized instructions
for built-ins.  These clusters have consistently better (and, in some
cases, considerably better) speedups than the clusters which do not
have it activated.  It is, therefore, a candidate to be part of the
set of ``best options.'' 
A similar pattern, although less acute, appears in the results of the
PowerPC experiments (Figures~\ref{fig:geom-mean-ppc}
and~\ref{fig:arith-mean-ppc}).

The two leftmost clusters of the group of four at the bottom
correspond to exe\-cu\-tions of emulators generated with the
\emph{\textbf{rw}} specialization activated, and the two clusters at
their right do not have it activated.
It can come as a surprise that using separate switches for read/write
modes, instead of checking the mode in every instruction which needs
to do so, does not seem to bring any advantage in the Intel processor.
Indeed, a similar result was already observed
in~\cite{demoen00:wam-variations}, and was attributed to modern
architectures performing branch prediction and speculative work with
redundant units. 

\ppcpicttwo{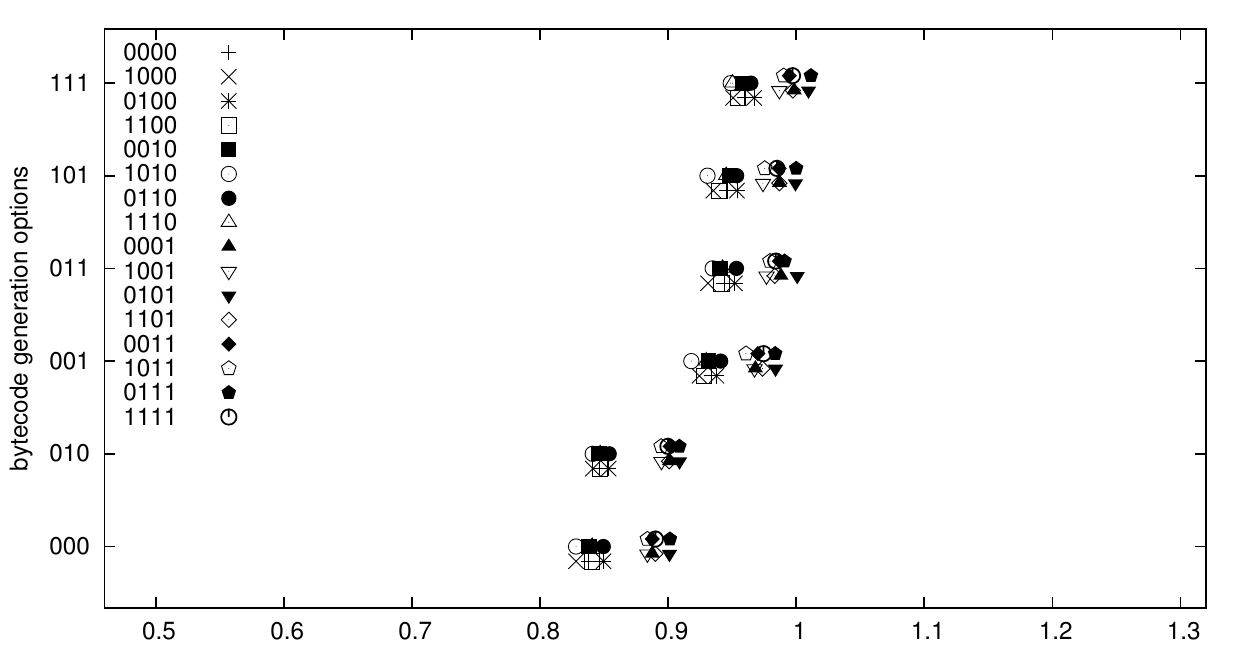}
{Geometric average of all benchmarks (with a dot per
  emulator)}{fig:geom-mean-ppc}
{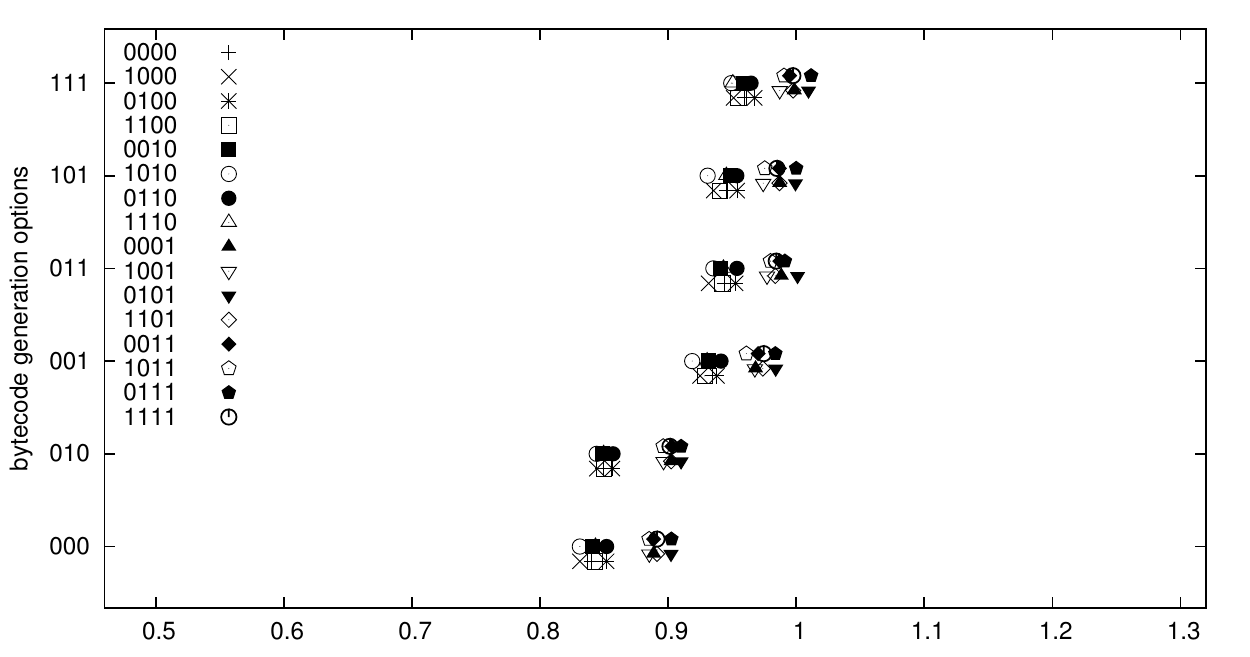} {Arithmetic average of all benchmarks (with a dot
  per emulator)}{fig:arith-mean-ppc}

\ppcpicttwo{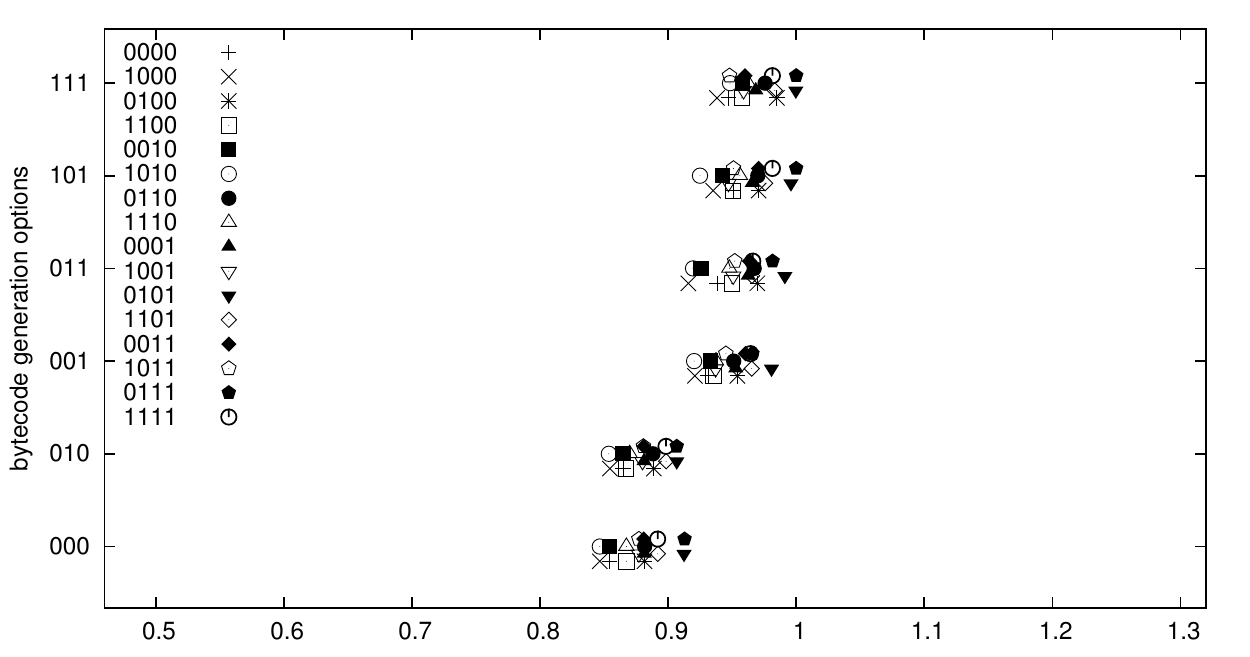}{Factorial involving large
  numbers}{fig:factorial-ppc}
{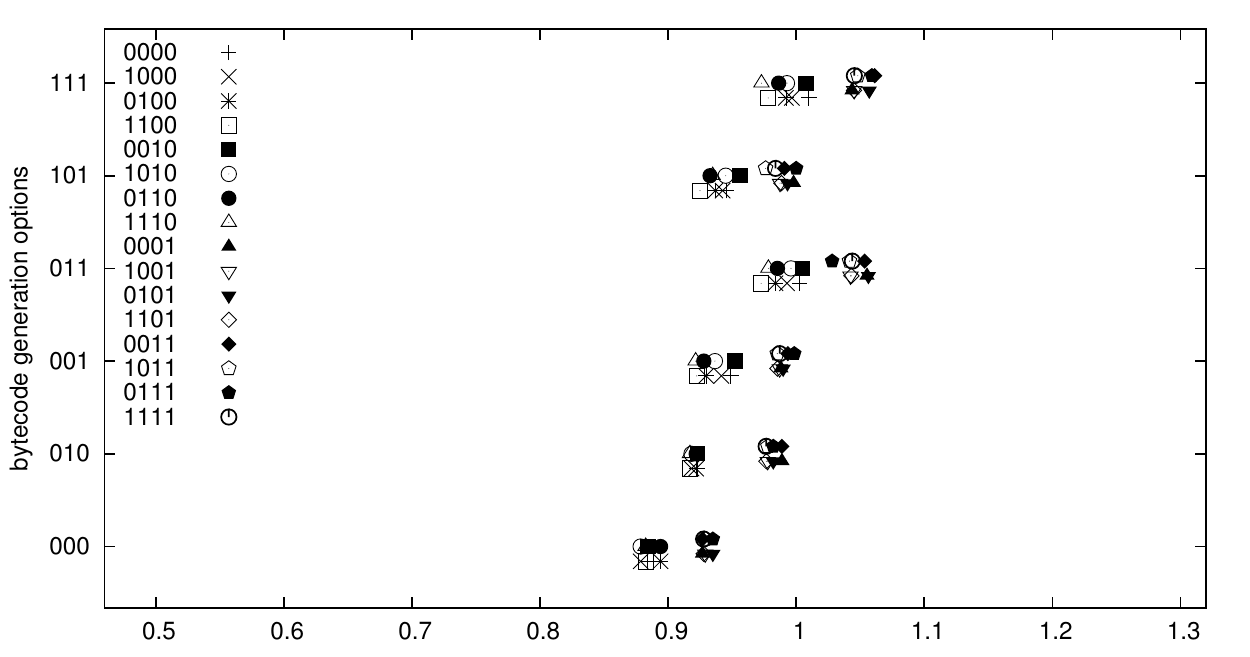}{Queens (with 11 queens to place)}{fig:queens11-ppc}

\ppcpicttwo{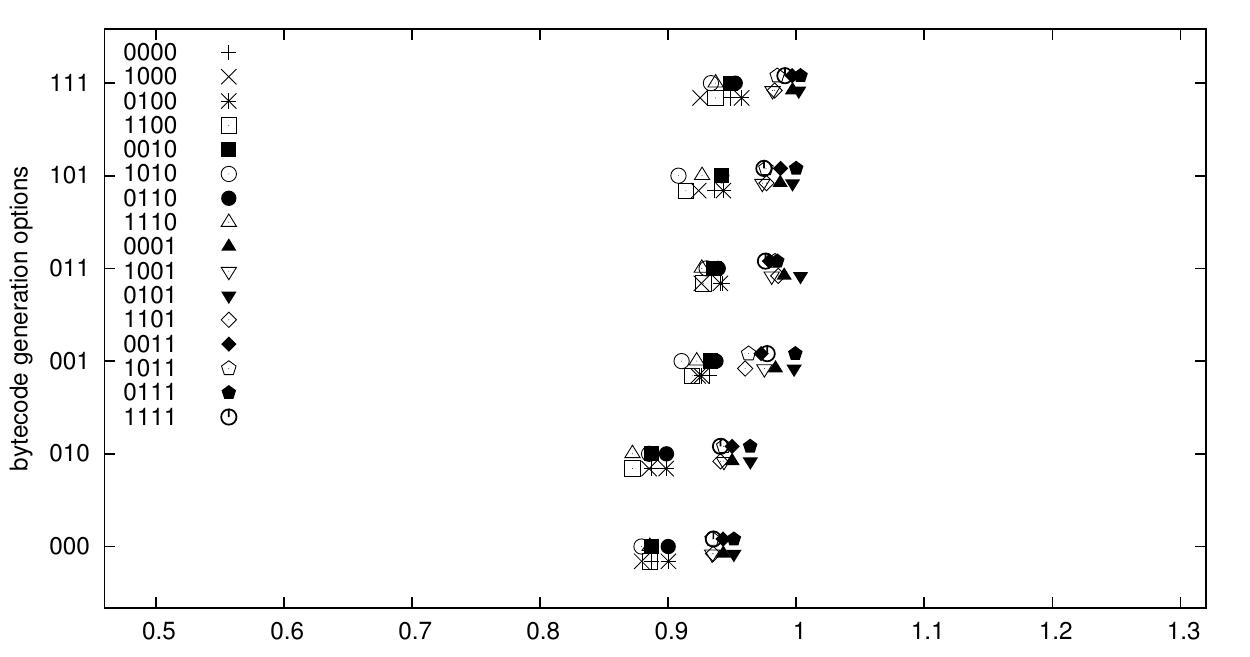}{Cryptoarithmetic puzzle}{fig:crypt-ppc}
{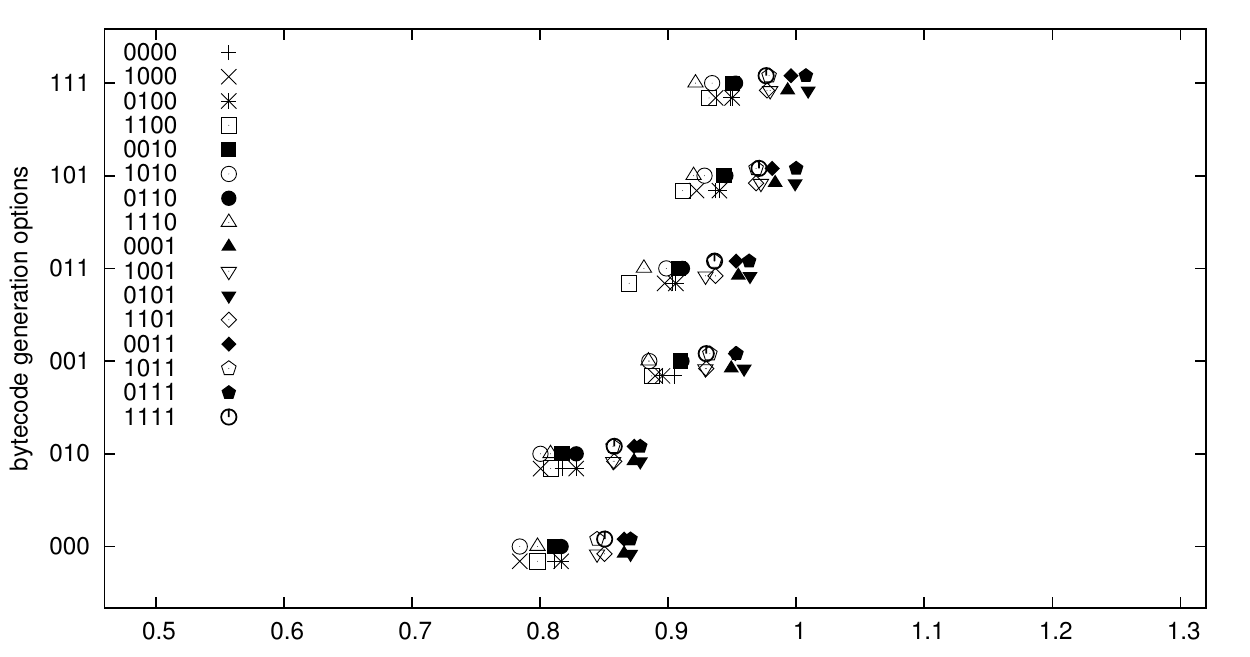}{Computation of the Takeuchi function}{fig:tak-ppc}

\ppcpicttwo{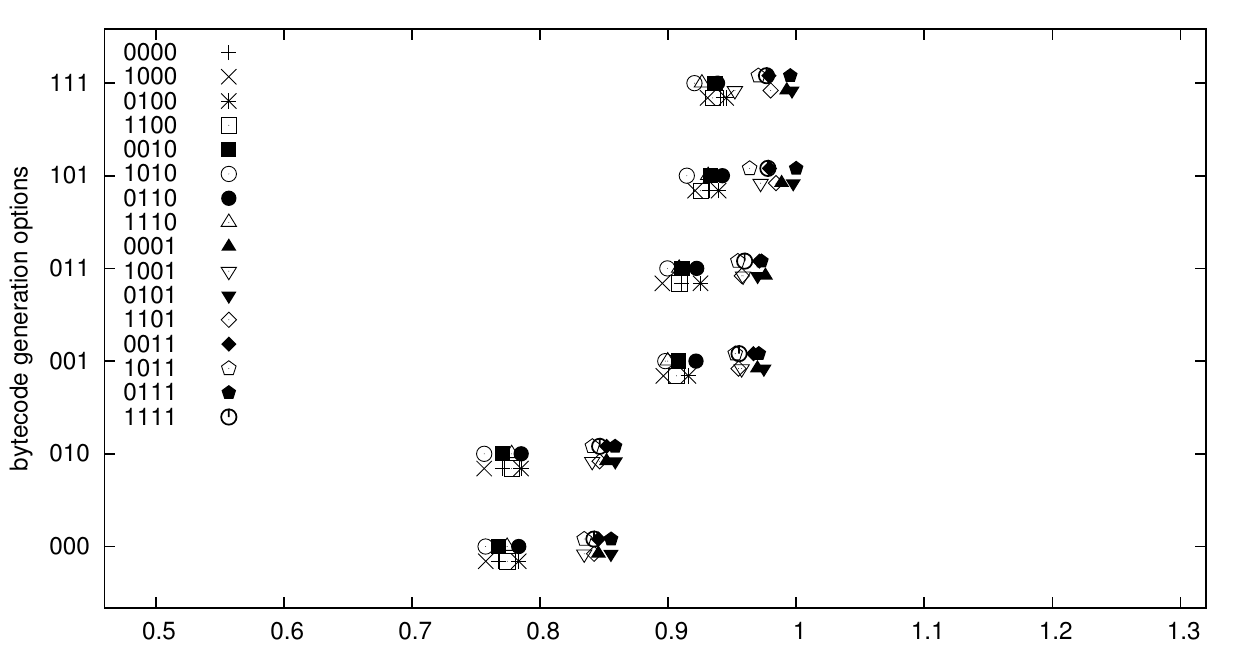}{Symbolic derivation of polynomials}{fig:deriv-ppc}
{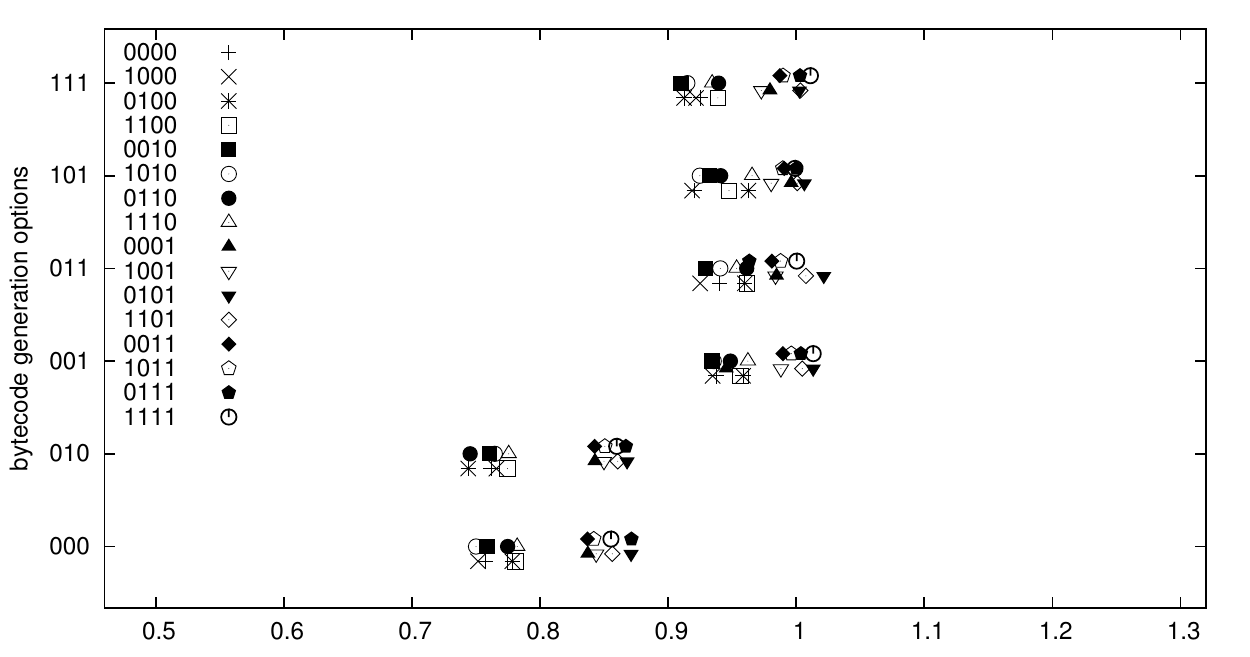}{Naive reverse}{fig:nrev-ppc}

\ppcpicttwo{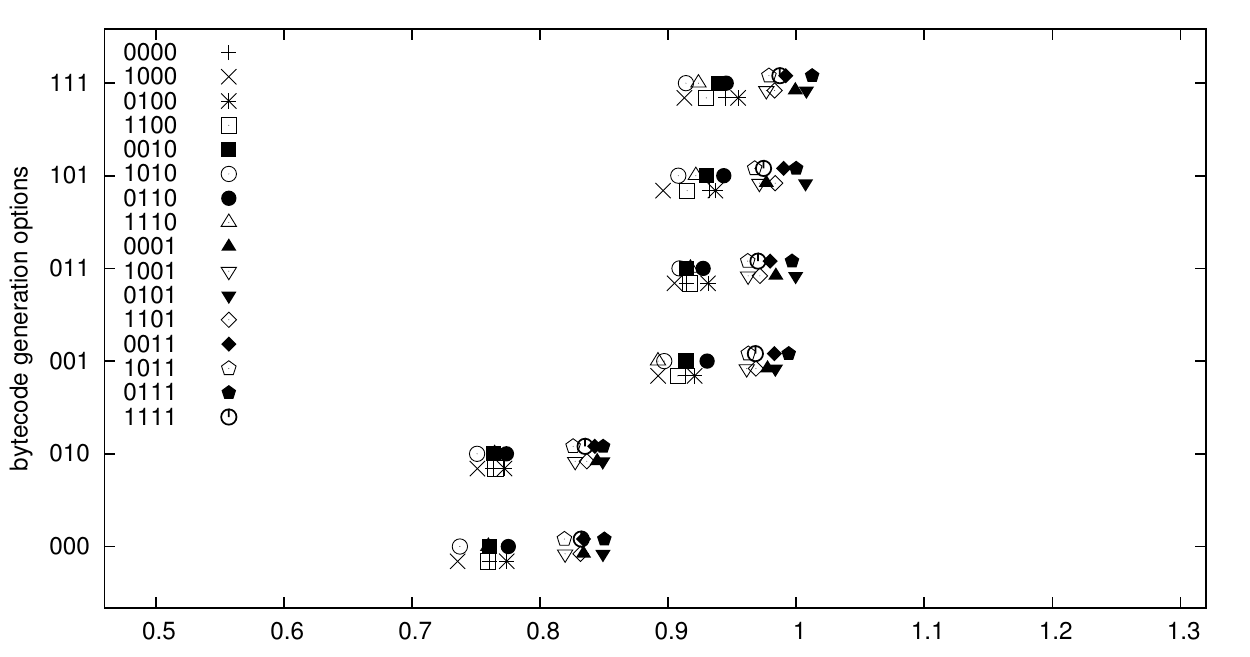}{Symbolic exponentiation of a polynomial}{fig:poly-ppc}
{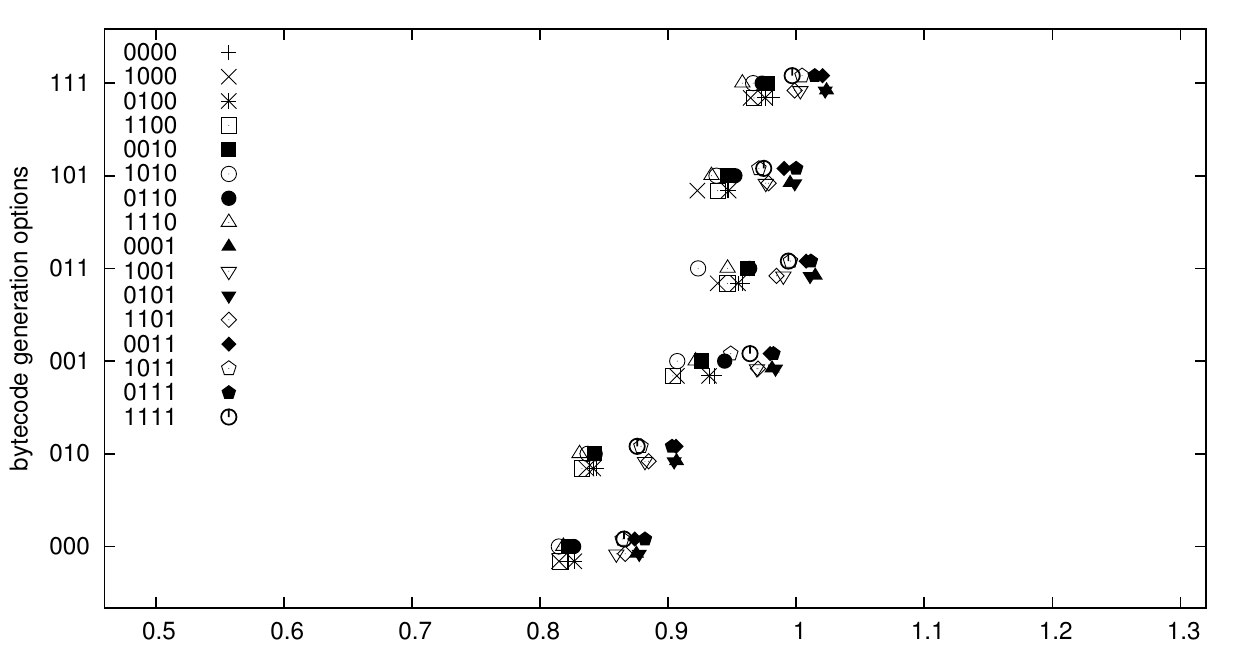}{Version of Boyer-Moore theorem prover}{fig:boyer-ppc}

\ppcpicttwo{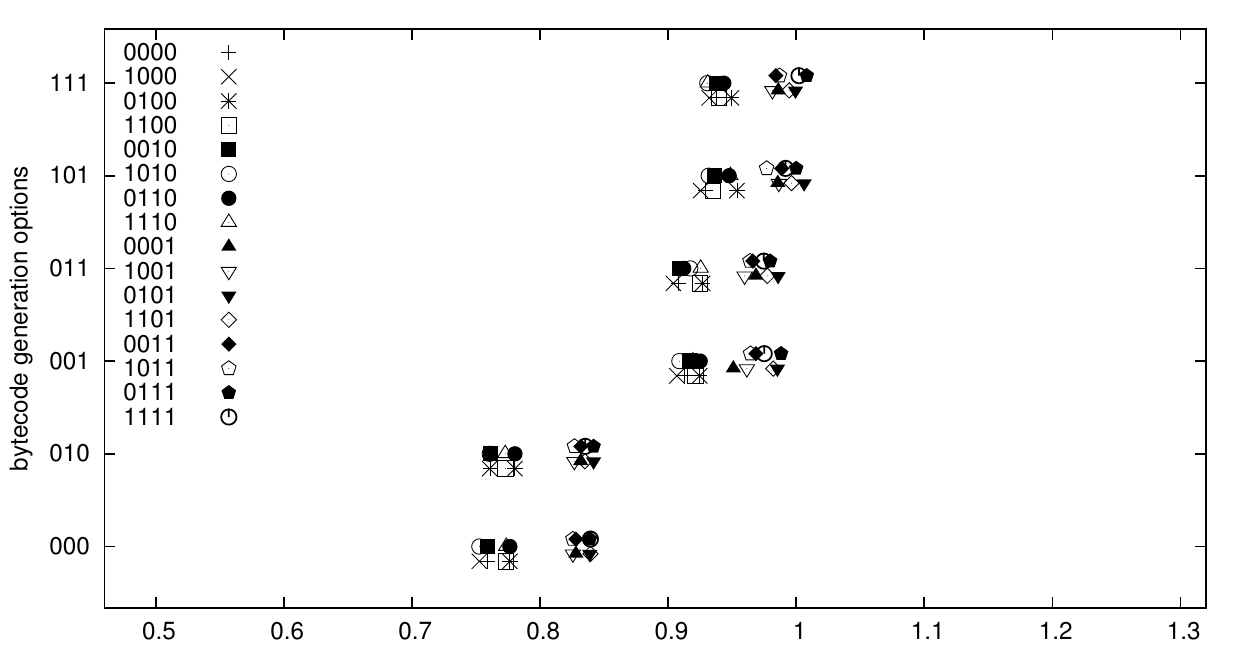}{QuickSort}{fig:qsort-ppc}
{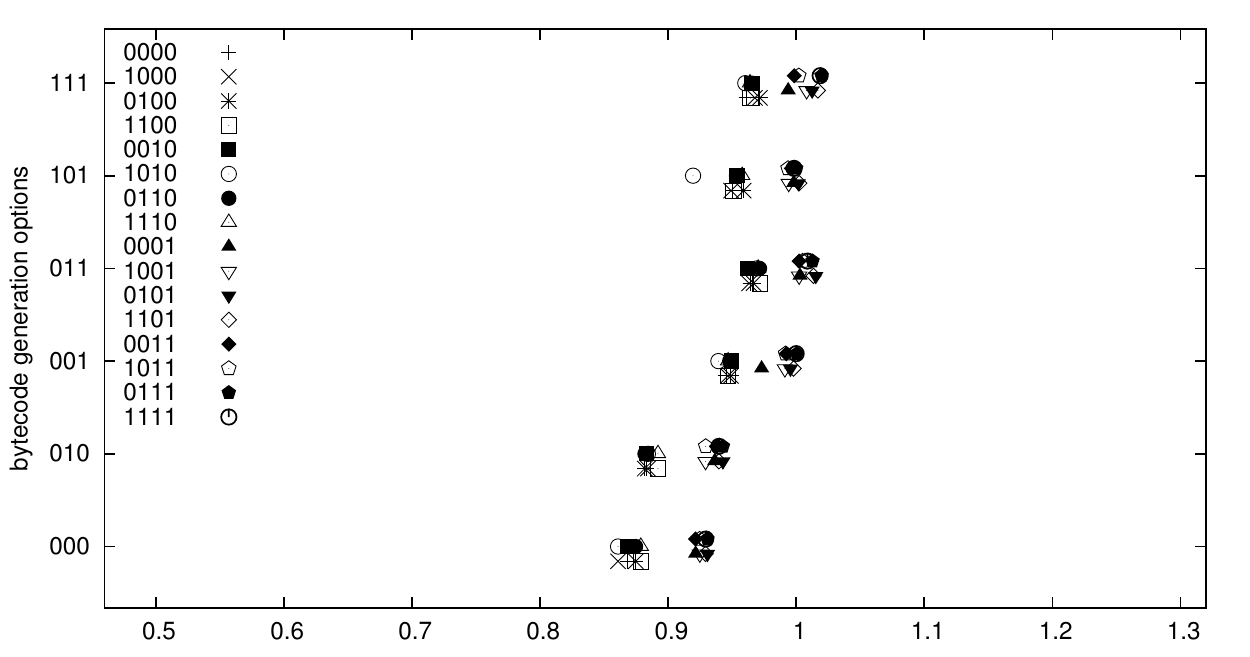}{Calculate primes using the sieve of Eratosthenes}{fig:primes-ppc}

\ppcpicttwo{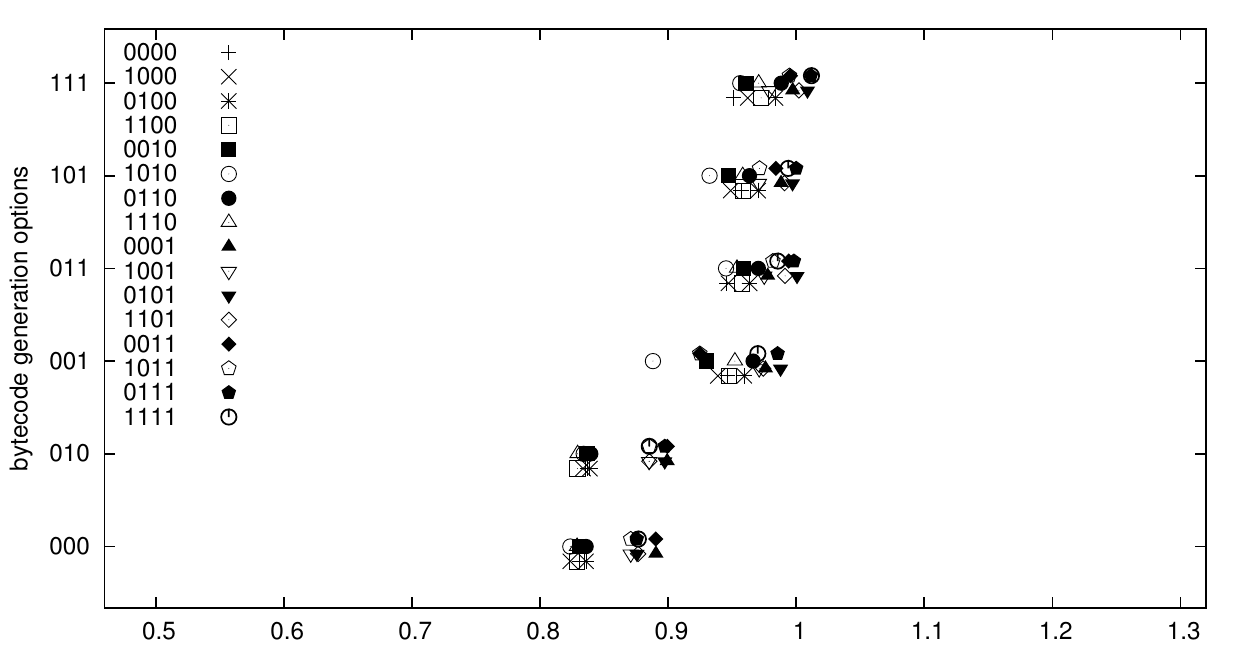}{Natural language query to a geographical database}{fig:query-ppc}
{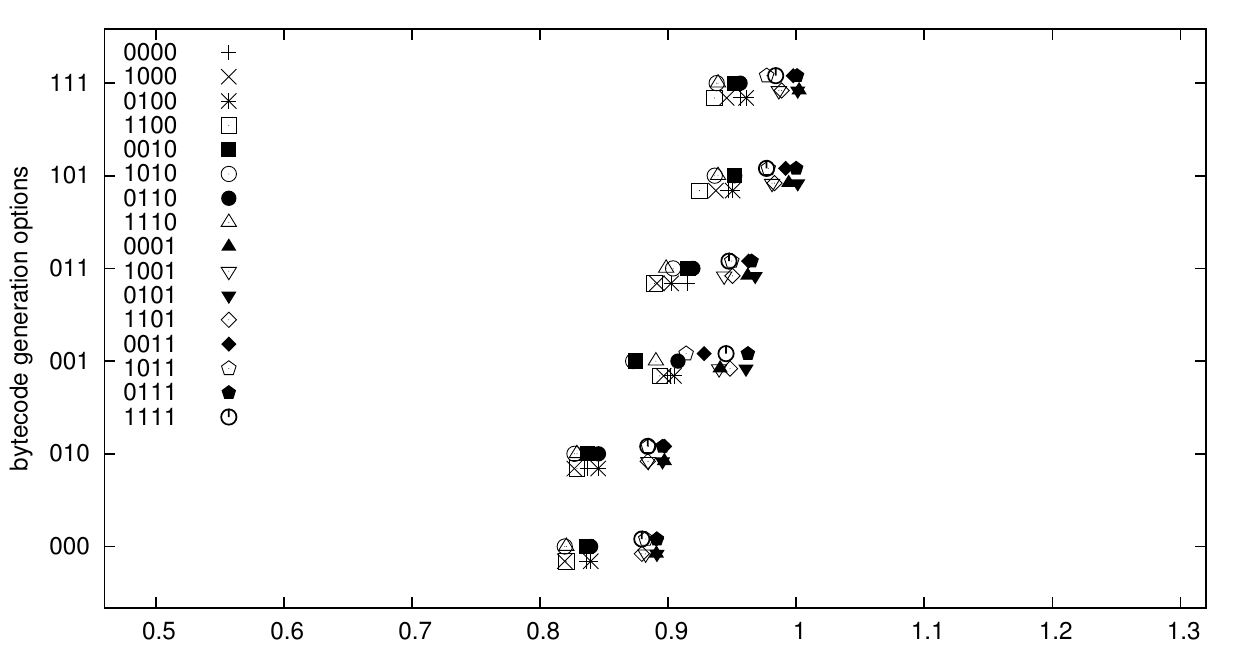}{Chess knights tour}{fig:knights-ppc}

\ppcpict{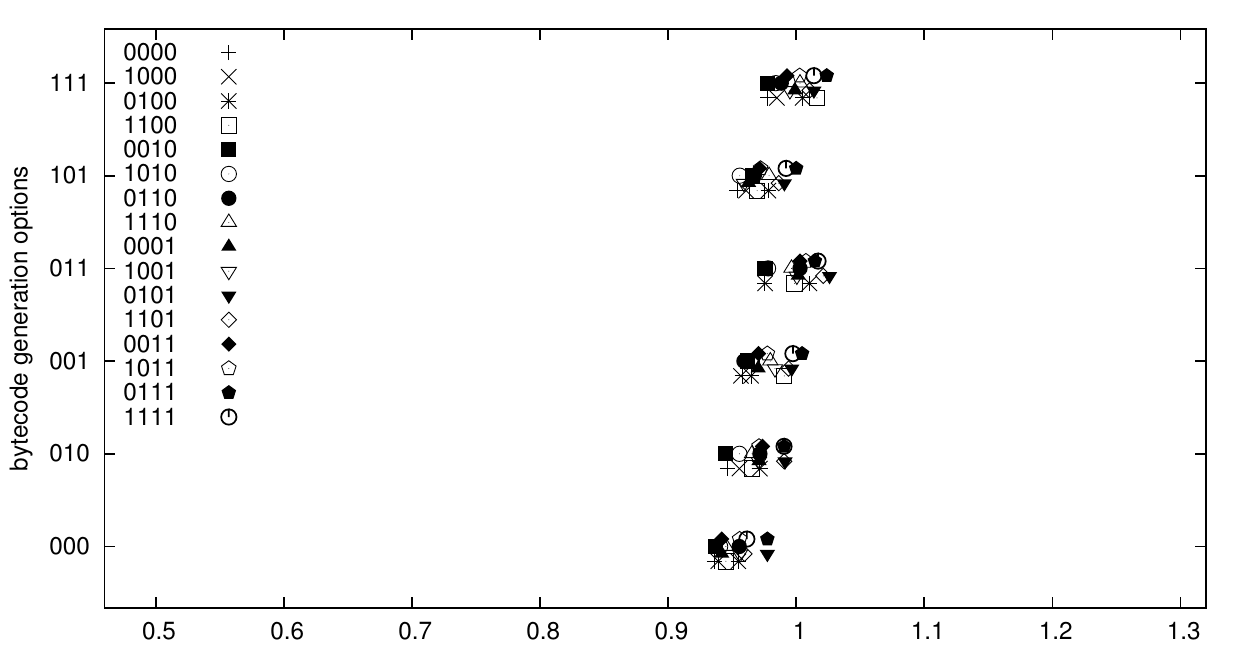}{Simply recursive Fibonacci}{fig:fib-ppc}

\begin{table}[t]
  \centering
  \begin{tabular}{|l||ccc|cccc|cc|}
    \cline{1-10}
    \raisebox{-2ex}[0em][-2ex]{\textbf{Benchmark}}
    & \multicolumn{9}{c|}{\textbf{Best performance}} \\\cline{2-10}
    & \textbf{ie} & \textbf{ib} & \textbf{im} & \textbf{ts} &
    \textbf{cc} & \textbf{ur} & \textbf{rw} & 
    \multicolumn{2}{c|}{\textbf{Speed-up}}\\
    \cline{1-8}
    \emph{baseline} & x &   & x &   & x & x & x & w.r.t.\ def. & w.r.t.\ base \\
    \cline{1-10}
    \emph{all} (geom.) & x & x & x &   & x &   & x & 1.05 & 1.28 \\
    \cline{1-10}
    boyer     & x &   & x &   & x &   & x & 1.18 & 1.52 \\
    crypt     &   & x & x & x &   &   &   & 1.22 & 1.07 \\
    deriv     & x & x & x &   & x &   &   & 1.10 & 1.46 \\
    factorial & x &   & x &   &   &   & x & 1.02 & 1.21 \\
    fib       & x & x & x & x & x &   & x & 1.02 & 1.32 \\
    knights   & x & x & x &   &   &   & x & 1.06 & 1.39 \\
    nreverse  & x & x & x &   &   &   & x & 1.03 & 1.34 \\
    poly      & x & x & x &   & x &   & x & 1.02 & 1.52 \\
    primes    & x &   & x &   & x &   & x & 1.10 & 1.26 \\
    qsort     &   & x & x & x &   &   & x & 1.05 & 1.46 \\
    queens11  & x & x & x & x & x & x & x & 1.26 & 1.46 \\
    query     &   & x & x & x & x & x & x & 1.06 & 1.21 \\
    tak       & x & x & x & x &   &   &   & 1.23 & 1.62 \\
    \cline{1-10}
    \multicolumn{10}{}{} \\
    \multicolumn{10}{}{} \\
    \cline{1-10}
    \raisebox{-2ex}[0em][-2ex]{\textbf{Benchmark}}
    & \multicolumn{9}{c|}{\textbf{Worst performance}} \\\cline{2-10}
    & \textbf{ie} & \textbf{ib} & \textbf{im} & \textbf{ts} &
    \textbf{cc} & \textbf{ur} & \textbf{rw} &
    \multicolumn{2}{c|}{\textbf{Speed-up}}\\
    \cline{1-8}
    \emph{baseline} & x &   & x &   & x & x & x & w.r.t. def. & w.r.t. base \\
    \cline{1-10}
    \emph{all} (geom.) &   &   &   &   & x &   & x & 0.70 & 0.88 \\
    \cline{1-10}
    boyer     &   &   &   &   &   &   & x & 0.70 & 0.90 \\
    crypt     &   &   &   & x &   &   & x & 0.86 & 0.75 \\
    deriv     &   & x &   &   &   &   & x & 0.62 & 0.82 \\
    factorial &   &   &   &   &   & x & x & 0.76 & 0.99 \\
    fib       &   & x & x &   &   &   & x & 0.75 & 0.91 \\
    knights   &   &   &   & x & x & x & x & 0.72 & 0.97 \\
    nreverse  &   & x &   &   &   & x & x & 0.57 & 0.95 \\
    poly      &   & x &   &   &   &   & x & 0.56 & 0.74 \\
    primes    &   &   &   &   & x & x & x & 0.73 & 0.84 \\
    qsort     &   &   &   &   & x &   & x & 0.54 & 0.84 \\
    queens11  &   &   &   &   & x & x & x & 0.77 & 0.75 \\
    query     &   &   &   & x &   &   & x & 0.71 & 0.89 \\
    tak       &   &   &   & x & x & x & x & 0.69 & 0.92 \\
    \cline{1-10}
  \end{tabular}
  \caption{Options which gave best/worst performance (x86).}
  \label{tab:performance-x86}
\end{table}

\begin{table}[t]
  \centering
  \begin{tabular}{|l||ccc|cccc|cc|}
    \cline{1-10}
    \raisebox{-2ex}[0em][-2ex]{\textbf{Benchmark}}
    & \multicolumn{9}{c|}{\textbf{Best performance}} \\\cline{2-10}
    & \textbf{ie} & \textbf{ib} & \textbf{im} & \textbf{ts} &
    \textbf{cc} & \textbf{ur} & \textbf{rw} &
    \multicolumn{2}{c|}{\textbf{Speed-up}}\\
    \cline{1-8}
    \emph{baseline} & x &   & x &   & x & x & x & w.r.t.\ def. & w.r.t.\ base \\
    \cline{1-10}
    \emph{all} (geom.) & x & x & x &   & x & x & x & 1.01 & 1.21 \\
    \cline{1-10}
    boyer     & x & x & x &   &   &   & x & 1.02 & 1.25 \\
    crypt     &   & x & x &   & x &   & x & 1.00 & 1.13 \\
    deriv     & x &   & x &   & x & x & x & 1.00 & 1.30 \\
    factorial & x &   & x &   & x & x & x & 1.00 & 1.02 \\
    fib       &   & x & x &   & x &   & x & 1.03 & 1.17 \\
    knights   & x & x & x &   &   &   & x & 1.00 & 1.10 \\
    nreverse  &   & x & x &   & x &   & x & 1.02 & 1.20 \\
    poly      & x & x & x &   & x & x & x & 1.01 & 1.35 \\
    primes    & x & x & x &   & x & x & x & 1.02 & 1.33 \\
    qsort     & x & x & x &   & x & x & x & 1.01 & 1.17 \\
    queens11  & x & x & x &   &   & x & x & 1.06 & 1.33 \\
    query     & x & x & x & x & x & x & x & 1.01 & 1.20 \\
    tak       & x & x & x &   & x &   & x & 1.01 & 1.22 \\
    \cline{1-10}
    \multicolumn{10}{}{} \\
    \multicolumn{10}{}{} \\
    \cline{1-10}
    \raisebox{-2ex}[0em][-2ex]{\textbf{Benchmark}}
    & \multicolumn{9}{c|}{\textbf{Worst performance}} \\\cline{2-10}
    & \textbf{ie} & \textbf{ib} & \textbf{im} & \textbf{ts} &
    \textbf{cc} & \textbf{ur} & \textbf{rw} &
    \multicolumn{2}{c|}{\textbf{Speed-up}}\\
    \cline{1-8}
    \emph{baseline} & x &   & x &   & x & x & x & w.r.t.\ def. & w.r.t.\ base \\
    \cline{1-10}
    \emph{all} (geom.) &   &   &   & x &   &   &   & 0.82 & 0.99 \\
    \cline{1-10}
    boyer     &   &   &   & x &   & x &   & 0.81 & 0.99 \\
    crypt     &   & x &   & x & x & x &   & 0.87 & 0.98 \\
    deriv     &   & x &   & x &   & x &   & 0.76 & 0.99 \\
    factorial &   &   &   & x &   & x &   & 0.85 & 0.97 \\
    fib       &   &   &   &   &   &   &   & 0.94 & 0.99 \\
    knights   &   &   &   & x &   & x &   & 0.82 & 1.00 \\
    nreverse  &   & x &   &   & x &   &   & 0.74 & 0.98 \\
    poly      &   &   &   & x &   &   &   & 0.74 & 0.98 \\
    primes    &   &   &   & x &   & x &   & 0.86 & 0.97 \\
    qsort     &   &   &   & x &   &   &   & 0.75 & 0.99 \\
    queens11  &   &   &   & x &   &   &   & 0.88 & 0.99 \\
    query     &   &   &   & x &   &   &   & 0.82 & 0.99 \\
    tak       &   &   &   & x &   & x &   & 0.78 & 0.99 \\
    \cline{1-10}
  \end{tabular}
  \caption{Options which gave best/worst performance (PowerPC).}
  \label{tab:performance-ppc}
\end{table}

\subsubsection{Best Generation Options and Overall Speedup }
\label{sec:overall-best}

An important general question is \emph{which options should be used
  for the ``stock'' emulator to be offered to general users}.
Our experimental results show that
options cannot be considered in isolation --- i.e., the overall option
set constructed by taking separately the best value for every option
does not yield a \emph{better set} (defined as the best options
obtained by averaging speedups for every option set).  As we have
seen, there is some interdependence among options.
A more realistic answer is that the average best set of options should
come from selecting the rightmost point in the plot corresponding to
average speedups.  We must however bear in mind that averages always
suffer the problem that a small set of good results may bias the
average and, in this case, force the selection of an option set which
performs worse for a larger set of benchmarks

In order to look more closely at the effects of individual options
(without resorting to extensively listing them and the obtained
performance), Tables~\ref{tab:performance-x86}
and~\ref{tab:performance-ppc} show which options produced the best and
the worst results time-wise for each benchmark.  We include the
geometric average as an specific case and the Ciao-1.10 baseline options
as reference.

It has to be noted that the best/worst set of
options is not the negation of the worst/best options:
there are plenty of cases where
the same option was (de)activated both for the best and for the worst
executions.  The observed situation for the PowerPC architecture
(Table~\ref{tab:performance-ppc}) is more homogeneous: at least some
better/worst behaviors really come from combinations which are
complementary, and, in the cases where this is not so, the amount of
non-complementary options goes typically from 1 to 3 --- definitely
less than in the x86 case.

Despite the complexity of the problem, some conclusions can be drawn:
instruction
merging (\textbf{\emph{im}}) is a winner for the x86, probably
followed by having a variable number of operands (\textbf{\emph{ie}}),
and then by specialized calls to built-ins (\textbf{\emph{ib}}).  The
first and second options save fetch cycles, while the third one saves
processing time in general.  It is to be noted that some options
appear both in the best and worst cases: this points to
interdependencies among the different options.

The performance table for the PowerPC
(Table~\ref{tab:performance-ppc}) also reveals that instruction
merging, having a variable number of operands, and generating
specialized instructions for built-ins, are options which bring
performance advantages.  However, and unlike the x86 table, the
\emph{read/write mode} specialization is activated in all the lines of
the ``best performance'' table, and off in the ``worst performance.'' 
A similar case is that of the \emph{tag switching schema}, in the
sense that the selection seems clear in the PowerPC case.

The transformation rules we have applied in our case are of course not
the only possible ones, and we look forward to enlarging this set of
transformations by, for example, performing a more aggressive merging
guided by profiling.\footnote{Merging is right now limited in depth to
  avoid a combinatorial explosion in the number of instructions.}
Similar work, with more emphasis on the production of languages for
microprocessors is presented in~\cite{holmer93:design_inst_sets},
where a set of benchmarks is used to guide the (constrained) synthesis
of such a set of instructions.

We want to note that although exceeding the speed of a hand-crafted
emulator is not the main concern in this work,\footnote{In order to do
  that, a better approach would probably be to start off by finding performance
  bottlenecks in the current emulator and redesigning / recoding it.
  We want to note, however, that we think that our approach can
  greatly help
  in making this redesign and recoding easier.} the performance
obtained by the implementation of the emulator in \ip\
allows us to conclude that the \ip\ approach can match the performance of lower-level
languages, while making it possible to apply non-trivial program
transformation techniques.

Additional experiments carried out in a very different context (that
of embedded systems and digital signal processing using Ciao
Prolog~\cite{carro06:stream_interpreter_cases}, which pertains to a
realm traditionally considered disadvantageous for symbolic languages)
showed also very good performance ---only 20\% slower than a
comparable C program--- and also very good speedups (up to 7-fold
compared with a bytecode-based implementation).
Analysis and compilation techniques similar to those applied in this
paper were used, but put to work in a program using the full Prolog
language.

\section{Conclusions}
\label{sec:conclusions}

We have designed a language (\ip, a variation of Prolog with some
imperative features) and its compiler, and we have used this language to
describe the semantics of the instructions of a bytecode interpreter
(in particular, the Ciao engine).
The \ip\ language, with the proposed constraints and
extensions, 
is semantically close enough to Prolog to share analysis,
optimization and compilation techniques, but at the same time 
it is designed to make translation into very efficient C code possible.
The low-level code for each instruction and the definition of the
bytecode is taken as input by a previously developed emulator
generator to assemble full high-quality emulators.  Since the process
of generating instruction code and bytecode format is automatic, we
were able to produce and test different versions thereof to which
several combinations of code generation options were applied.

Our main conclusion is that indeed the proposed approach can produce
emulators that are as efficient as the best state-of-the-art emulators
hand-coded in C, while starting from a much higher-level
description. This high-level nature of the language used allows
avoiding some of the typical problems that hinder the development and
maintenance of highly-tuned emulators.

In our experience, in addition to greatly increased clarity and
readability the port allowed replacing many duplicated code
structures, redundant hand-made specializations, and a large number of
C macros (which notwithstanding do help in writing less code, but they
are not easily recognized in the source --leading often to programmer
confusion--, they are prone to subtle scoping-related errors, and they
are typically not understood by automatic compilation tools), with
more abstract predicates, sometimes adorned with some annotations to
guide the transformations.

The proposed approach makes it possible to perform non-trivial
transformations on both the emulator and the instruction level (e.g.,
unfolding and partial evaluation of instruction definitions,
instruction merging or specialization, etc.).
The different transformations and code generation options, result in
different grades of optimization / specialization and different
bytecode languages from a single (higher-level) abstract machine
definition.

We have also studied how these combinations perform with a series of
benchmarks in order to find, e.g., what is the ``best'' average
solution and how independent coding rules affect the overall speed.
We have in this way as one case the regular emulator we started with
(and which was decomposed to break complex instructions into basic
blocks).  However, we also found out that it is possible to outperform
it slightly by using some code patterns and optimizations not explored in the
initial emulator, and, what is more important, starting from abstract
machine definitions in \ip.

Performance evaluation of non-trivial variations in the emulator code
showed that some results are hard to predict and that there is no
absolute winner for all architectures and programs. On the other hand,
it is increasingly difficult to reflect all the variations in a single
source using more traditional methods like \texttt{m4} or \texttt{cpp}
macros. Automatic program manipulation at the emulator level
represents a very attractive approach, and although difficult, the
problem becomes more tractable when the abstraction level of the
language to define the virtual machine is raised and enriched with
some problem-specific declarations.

As future work, it would be interesting to study the generation of
efficient code for full Prolog with \ip\ features.  A partial
application of such compilation techniques was already put to work
in the real-time digital sound processing application written in
Prolog mentioned before, with very successful results.

\section{Acknowledgments}
\label{sec:acks}

\paragraph{\textbf{Acknowledgments:}} 

This work was funded in part by the Information Society Technologies
program of the European Commission, through EU project FP7 318337
\emph{ENTRA}, by the Spanish Ministry of Economy and Competitiveness
through projects TIN2012-39391 \emph{StrongSoft} and TIN2008-05624
\emph{DOVES}, and by the Madrid Regional Government through project
S-0505/TIC/0407 {\em PROMESAS}.

\bibliographystyle{unsrt}

\end{document}